\documentclass[12pt]{iopart}
\usepackage{iopams}
\usepackage{graphicx}
\usepackage{cite}
\usepackage{mathrsfs}
\usepackage{setstack}
\renewcommand{\vec}[1]{\ensuremath{{\boldsymbol{#1}}}}
\newcommand{\stackunder}[2]{\underset{#1}{#2}}
\newcommand{\bra}[1]{\ensuremath{\langle #1 |}}
\newcommand{\ket}[1]{\ensuremath{| #1 \rangle}}
\newcommand{\braket}[2]{\ensuremath{\langle #1|#2 \rangle}}
\newcommand{\x}{{\vec{\rm x}}}
\newcommand{\ka}{{\vec{\rm k}}}
\newcommand{\y}{{\vec{\rm y}}}
\newcommand{\q}{{\vec{\rm q}}}
\newcommand{\p}{{\vec{\rm p}}}
\newcommand{\us}{{\vec{\rm u}}}
\newcommand{\vs}{{\vec{\rm v}}}
\newcommand{\n}{{\vec{\rm n}}}

\newcommand{\lv}{{\vec{\rm l}}}

\newcommand{\Pop}{{\mbox{\bf P}}}  
\newcommand{\Xop}{{\mbox{\bf X}}}  
\newcommand%
{\MVV}%
[1]%
{{\langle\!\langle #1\rangle\!\rangle}}%
\newcommand%
{\MMV}%
[1]%
{{\left\langle\!\!\!\left\langle #1\right\rangle\!\!\!\right\rangle}}%
\newcommand{\bea}{\begin{eqnarray}}
\newcommand{\eea}{\end{eqnarray}}
\eqnobysec
\begin{document}
\title[Interpolating wave packets and composite wave functions]
{Interpolating wave packets and composite wave functions in QFT and neutrino oscillation 
problem}
\author{S. E. Korenblit$^{1,2}$, 
D. V. Taychenachev$^{1}$, \\ and M. V. Petropavlova$^{1,2}$ }

\address{$^{1}$Department of Physics, Irkutsk State University, 20 Gagarin blvd,  
RU-664003, Irkutsk, Russia.} 

\address{$^{2}$Dzhelepov Laboratory of Nuclear Problems, Joint Institute for Nuclear 
Research, RU-141980, Dubna, Russia.}

\ead{korenb@ic.isu.ru} 
\begin{abstract}   
A consistent constructive covariant description of neutrino flavour transition 
amplitude in vacuum is presented. 
To this end a special generalized relativistic wave packet is constructed with correct 
extension onto the higher spins.  
This packet is uniquely defined as an `interpolating' wave packet, which by means of 
relativistically invariant `width' accurately interpolates between the states localized 
in momentum space and in coordinate space. 
The wave packet is unambiguously determined by analytical properties of Wightman 
functions in complex coordinate space naturally connected with its minimization properties. 
The packet gives natural relativistic generalization of non relativistic Gaussian 
wave packet but it contains covariant states of particle (antiparticle) only with positive 
(negative) energy sign and propagates without their mixing and without changing of its 
relativistically invariant width. 
An application to the intermediate wave-packet picture of neutrino oscillations is given with 
the generalized expression for two-flavour oscillations of leptonic charge of electronic 
neutrino. 
For the diagrammatic treatment of oscillation with the use of these wave packets for 
external particles, the notion of covariant composite wave function for intermediate 
neutrino is introduced. It strictly and naturally connects both oscillation 
pictures, giving an effective language for detailed description of this process, and 
resolves the problems with causality and with covariant equal time prescription for 
the intermediate neutrino picture. 
It is closely related to overlap function of neutrino creation/detection vertices,   
elucidating a covariant meaning of the `pole integration' procedure, and possess an exact 
one-packet covariant integral representation for on-shell case, as superposition of packets 
with only different widths and centers. 
The time asymptotic regime already reduces the off-shell composite wave function to 
the on-shell one, and their space-time asymptotics regimes coincide. 
Their space-time asymptotic behaviour in narrow-packets approximation naturally conforms 
with such approximation of one-packet state and with the asymptotic behaviour of 
oscillation amplitude. 
The respective overlap function is explicitly calculated for two-packet example of pion 
decay vertex. Its correspondence and difference with previous approximate calculations 
is analyzed. 
\end{abstract}
\submitto{\JPG}
\maketitle
\section{Introduction}\label{sec:intro}

The notion of a wave packet permeates the entire construction of modern quantum 
mechanics (QM) and quantum field theory (QFT) 
\cite{nn,nsh,dn,n_sh,NN_shk,kt_1,LL,Mess,Tl,g_w,feyn,blp,tir,p_s,vrg,schw,blt,oksak,jost,
s_w,stroc,Dvt,b_d,i_z,Nvj,Bil,Wnb}.
It plays a central role in achieving their mathematical rigor and provide their 
interpretation in terms of particles. 
Wave packets are necessary to obtain the weak asymptotic conditions for the in- or out-   
asymptotic fields and states, and for the use of LSZ formalism 
\cite{Tl,g_w,feyn,blp,tir,p_s,vrg,schw,blt,oksak,jost,s_w,stroc,Dvt,b_d,i_z,Nvj,Bil,Wnb}. 
The dominant paradigm governing the use of this notion came out from optics \cite{kil},  
in fact dealing with the wave packets for massless particles with infinite Compton 
wave length $\lambda=\hbar/(mc)$ only. It declares the arbitrariness of 
amplitude of profile function of wave packet in the suitable functional space 
$L^2(\vec{\rm R}^3), {\cal S}(\vec{\rm R}^3)$, etc. This is also enough for mathematical 
rigor \cite{blt,oksak,jost,s_w,stroc,Dvt} for the massive case but it is not enough for 
consistent physical interpretation \cite{nn,nsh,dn,n_sh,NN_shk,kt_1}. 

Indeed, the local nature of the basic quantum-field interactions, jointly with the 
hypothesis about their adiabatic switching on and switching off, allows in a 
final version of the asymptotic formalism of $S$- matrix for these interactions to 
replace wave packets of asymptotic states by the plane waves that are elements of equipped 
Hilbert space \cite{blt} for given QFT. 
Having been successfully applied to description of the vast majority of phenomena in 
particle physics 
\cite{feyn,blp,tir,p_s,vrg,schw,blt,oksak,jost,s_w,stroc,Dvt,b_d,i_z,Nvj,Bil,Wnb}, 
this formalism is not suitable for a consistent description of the phenomenon of neutrino 
oscillations in vacuum 
\cite{nn,nsh,dn,n_sh,NN_shk,kt_1,bern,gunt,rich,bth,ahm,Dl_1,Nis,kbz,GS,NN_anp,bGG,Dvor,
bl1,bl2,bl3,fujii} and surely for their oscillations in matter. 
This fact, being basically a manifestation of non trivial vacuum structure of QFT of 
electro-weak interactions and the existence of non equivalent representations for its 
(anti) commutation relations \cite{bl1,bl2,bl3}, reflects also an inability of adiabatic 
switching of the above mechanism of oscillations. 

As a consequence, the intriguing feature of the neutrino oscillation problem is the 
necessity to describe the phenomenon on two or more very different scales simultaneously. 
The first one is the microscopic scale of the process which fixes the effective points of 
the neutrino creation/detection. 
The second is the macroscopic base line $L$ as the distance separating these points.
The third scale is the coherence length that allows to observe the oscillation phenomenon. 
Since the description of the first scale implies the use of relevant Lorentz invariant and 
consequently local Fermi or electro-weak interaction, the exhaustive description of the whole 
problem should also be Lorentz invariant. 

One of the most convenient possibility is given by the method of macroscopic Feynman 
diagrams with covariant causal neutrino propagator as main ingredient 
\cite{nn,nsh,bth,ahm}.  
But the question arises about the covariance and localization of the initial and final 
states of external particles in the creation and detection vertices of this diagrams. 
It is well known \cite{gunt,bth,ahm} (and the references therein), that the use of plane 
wave for the states of external particles, unlike its previous use \cite{bGG} for the 
states of internal intermediate neutrino, fully disables the oscillation pattern. 
The reason is the smearing of plane wave over entire space-time \cite{ahm}. 
So, only localized or localizable states may be appropriate for external particles of 
macroscopic Feynman diagrams. There are two physicaly different possibilities of such 
states: the bound state \cite{GS} and the wave packet of free particle states 
\cite{nn,nsh,dn,n_sh,NN_shk,kt_1,bern,gunt,rich,bth,ahm,Dl_1,Nis}. 
Therefore, an attempt to describe the mechanism of oscillations on the language of 
ordinary $S$- matrix diagram technique requires to return the wave packet into the theory, 
as a such basic element of that description, as the plane wave previously was. 

Thus, trying to understand the neutrino oscillation, we face the problem of construction 
a consistent relativistic covariant analog \cite{nn,nsh,dn,n_sh,NN_shk,kt_1},
of Gaussian wave packet 
\cite{bern,gunt,rich,bth,ahm,Dl_1,Nis,kbz,GS,NN_anp,bGG,Dvor,bl1,bl2,bl3,fujii}, 
which in fact corresponds to the drastically changed meaning of Heisenberg uncertainty 
conditions in the relativistic QFT \cite{blp,tir,p_s,vrg,kil}. 
Furthermore, for some {\sl invariant parameter of width} $\sigma$ it happens that this 
relativistic wave packet, unlike the non relativistic one, {\sl directly interpolates} 
between the covariant state with definite 4-momentum at $\sigma\to 0$ and the covariant 
state with definite 4-coordinate in relativistically covariant sense at $\sigma\to\infty$. 
The inevitable appearance of {\sl two} independent invariant dimensional parameters 
for $a$- th particle: mass $m_a$ and width $\sigma_a$ gives rise to non trivial dimensionless 
function of their dimensionless combination $\tau_a\simeq(m_ac/\sigma_a)^\epsilon$, 
$\epsilon>0$. 

The aim of this paper is to demonstrate also, that up to this non trivial function in 
normalization factor, the constraints from relativistic QFT joined with minimization 
properties unambiguously determine such interpolating relativistic covariant 
wave-packet state with positive mass for arbitrary spin, and that the above mentioned 
limiting properties have unambiguously fix here the only 
important asymptotic behaviour of this dimensionless function in normalization factor. 
Whereas its remaining ambiguity defines the form of averages corresponding to observables. 
Constancy of invariant width $\sigma_a$ is evident merit of Lorentz-covariant propagation. 

In sections \ref{sec:wp_QM}, \ref{sec:wp_sf_QFT}, \ref{sec:wp_ff} we show, how the different 
role of time and the different meaning of localization in QM and QFT leads to corresponding 
difference between localizable states in QM and QFT. 
In section \ref{sec:wp_QM} we remind the properties of wave packets in quantum 
mechanic. In section \ref{sec:wp_sf_QFT} the relativistic Lorentz-covariant interpolating 
wave packet for massive (pseudo) scalar field is constructed. Its connection with 
analytically continued Wightman function for this field is shown. Its interpolating and 
limiting properties are demonstrated and its uniqueness is advocated. 
In section \ref{sec:wp_ff} this construction is generalized onto the massive fields with 
spins $1/2$ and $1$. 
The general formulaes for averages corresponding to different observables are justified. 
Unlike \cite{nn,nsh,dn,n_sh,NN_shk}, the states with definite mass and definite 
spin are constructed as irreducible representation of extended Lorentz group 
\cite{Nvj,Bil,Wnb}.

In section \ref{sec:Disc} the physical meaning of parameter $\tau_a$ for $a$- th particle 
is illuminating by special limit of the wave packet, when $\sigma_a\to 0$, $m_a\to 0$ 
with fixed $\tau_a$. In fact this limit leads to 1+1 - dimension covariant wave packet, 
propagating on the light cone. 

Sections \ref{sec:intrm_wp} and \ref{sec:diagr_tr} contain applications of suggested wave 
packets to both treatments of neutrino oscillations: using the states of intermediate 
neutrino or the macroscopic Feynman diagrams. 
In section \ref{sec:intrm_wp} the deduced above covariant single-particle interpolating 
wave packet is applied for intermediate neutrino treatment by using the known Blasone 
oscillation picture \cite{bl1,bl2,bl3} basing on inequivalent representations of commutation 
relations. 
An inconsistency with general QFT principles of the conventional ``plane-wave'' mixing 
relations for these wave packets is elucidated. The respectively generalized expression 
for two-flavour oscillations of leptonic charge of electronic neutrino is given.  

In section \ref{sec:diagr_tr} the notion of composite wave function for the neutrino 
creation/detection $\{C/D\}$ vertices is shown to be convenient in order to analyse the 
connection between both approaches. It is shown its natural correspondence with overlap 
function \cite{nn,bth,ahm} and how it naturally resolves the 
problems with causality and with covariant equal time prescription \cite{bth}. 
The scalar product of composite wave functions is unambiguously defined in accordance 
with Huygens' principle. 
The well known procedure of pole integration \cite{bth,ahm,fujii} for oscillation 
amplitude is related to the pole approximation for the off-shell composite wave 
function, converting the latter to the on-shell one. 
It is shown the coincidence of their space-time asymptotic behavior and its correlation 
with asymptotic of oscillation amplitude. 
The narrow-packet approximation of the explicitly calculated for two-packet case overlap 
function is shown to be similar but not exactly the same as was approximately obtained in 
\cite{nn,nsh,dn,n_sh,NN_shk}. The conclusions are given in section \ref{sec:concl}. 

In \ref{ap:1_A} the minimization properties of suggested wave packet are discussed.  
In  \ref{ap:A} the plane-wave limit, the non relativistic limit and 
zero-mass-width limit of suggested wave packet are traced in some details. 
Some intermediate useful formulas and definitions are collected in  \ref{ap:B}. 
In \ref{ap:C} an explicit form of two-packet overlap function is calculated, 
whose narrow-packet approximation is obtained and analyzed in \ref{ap:D}. 
In \ref{sec:one_p_rep} the exact integral representation of two-packet on-shell composite wave 
function is obtained as a linear superposition of one-packet functions. 
Its correct localization properties are traced. Its zero-mass limit is investigated. 
\section{Wave packets in quantum mechanics}\label{sec:wp_QM}
As it is well known \cite{LL,Mess,Tl,g_w}, time in non relativistic quantum mechanics  
plays the role of parameter. So, both momentum- and coordinate- ket states make 
sense {\sl at any instant of time} $t$ as formal eigenstates of three-dimensional 
momentum $\Pop$- operator and coordinate $\Xop$- operator: $\Pop |\ka>\,=\ka|\ka>$, 
$<\ka\mid\lv>\,=\delta_3(\ka-\lv)$ and $\Xop|\x>\,=\x |\x>$, 
$<\x|\y>\,=\delta_3(\x-\y)$, with $[{\rm X}_n,{\rm P}_l]=i\hbar\delta_{nl}$. 
Then for any state $|f>\,$: $\Xop = \x$, 
$\Pop = {\Pop}_{\x}=-i\hbar\vec{\nabla}_{\x}$ on the wave function  
$f(\x)=\, <\x|f>$ of its $\x$ -representation; and 
$\Xop = \Xop_{\rm p}=i\hbar\vec{\nabla}_{\rm p}$, $\Pop = \p$ on the 
wave function $\widetilde{f}(\p)=\, <\p|f>$ of its $\p$ -representation, also 
{\sl at any instant of time} $t$, regardless of the Hamiltonian and the non-relativistic 
quantum pattern used: Schr\"odinger, Heisenberg, etc. \cite{LL,Mess,vrg}. 
Sinse for $\hbar=1$ the completeness and the operator algebraic relations take place: 
\bea
&&\!\!\!\!\!\!\!\!\!\!\!\!\!\!\!\!\!\!
\int\! d^3{\rm k}|\ka><\ka|=\vec{\rm I},\quad \; 
\int\! d^3{\rm x}|\x><\x|=\vec{\rm I}, 
\label{0_1} \\
&&\!\!\!\!\!\!\!\!\!\!\!\!\!\!\!\!\!\!
e^{-i(\ka\cdot\Xop)}\,\Pop\, e^{i(\ka\cdot\Xop)}=\Pop +\ka, \quad \;
e^{i(\y\cdot\Pop )}\,\Xop\, e^{-i(\y\cdot\Pop)}=\Xop +\y,
\label{0_2} \\
&&\!\!\!\!\!\!\!\!\!\!\!\!\!\!\!\!\!\!
\mbox{whence: }\;
e^{i(\ka\cdot\Xop)}|\p>\,=|\p+\ka> , \quad \;
e^{-i(\y\cdot\Pop)}|\x>\,=|\x+\y>, 
\label{0_3} \\
&&\!\!\!\!\!\!\!\!\!\!\!\!\!\!\!\!\!\!
\mbox{or: }\,
<\p|e^{-i(\ka\cdot\Xop)}|f>\,=
\exp\left\{-i\left(\ka\cdot\Xop_{\rm p}\right)\right\}\widetilde{f}(\p)=
\widetilde{f}(\p+\ka),
\label{0_4_0} \\
&&\!\!\!\!\!\!\!\!\!\!\!\!\!\!\!\!\!\!
\mbox{and: }\,
<\x|e^{i(\y\cdot\Pop )}|f>\,=
\exp\left\{i\left(\y\cdot\Pop_{\x}\right)\right\}f(\x)=f(\x+\y),
\label{0_4}
\eea
then the wave functions are connected by direct and inverse 3- dimension Fourier 
transformation also {\sl at any instant of time} $t$ as:
\bea
\fl 
f(\x)=\!\int\! d^3{\rm k}<\x|\ka>\widetilde{f}(\ka),\qquad 
\widetilde{f}(\ka)=\!\int\! d^3{\rm x}<\ka|\x> f(\x), \; \mbox{ where:}
\label{2_1} \\ 
\fl 
<\x|\ka>\ = \frac {e^{i(\ka\cdot\x)}}{(2\pi)^{3/2}},\quad  
|\x>\ =\!\!\int\!\frac{d^3{\rm k}|\ka> e^{-i(\ka\cdot\x)}}{(2\pi)^{3/2}}, \quad 
<\x|\ka>^*=\frac {e^{-i(\ka\cdot\x)}}{(2\pi)^{3/2}}=\ <\ka|\x>, 
\label{2_2}  
\eea
is $\x$ -representation of the state $|\ka>$ with definite momentum $\ka$, and vice 
versa \cite{LL,Mess,Tl,g_w}. 
For Gaussian wave packet, which is localized in the domain $\sigma_x$ around the point 
$\x_a$ in coordinate space and contains respectively the plane-wave modes 
$|\ka>$ with $\ka$ in the domain $\sigma_p = 1/\sigma_x$ near momentum $\p_a$, for 
$\Delta\ka=\ka - \p_a$, $\Delta\x=\x - \x_a$, 
$f(\x)=\Psi_{\p_a,\x_a,\sigma}(\x)$, 
$\widetilde{f}(\ka)=\widetilde{\Psi}_{\p_a,\x_a,\sigma}(\ka)$, with 
\cite{LL,Mess,Tl,g_w}: 
\bea
\fl 
\widetilde{\Psi}_{\p_a,\x_a,\sigma}(\ka)=\,<\ka|\{\p_a, \x_a, \sigma\}>\, 
= \left(\frac{1}{\pi\sigma_p^2}\right)^{3/4}
\exp\left[-\,\frac{(\Delta\ka)^2}{2\sigma_p^2}-i(\Delta\ka\!\cdot\!\x_a)\right]\!, 
\label{eq:gauss_p} \\
\fl 
\Psi_{\p_a, \x_a, \sigma}(\x)=\,<\x|\{\p_a,\x_a,\sigma\}>\,=
e^{i(\p_a\!\cdot\!\x_a)}
\left(\frac{1}{\pi\sigma_x^2}\right)^{3/4}\!
\exp\left[-\,\frac{(\Delta\x)^2}{2\sigma_x^2}+i(\Delta\x\!\cdot\!\p_a)\right]\!,
\label{eq:gauss_x} 
\eea 
these wave functions (\ref{eq:gauss_p}), (\ref{eq:gauss_x}) define {\sl the same} wave 
packet state $|\{\p_a, \x_a, \sigma\}>$, normalized to unity 
$<\{\p_a,\x_a,\sigma\}|\{\p_a,\x_a,\sigma\}>\,=1$ in fact for arbitrary 
initial instant $t_a$ (\ref{D_6}) at instant $t=t_a$. Both its limits onto above 
eigenstates make sense also {\sl for arbitrary initial instant} $t_a$:
\bea
&&\!\!\!\!\!\!\!\!\!\!\!\!\!\!\!\!\!\!
\left(2\sigma_p\sqrt{\pi}\right)^{-3/2}
|\{\p_a, \x_a, \sigma\}>\, \longrightarrow\, |\p_a>, 
\; \mbox { for }\;\sigma_p\to 0, \;\;(\sigma_x\to\infty),
\label{sgm_p} \\
&&\!\!\!\!\!\!\!\!\!\!\!\!\!\!\!\!\!\!
\left(2\sigma_x\sqrt{\pi}\right)^{-3/2}
|\{\p_a, \x_a, \sigma\}>\, \longrightarrow \,\exp\{i(\p_a\!\cdot\!\x_a)\}|\x_a>,
\;\; \sigma_x\to 0\;\;(\sigma_p\to\infty),   
\label{sgm_x}
\eea
also regardless the Hamiltonian and the quantum pattern used. The further 
fate of the initial state (\ref{eq:gauss_p}), (\ref{eq:gauss_x}) at $t>t_a$ surely  
depends on the Hamiltonian. For the case of harmonic oscillator this wave packet 
propagates without spreading in coordinate space, while for the free non 
relativistic case $\p^2/(2m)$ it spreads with time with the effective width 
$\sigma^2_x(t)=\sigma^2_x+t^2\sigma^2_p/m^2$ \cite{dn,LL,Mess,Tl,g_w} in any fixed reference 
frame.

As it is well known, the Gaussian wave packet (\ref{eq:gauss_p}), (\ref{eq:gauss_x}) 
minimizes the Heisenberg uncertainty condition \cite{LL,Mess}. 
Nevertheless this profile of wave packet is not the unique possible one as in the 
non-relativistic quantum mechanics \cite{Tl,g_w} as well in optics \cite{kil}:  
\bea  
&&\!\!\!\!\!\!\!\!\!\!\!\!\!\!\!\!\!\!
\frac 13 \MVV{(\Delta \Pop)^2}_{\widetilde{f}}\,\,
\frac 13 \MVV{(\Delta \Xop)^2}_f\geqslant
\frac {\hbar^2}9 \MVV{(\Delta\ka)^2}_{\widetilde{\Psi}}\, \MVV{(\Delta\x)^2}_\Psi
=\frac{\hbar^2}4 \sigma^2_p\sigma^2_x=\frac{\hbar^2}4.
\label{hein}
\eea 
In the next sections is shown how the constraints of relativistic QFT up to non trivial 
normalization factor unambiguously determine a relativistically covariant wave packet state 
for arbitrary spin. It has the limiting properties similar to (\ref{sgm_p}), (\ref{sgm_x}), 
and in non relativistic limit it recasts into the Gaussian wave packet (\ref{eq:gauss_p}), 
(\ref{eq:gauss_x}) exactly regardless of the remained ambiguities. 

\section{Relativistic wave packet. Scalar field.}\label{sec:wp_sf_QFT}
A free real massive quantum scalar field $\varphi(x)$, for $x^\nu=(x^0,\x)$, 
$x^0=ct$, $k_\mu=(k^0,-\ka)$ satisfying Klein-Gordon (KG) equation of motion has the form 
\cite{blp,tir,p_s,vrg,schw,blt,oksak,jost,s_w,stroc,Dvt,b_d,i_z}: 
\bea
&&\!\!\!\!\!\!\!\!\!\!\!\!\!\!\!\!\!\!
\varphi(x)=\int\frac{d^3{\rm k}}{(2\pi)^3 2k^0}
\left(a_{\rm k}f_{\rm k}(x)+a^\dagger_{\rm k}f^*_{\rm k}(x)\!\right), 
\label{eq:scalar_field} \\
&&\!\!\!\!\!\!\!\!\!\!\!\!\!\!\!\!\!\!
(\partial^2+m^2)\varphi(x) = 0, \quad (\partial^2+m^2)f_{\rm k}(x) = 0, \quad 
{a}_\ka\ket{0} = 0, \quad {a}_\ka^\dagger\ket{0} = \ket{\ka},
\label{eq:klein_gordon} \\
&&\!\!\!\!\!\!\!\!\!\!\!\!\!\!\!\!\!\!
\mbox{where: }\;
k^0 =\frac{E_{\rm k}}{c}=+\sqrt{\ka^2+(mc)^2}\longmapsto +\sqrt{\ka^2+m^2}=E_{\rm k}>0, 
\label{2_3} \\ 
&&\!\!\!\!\!\!\!\!\!\!\!\!\!\!\!\!\!\!
\mbox{with: }\;
\frac{mc}{\hbar}\equiv \frac 1\lambda \longmapsto mc \longmapsto m, 
\;\mbox{ for }\; 
\hbar \longmapsto 1, \;\mbox{ and }\;  c \longmapsto 1. 
\label{2_3_0}
\eea 
The creation operator $a^\dagger_{\rm k}$ creates a state with definite momentum 
$\ket{\ka}$ (\ref{eq:klein_gordon}) by acting on {\sl vacuum state} $\ket{0}$ and obeys 
the commutation relation with annihilation operator $a_{\rm k}$:
\bea
&&\!\!\!\!\!\!\!\!\!\!\!\!\!\!\!\!\!\!
 \left[a_{\rm q},a^\dagger_{\rm k}\right]=
\langle\q|\ka\rangle=(2\pi)^3 2k_0\delta_3(\ka-\q),\; \mbox{ whence:}
\label{2_4} \\
&&\!\!\!\!\!\!\!\!\!\!\!\!\!\!\!\!\!\! 
\langle 0|\varphi(x)|\ka\rangle=f_{\rm k}(x)=e^{-i(kx)}\mapsto\langle x|\ka\rangle,
\label{2_04} \\
&&\!\!\!\!\!\!\!\!\!\!\!\!\!\!\!\!\!\! 
\langle\ka|\varphi(x)|0\rangle=
f^*_{\rm k}(x)=f_{-\rm k}(x)=e^{i(kx)}\mapsto\langle\ka|x\rangle, 
\label{2_004}
\eea 
for $(kx)=k_\nu x^\nu$ is a plane-wave solution to KG equation (\ref{eq:klein_gordon}) 
and represents by definition 
\cite{blp,tir,p_s,vrg,schw,blt,oksak,jost,s_w,stroc,Dvt,b_d,i_z} 
the coordinate wave function of state with the definite mass $m$, momentum $\ka$ and 
energy $ck^0=E_\ka$. 
The state $\ket{\ka}$ (\ref{eq:klein_gordon}) differs from the states $|\ka>$ of  
(\ref{0_1}), (\ref{0_3}), (\ref{2_2}) not only by its Lorentz covariant normalization 
condition (\ref{2_4}) but also by its meaning \cite{p_s,ahm}, because for 
\bea
&&\!\!\!\!\!\!\!\!\!\!\!\!\!\!\!\!\!\!
|\ka\rangle=(2\pi)^{3/2}\sqrt{2k_0}\,|\ka >, \quad k_0=E_{\rm k}/c>0, 
\;\mbox{ the condition:}
\label{2_5} \\
&&\!\!\!\!\!\!\!\!\!\!\!\!\!\!\!\!\!\!
\int \frac{d^3{\rm k}\,|\ka\rangle\langle\ka|}{(2\pi)^3 2k_0}=\!\vec{\rm I}_1,\; 
\mbox{ now is the one-particle completeness only}.
\label{2_6}
\eea 
The well known inability in relativistic QFT to localize only one particle at a space 
domain and time interval, respectively less than $\lambda=\hbar/(mc)$ and 
$\lambda/c=\hbar/(mc^2)$, {\sl makes meaningless also the non covariant states} 
$|\x>, |\ka>$ of (\ref{0_1}), (\ref{2_2}) {\sl for any instant of 
time} $t=x^0/c$ as well as corresponding definition of non covariant packet state 
(\ref{eq:gauss_p}) -- (\ref{sgm_x}), and {\sl gives no chances} to define a covariant 
self-adjoint operator of {\sl four-dimensional position} for this particle
\cite{blp,p_s,vrg,tir,schw,blt,oksak,jost,s_w,stroc,Dvt,b_d,i_z,Nvj,Bil,Wnb}. 

The change of normalization (\ref{2_5}) is not an unitary transformation rather 
dilatation, which therefore change the representation of other operators.   
The respective relativistic generalization of self-adjoint 
{\sl three-dimensional position operator} \cite{tir,p_s,vrg,schw,blt}: 
$\widehat{\Xop}=\sqrt{E(\Pop)}\Xop(\sqrt{E(\Pop)}\,)^{-1}$ as  
$\widehat{\Xop}_{\rm p}=
\sqrt{E_{\rm p}}i\hbar\vec{\nabla}_{\rm p}(\sqrt{E_{\rm p}})^{-1}$ 
only for $x^0=0$ has the function $\langle\p|\y>\,=\sqrt{2E_{\rm p}}\,e^{-i(\p\cdot\y)}$ 
as eigenfunction with eigenvalue $\y$:  
$\widehat{\Xop}_{\rm p}\langle\p|\y>\,= \y \langle \p|\y>$. 
In spite of the validity of the same commutation relation with momentum operator 
$[\widehat{\rm X}_n,{\rm P}_l]=i\hbar\delta_{nl}$, and of the relations 
(\ref{0_2})--(\ref{0_4}) with $\Xop_{\rm p}\mapsto\widehat{\Xop}_{\rm p}$ and  
$<\p|\mapsto\langle\p|$, and the completeness (\ref{2_6}) instead of (\ref{0_1}), 
this operator loses the above localization propertie (\ref{2_1}) of functions (\ref{2_2}) 
for its eigenfunctions $\langle\p|\y>$ already at $x^0=0$ \cite{tir,schw}. 

As shown out by the Reeh-Schlieder theorem completed with the `edge of wedge' theorem 
\cite{blt,oksak,jost,s_w,stroc}, the localization of particles is a subtle issue in quantum 
field theory. 
According to (\ref{2_04}), the operator creating from the {\sl vacuum state} $\ket{0}$ 
{\sl the covariant one-particle state} with definite 4-coordinate $|x\rangle=\ket{x^0,\x}$   
is the operator (\ref{eq:scalar_field}) of quantized free field itself. 
It is accepted as a state of free particle {\sl localized in the 3-point $\x$ only at the 
instant} $x^0 = 0$ \cite{tir,p_s,vrg,schw}, but by means of 
(\ref{2_6}) one has single-particle complete system of states at any instant $x^0$ 
(hereafter $c=1$, where it is possible): $k^0= E_{\rm k}>0$, 
\bea
\fl  
\varphi(x)|0\rangle\!=\!
\int\!\frac{d^3{\rm k} |\ka\rangle e^{i(kx)} }{(2\pi)^3 2k^0 }= 
\!\int\!\frac{d^3{\rm k}\,|\ka\rangle\langle\ka|x\rangle}{(2\pi)^3 2k^0}= |x\rangle,  
\;\mbox{ whence: }\int\! d^3{\rm x}|x\rangle 
(i\!\stackrel{\leftrightarrow}{\partial^0_{x}})\langle x|=\!\vec{\rm I}_1,
\label{eq:ktp_coord} \\
\fl 
|x\rangle  \stackunder{x_0\to 0}{\longrightarrow}
\int\!\frac{d^3{\rm k} |\ka\rangle e^{-i(\ka\cdot\x)}}{(2\pi)^{3} 2k^0 }=\!
\int\!\frac{d^3{\rm k} |\ka> e^{-i(\ka\cdot\x)}}{(2\pi)^{3/2}\sqrt{2k^0}}=|0,\x\rangle 
\neq 
|\x>=\!\int\!\frac{d^3{\rm k}|\ka\rangle e^{-i(\ka\cdot\x)}}{(2\pi)^{3}\sqrt{2k^0} },
\label{2_8} \\
\fl  
\mbox{or: }\;
\langle\p|x\rangle=f^*_{\rm p}(x)=e^{i(px)}\stackunder{x_0\to 0}{\longrightarrow}
\langle\p|0, \x\rangle=e^{-i(\p\cdot\x)} \neq 
\langle\p|\x>\,=\sqrt{2E_{\rm p}}\,e^{-i(\p\cdot\x)}.
\label{2_8_0} 
\eea 
Only the time dependent covariant states in the l.h.s. of inequalities (\ref{2_8}), 
(\ref{2_8_0}), unlike the non covariant ones in the r.h.s. \cite{tir,p_s,vrg,schw}, make a 
safe ground for construction of covariant wave-packet states. 
Similarly (\ref{sgm_p}), (\ref{sgm_x}) for the packet (\ref{eq:gauss_p}), 
(\ref{eq:gauss_x}), the desired relativistic wave packet $\ket{\{p_a, x_a,\sigma\}}$ 
{\sl has to interpolate} between the {\sl covariant state} $\ket{x_a}$ (\ref{eq:ktp_coord}) 
with definite 4-coordinate for some $\sigma\to\infty$ and the {\sl covariant state} 
$\ket{\p_a}$ (\ref{eq:klein_gordon}), (\ref{2_5}) with definite 4-momentum for the 
$\sigma \to 0$. 
So, its similar to (\ref{2_04}) Lorentz covariant (or invariant) wave function in this  
coordinate-space representation due to conditions of translation symmetry, 
for $\underline{x}=x_a-x$, looks like:
\bea
\fl  
F_{p_ax_a}(x)=
\bra{0}\varphi(x)\ket{\{p_a, x_a,\sigma\}}=e^{-i(p_a x)}{\Phi}_\sigma(\p_a,x-x_a), \;\;
\mbox{ or:}
\label{2_9} \\
\fl  
F_{p_ax_a}(x)=\langle x\ket{\{p_a, x_a,\sigma\}}
=e^{-i(p_ax_a)}\psi_\sigma (\p_a,x_a-x), \;\; \mbox { with:}
\label{2_10} \\
\fl  
\psi_\sigma(\p_a,\underline{x})=
\!\int\!\!\frac{d^3{\rm k}}{(2\pi)^3 2E_{\rm k}}\,
\phi^\sigma(\ka,\p_a)\,e^{i(k\underline{x})}=
\!\int\!\!\frac{d^4 k}{(2\pi)^3}\,\phi^\sigma(\ka,\p_a)\,e^{i(k\underline{x})}
\theta(k^0)\,\delta(k^2-m^2_a),
\label{eq:scalar_wp_wf} \\
\fl  
\psi_\sigma(\p_a,\underline{x})=
e^{i(p_a\underline{x})}{\Phi}_\sigma(\p_a,-\underline{x}),\quad 
\ket{\{p_a, x_a,\sigma\}}=
\!\int\!\!\frac{d^3{\rm k}\,|\ka\rangle }{(2\pi)^32E_{\rm k}}\,
\phi^\sigma(\ka,\p_a)\,e^{i((k-p_a)x_a)}, 
\label{2_11}
\eea 
where $k^0 = E_{\rm k}$, $p^0_a= E_{\rm p_a}$, and, unlike the function 
$\Phi_\sigma(\p_a,-\underline{x})$, the {\sl Lorentz invariant function} 
$\psi_\sigma(\p_a,x_a-x)$ (\ref{eq:scalar_wp_wf}) satisfies KG equation 
(\ref{eq:klein_gordon}) relative to both variables $x,\,x_a$. 

The scalar function $\phi^\sigma(\ka,\p_a)$ is to be determined as Lorentz-invariant
due to invariance of the measure in (\ref{eq:scalar_wp_wf}), (\ref{2_11}). It depends on 
the invariants: $\sigma=\sigma_a$, $m=m_a$, $\zeta^2_a$, $(k\zeta_a)$, where the 
time-like for $m_a>0$ 4-vector $\zeta_a(p_a,\sigma_a)$ carries all the basic properties 
and ``particle-like'' vector quantum numbers of the wave packet. Thus in general it 
has to be a linear combination of the 4-momentum $p_a$ and the ``4-spin''
$\widehat{w}_a=\widehat{s}_a m_a\sqrt{S(S+1)}$ of wave packet\footnote{Here the product of 
truth (polar) vectors $(\widehat{w}_a\widehat{s}_a)$ only parametrizes the eigenvalue of 
square of Pauli-Lubanski pseudo vector operator ${\cal W}^\mu$ of type (\ref{3_7_04}), 
unlike the operator product (\ref{3_2_1_0}) below (see \cite{i_z,Nvj,Bil}).}, 
that for $p^2_a=m^2_a$, 
$(p_a \widehat{s}_a)=0$, $\widehat{s}^2_a=-1$ fully characterize the relativistic 
one-particle state in the rest frame by its mass $p^\mu_{*a}=(m_a,\vec{0})$ and by the 
axis of its spin quantization $\widehat{s}^\mu_{*a}=(0,\widehat{\vec{\rm s}}_a)$ 
\cite{Bil}:  
\bea
\fl  
\zeta_a(p_a,\sigma_a)=p_a g_1(m_a,\sigma_a)+ \widehat{w}_a g_2(m_a,\sigma_a), \quad \;
\zeta^2_a=m^2_a\left[g^2_1-S(S+1)g^2_2\right], 
\label{z_1} 
\eea
where $\zeta^2_a>0$, $\zeta^0_a>0$, if without loss of generality is imposed 
$\forall\,\sigma$: $g_1(m,\sigma)\gg |g_2(m,\sigma)|$. This is meaningful because the 
spin degrees of freedom should not change drastically the properties of the scalar 
amplitude  $\phi^\sigma(\ka,\p_a)$ of wave packet with any spin, and because for the 
scalar field (\ref{eq:scalar_field}): $S=0$, $\widehat{w}_a=0$, which is equivalent to 
choosing here $g_2\equiv 0$. The further requirements onto universal invariant  
functions $g_1,g_2$ will be given below. 

The case of full delocalization in coordinate space is expressed by the limit 
$\sigma \to 0$. So the wave packet (\ref{2_11}) reduces precisely to the state 
with definite momentum $\ket{\p_a}$:
\bea
&&\!\!\!\!\!\!\!\!\!\!\!\!\!\!\!\!\!\! 
|\{p_a,x_a,\sigma\}\rangle\stackunder{\sigma\to 0}{\longrightarrow}|\p_a\rangle,\quad 
\Phi_\sigma(\p_a,-\underline{x})\stackunder{\sigma\to 0}{\longrightarrow}1,
\label{2_12} \\
&&\!\!\!\!\!\!\!\!\!\!\!\!\!\!\!\!\!\!
F_{p_ax_a}(x)\stackunder{\sigma\to 0}{\longrightarrow}e^{-i\left(p_ax\right)}, 
\quad \psi_\sigma(\p_a,x_a-x)
\stackunder{\sigma\to 0}{\longrightarrow}
e^{i\left(p_a(x_a-x)\right)}, 
\label{eq:limit_sigma_0} \\
&&\!\!\!\!\!\!\!\!\!\!\!\!\!\!\!\!\!\!
\phi^\sigma(\ka,\p_a) =e^{i((p_a-k)x_a)}\langle\ka\ket{\{p_a, x_a,\sigma\}}
\stackunder{\sigma\to 0}{\longrightarrow}
(2\pi)^3\sqrt{2E_{\rm k}2E_{\rm p_a}}\,\delta_3(\ka-\p_a), 
\label{2_13} \\
&&\!\!\!\!\!\!\!\!\!\!\!\!\!\!\!\!\!\!
\mbox{for: }\;
\left\{g_1(m,\sigma)\gg |g_2(m,\sigma)|,\,\mbox{ and }\,\zeta^2_a,\,\zeta^0_a\right\}
\stackunder{\sigma\to 0}{\longrightarrow} +\infty,\;\mbox{ with }\; 
g_2/g_1\stackunder{\sigma\to 0}{\longrightarrow} 0.
\label{2_14}
\eea 
The function ${\Phi}_\sigma(\p_a,x-x_a)$ in (\ref{2_9}) varies with $x$ very smoothly in 
compare with the plane wave $e^{-i(p_a x)}$ and in the limit (\ref{2_12}) does not 
contribute to the flux density \cite{i_z} of the state (\ref{2_10}): 
\bea
&&\!\!\!\!\!\!\!\!\!\!\!\!\!\!\!\!\!\!
{\cal J}^\mu_{p_ax_a}(x)= F^*_{p_ax_a}(x)\,
(i\!\stackrel{\leftrightarrow}{\partial^\mu_{x}})F_{p_ax_a}(x)
\stackunder{\sigma\to 0}{\longrightarrow}2p^\mu_a, \;\;\,\mbox { with: }\;
\stackrel{\leftrightarrow}{\partial^\mu_{x}}=
\stackrel{\rightarrow}{\partial^\mu_{x}}-\stackrel{\leftarrow}{\partial^\mu_{x}}.  
\label{2_17_0}
\eea
For the state (\ref{2_11}) the space integration over the wave packet center $\x_a$ for 
arbitrary time $x^0_a$ and $\sigma$ gives again the momentum state (\ref{2_12}), where 
for $\phi^\sigma(\p_a,\p_a)=\phi^\sigma_m=$ const $\neq 0$: 
\bea
&&\!\!\!\!\!\!\!\!\!\!\!\!\!\!\!\!\!\!
|\p_a\rangle=
\frac{2E_{{\rm p}_a} }{\phi^\sigma_m}\int d^3{\rm x}_a|\{p_a,x_a,\sigma\}\rangle. 
\label{2_15_14} 
\eea
The conditions (\ref{2_12})--(\ref{2_15_14}) are implied for any wave packets in any 
scattering theory 
\cite{nn,nsh,dn,n_sh,NN_shk,kt_1,LL,Mess,Tl,g_w,feyn,blp,p_s,vrg,tir,schw,blt,oksak,jost,s_w,
stroc,Dvt,b_d,i_z,Nvj,Bil,Wnb}. 

The opposite case $\sigma \to \infty$ corresponds to field excitation fully localized 
in the point $x_a$ in above relativisticaly covariant sense (\ref{eq:ktp_coord}),
(\ref{2_8}). The packet state should transforms to the state (\ref{eq:ktp_coord}) up to 
some normalization factor ${ N}_\infty$ and up to inessential now again phase factor 
$e^{-i\Theta_a}$ with invariant phase $\Theta_a = (p_ax_a)$ instead of        
$\Theta_a \mapsto -(\p_a\!\cdot\!\x_a)$ in   (\ref{sgm_x}). 
So, up to the same factors its wave function (\ref{2_10}), (\ref{eq:scalar_wp_wf}) recasts   
into the matrix element (\ref{eq:limit_sigma_inf}), because in accordance with relativity 
of time it assumes the sense of the only meaningful now {\sl transition amplitude}  
\cite{tir,p_s,vrg} from point $\x_a$ to the point $\x$ during the time $T=x^0-x^0_a>0$: 
\bea
&&\!\!\!\!\!\!\!\!\!\!\!\!\!\!\!\!\!\! 
|\{p_a,x_a,\sigma\}\rangle\stackunder{\sigma\to\infty}{\longrightarrow}
{ N}_\infty e^{-i\Theta_a}|x_a\rangle=
{ N}_\infty e^{-i\Theta_a} \varphi(x_a)|0\rangle,
\label{2_15} \\
&&\!\!\!\!\!\!\!\!\!\!\!\!\!\!\!\!\!\! 
\psi_\sigma(\p_a,x_a-x)\stackunder{\sigma\to\infty}{\longrightarrow}
\psi_\infty(\p_a,x_a-x)={ N}_\infty \langle 0|\varphi(x)\varphi(x_a)|0\rangle, 
\label{eq:limit_sigma_inf} \\
&&\!\!\!\!\!\!\!\!\!\!\!\!\!\!\!\!\!\!
\phi^\sigma(\ka,\p_a)=e^{i((p_a-k)x_a)}\langle\ka\ket{\{p_a, x_a,\sigma\}}
\stackunder{\sigma\to\infty}{\longrightarrow} \phi^\infty(\ka,\p_a)={ N}_\infty,
\label{2_16} \\
&&\!\!\!\!\!\!\!\!\!\!\!\!\!\!\!\!\!\! 
\mbox{for: }\;
\left\{g_1(m,\sigma)\gg |g_2(m,\sigma)|,\mbox{ and }\,\zeta^2_a,\, \zeta^0_a\right\}
\stackunder{\sigma\to\infty}{\longrightarrow} +0, \;\mbox{ with }\; 
g_2/g_1\stackunder{\sigma\to\infty}{\longrightarrow} 0.
\label{2_17}
\eea 
The conditions (\ref{2_15})--(\ref{2_17}) are essentially new. Since the Wightman 
function in the r.h.s of   (\ref{eq:limit_sigma_inf}) is the boundary value 
(\ref{2_23_0}) for $\zeta_a$ (\ref{2_17}) of analytic function (\ref{2_19}) of complex 
4-vector variable $z_a\!=x_a+i\zeta_a$, holomorphic in future tube $V^+$: $\zeta^2_a,\zeta^0_a>0$ 
\cite{schw,blt,oksak,jost,s_w,stroc}, and because the existence of asymptotic 
fields in Haag-Ruelle scattering theory and the reduction formulas for $S$ -matrix in 
QFT \cite{schw,blt,oksak,jost,s_w,stroc,Dvt,b_d,i_z,Nvj} requires an infinite smoothness of 
wave packet relative to $\x$: $F_{p_ax_a}(x^0,\x)\in{\cal S}(\vec{\rm R}^3_{\x})$ for 
fixed $x^0$, let us consider the function 
$\phi^\sigma(\ka,\p_a)\in{\cal S}(\vec{\rm R}^3_{\rm k})$, i.e. in the space of functions 
infinitely differentiable 
$\forall\,\ka\in \vec{\rm R}^3_{\rm k}$, decreasing faster than any power of $1/|\ka|$ 
together with all its derivatives \cite{blt,oksak}. This ensures all such properties of 
$F_{p_ax_a}(x)$ with the above vector $\zeta_a=\zeta_a(p_a,\sigma)$ (\ref{z_1}), due to 
$\zeta^2_a>0$, $k^2=m^2_a>0$, with the amplitude:
\bea
\fl 
\phi^\sigma(\ka,\p_a)={ N}_\sigma\!\left(m_a,\zeta^2_a\right) e^{-(k\zeta_a)},
\;\;\mbox { for: }\; \left(k\zeta_a\right)>0, \quad { N}_\sigma>0,\;\mbox{ whence:}
\label{eq:phidef} \\
\fl 
\psi_\sigma(\p_a,x_a-x)
={ N}_\sigma \langle 0|\varphi(x)\varphi(x_a+i\zeta_a(p_a,\sigma))|0\rangle
={ N}_\sigma \frac 1i D^-_{m_a}\!\left(x-x_a-i\zeta_a(p_a,\sigma)\right), 
\label{2_18} \\
\fl 
\mbox {where: }\,  \frac 1i D^-_{m}(y)=
\int\frac{d^3{\rm k}}{(2\pi)^3 2k^0}\,e^{-i(ky)},\;\mbox{ for: }\;
k^0=E_{\rm k}>0, \quad y=x-z_a, 
\label{2_19} 
\eea
is Wightman function (WF) \cite{schw,blt,oksak,jost,s_w,stroc,Dvt} of a free scalar field 
(\ref{eq:scalar_field}). Thus the wave-packet state (\ref{2_11}) now may be elegantly 
rewritten in such a form, that makes obvious the limit (\ref{2_15})--(\ref{2_17}):
\bea
&& \!\!\!\!\!\!\!\!\!\!\!\!\!\!\!\!\!\!
|\{p_a,x_a,\sigma\}\rangle ={ N}_\sigma e^{-i(p_a x_a)}
\varphi\left(x_a +i\zeta_a(p_a,\sigma)\right)|0\rangle\equiv 
{ N}_\sigma e^{-i(p_a x_a)}\varphi(z_a)|0\rangle, 
\label{eq:packet_state_scalar} \\
&& \!\!\!\!\!\!\!\!\!\!\!\!\!\!\!\!\!\!
\langle\{p_a,x_a,\sigma\}|={ N}_\sigma e^{i(p_a x_a)}  
\langle 0|\varphi\left(x_a -i\zeta_a(p_a,\sigma)\right)\equiv 
{ N}_\sigma e^{i(p_a x_a)} \langle 0|\varphi\left(z^*_a\right).
\label{2_19_0} 
\eea
The well known generalization of this vector-valued function of complex 4-vector variable 
$z_a$ onto the case of product of a few fields \cite{jost,s_w} defines a representation 
of nonhomogeneous complex Lorentz group. It allowes to analyze the main properties of 
higher WFs and to prove Bargmann\--Hall\--Wightman theorem, $PCT$-theorem and dispersion 
relations \cite{schw,blt,oksak,jost,s_w,stroc}. According to \cite{al_h_w} and \ref{ap:1_A}
in one-particle QM the exponential function in (\ref{eq:phidef}) in general provides the 
minimization properties of the wave packet together with the correct analytical continuation 
of corresponding Green function. Here the last is replaced by Wightman function in view of 
(\ref{E_01}).  

It turns out that the covariant packet (\ref{eq:phidef}) -- (\ref{2_19_0}) 
precisely conforms to the plane-wave limit (\ref{2_12})--(\ref{2_14}). In view of Lorentz 
invariance of $F_{p_ax_a}(x)$ and $\psi_\sigma(\p_a,\underline{x})$ (\ref{2_10}), 
(\ref{eq:scalar_wp_wf}) the local value of 
$F_{p_ax_a}(x_a)=e^{-i(p_ax_a)}\psi_\sigma(\vec{0},\underline{0})$ is finite.
It defines the constant ${ N}_\sigma\!\left(m,\zeta^2_a\right)>0$ up to some dimensionless 
function with fixed asymptotic behaviour over the unique {\sl invariant} dimensionless 
variable $\tau=\tau_a=m\sqrt{\zeta^2_a(p_a,\sigma)}$: 
\bea 
&& 
{\cal I}(\tau) = \int \frac{d^3{\rm k}\,\phi^\sigma(\ka,\p_a)}{(2\pi)^3 2E_{\rm k}} = 
{ N}_\sigma  \frac {1}{i}D^-_{m}\!\left(-i\zeta_a(p_a,\sigma)\right) 
\label{2_20} \\
&& 
=\psi_\sigma(\p_a,\underline{0})=\psi_\sigma(\vec{0},\underline{0})=
\frac{{ N}_\sigma m^2}{(2\pi)^2}\,
\frac{K_1(\tau)}{\tau}\equiv \frac{\aleph(\tau)}{(2\pi)^2}h\left(\tau^2\right),
\label{2_20_0}
\eea
for $\aleph(\tau)\equiv{N}_\sigma m^2$, $K_1(\tau)$ is Macdonald function 
(\ref{A_7}) \cite{b_er}. The conditions (\ref{2_13}), (\ref{2_14}) and (\ref{2_16}), 
(\ref{2_17}) define independently the same dimension of ${N}_\sigma$ and the 
corresponding to (\ref{2_15_36})) asymptotic behaviour of unknown smooth dimensionless 
function ${\cal I}(\tau)$ or $\aleph(\tau)$:  
\bea
&&\!\!\!\!\!\!\!\!\!\!\!\!\!\!\!\!\!\! 
\lim_{\tau\to\infty}{\cal I}(\tau)=1, \;\mbox{ or }\; 
\aleph(\tau)\stackunder{\tau\to\infty}{\longrightarrow}
2(2\pi)^{3/2}\tau^{3/2}e^\tau, \; \mbox{ whence }\; 
{ N}_\sigma \stackunder{\sigma\to 0}{\longrightarrow} +\infty,
\label{eq:sc_limit_0} \\
&&\!\!\!\!\!\!\!\!\!\!\!\!\!\!\!\!\!\! 
\lim_{\tau\to 0}\tau^2{\cal I}(\tau)=\frac{\aleph(0)}{(2\pi)^2}, \quad 
\aleph(0)> 0, \;\mbox{ or }\; 
{ N}_\sigma \stackunder{\sigma\to\infty}{\longrightarrow}{ N}_\infty=
\frac{\aleph(0)}{m^2}> 0. 
\label{2_21}
\eea 
Indeed, from the simple dimensional analysis with $c\neq 1$, in order to satisfy both  
conditions (\ref{2_14}) and (\ref{2_17}), up to rescaling of $\sigma$ by dimensionless 
constant, it is enough to suppose that: 
\bea
&&\!\!\!\!\!\!\!\!\!\!\!\!\!\!\!\!\!\! 
(mc)^2g_{1,\,2}(m,\sigma)\stackunder{\sigma\to 0}{\longrightarrow}
{\cal C}_{1,\,2}\left(\frac{mc}{\sigma}\right)^{\epsilon,\,\gamma}, \quad {\cal C}_1=1,
\quad \; \epsilon>\gamma>0, 
\label {2_301} \\
&&\!\!\!\!\!\!\!\!\!\!\!\!\!\!\!\!\!\! 
(mc)^2g_{1,\,2}(m,\sigma)\stackunder{\sigma\to\infty}{\longrightarrow}
C_{1,\,2}\left(\frac{mc}{\sigma}\right)^{A,\,B}, \quad C_1>0,\quad 0< A<B,\;
\mbox{ whence:}
\label {2_302} \\
&&\!\!\!\!\!\!\!\!\!\!\!\!\!\!\!\!\!\! 
\tau \mapsto (mc)^2g_{1}(m,\sigma)\to+\infty, \; \mbox{ with }\; 
\sigma\to 0, \;\mbox{ or with }\;  mc\to+\infty,
\label {2_303} \\
&&\!\!\!\!\!\!\!\!\!\!\!\!\!\!\!\!\!\! 
\tau \mapsto (mc)^2g_{1}(m,\sigma)\to 0, \; \mbox{ with }\; 
\sigma\to+\infty, \;\mbox{ or with }\;  mc\to 0.
\label {2_304} 
\eea
According to (\ref{2_14}) and (\ref{2_301}) the limit (\ref{2_13}) is defined 
only by properties of the function $g_1(m,\sigma)$ with 
$\tau\mapsto (mc)^2g_1(m,\sigma)\equiv(p_a\zeta_a)$, and from
(\ref{eq:phidef}), (\ref{eq:sc_limit_0}), for $c=1$, by means of: 
\bea
&&\!\!\!\!\!\!\!\!\!\!\!\!\!\!\!\!\!\! 
2(kp)=2m^2-\left(k^0-p^0\right)^2\!+\left(\ka-\p\right)^2>0, \quad 
k^0 = E_{\rm k}, \quad p^0 = E_{\rm p}, 
\label{2_22} \\
&&\!\!\!\!\!\!\!\!\!\!\!\!\!\!\!\!\!\! 
\mbox{for: }\,|\ka|={\rm k}, \quad \ka={\rm k}\n_{\rm k}, \quad |\p|={\rm p},\quad 
\p={\rm p}\n_{\rm p}, \quad \n^2_{\rm k}=\n^2_{\rm p}=1,
\label{2_23}
\eea
combining the well known definitions and expressions of delta-functions 
\cite{oksak,g_sh}: 
\bea
\fl 
\left\{\!\left(\frac{g_1}{2\pi}\right)^{3/2}\!
e^{-(g_1/2)(\ka-\p)^2}\!\right\}\!   
\stackunder{g_1\to\infty}{\Longrightarrow} \delta_3(\ka-\p), \quad 
\left\{\!\left(\frac{g_1}{2\pi}\right)^{1/2}\!
e^{-(g_1/2)({\rm k}-{\rm p})^2}\!\right\}\!
\stackunder{g_1\to\infty}{\Longrightarrow} \delta ({\rm k}-{\rm p}), 
\label{2_26} \\
\fl 
\delta_3(\ka-\p)=\frac{\delta_\Omega(\n_{\rm k},\,\n_{\rm p})}{{\rm k}{\rm p}}\,
\delta\left({\rm k}-{\rm p}\right), \,\mbox { where: }\,
\int d\Omega(\n_{\rm k}){\rm f}(\n_{\rm k})\delta_\Omega(\n_{\rm k},\,\n_{\rm p})=
{\rm f}(\n_{\rm p}),
\label{2_25} \\
\fl 
\mbox{with: }\;
k^0-p^0=E_{\rm k}-E_{\rm p}\approx\frac{{\rm k}}{E_{\rm k}}({\rm k}-{\rm p}),\;
\mbox{ and: }\;
1-\frac{{\rm k}^2}{E^2_{\rm k}}=\frac{m^2}{E^2_{\rm k}},\;\mbox{ for }\; 
p^\mu_a=(p^0,\p),
\label{2_28} \\
\fl 
\mbox {one finds: } \; 
\phi^\sigma(\ka,\p)=e^{i((p_a-k)x_a)}\langle\ka\ket{\{p_a, x_a,\sigma\}}=
{N}_\sigma  e^{-(k\zeta_a)}
\stackunder{\tau\to \infty}{\longmapsto}
{N}_\sigma  e^{-g_1(kp_a)} 
\label{2_28_1}  \\
\fl 
\stackunder{\tau\to\infty}{\longrightarrow}
2m(2\pi)^3 e^{(g_1/2)\left(k^0-p^0\right)^2}\!
\left\{\left(\frac{g_1}{2\pi}\right)^{3/2}e^{-(g_1/2)(\ka-\p)^2}\right\},
\label{2_28_2} 
\eea
that by the use of (\ref{D_7_0}), (\ref{D_7_0_0}) may be rewritten as:
\bea
\fl 
\stackunder{g_1\to\infty}{\longrightarrow}
2m(2\pi)^3\,\frac{\delta_\Omega(\n_{\rm k},\,\n_{\rm p})}{{\rm k}{\rm p}}\,
\lim_{g_1\to\infty}e^{(g_1/2)\left(E_{\rm k}-E_{\rm p}\right)^2} 
\left\{ \left(\frac{g_1}{2\pi}\right)^{1/2}
e^{-(g_1/2)({\rm k}-{\rm p})^2} \right\}  
\nonumber \\
\fl 
=2m(2\pi)^3\,\frac{\delta_\Omega (\n_{\rm k},\,\n_{\rm p})}{{\rm k}{\rm p}}\,
\lim_{g_1\to\infty}\left\{\left(\frac{g_1}{2\pi}\right)^{1/2}\!\!
\exp\left[-\frac {g_1}{2}\frac{m^2}{E^2_{\rm k}}\left({\rm k}-{\rm p}\right)^2\right]\right\} 
\label{2_28_4} \\ 
\fl 
=2m(2\pi)^3\,\frac{E_{\rm k}}{m}\, 
\frac{\delta_\Omega (\n_{\rm k},\,\n_{\rm p})}{{\rm k}^2}\,
\delta\left({\rm k}-{\rm p}\right)=(2\pi)^3 2E_{\rm k}\, \delta_3(\ka-\p).  
\;\mbox { (comp. \ref{ap:A})} 
\label {2_29}
\eea
Starting from quite different reasoning the wave packet like (\ref{eq:scalar_wp_wf}), 
(\ref{eq:phidef}) was introduced earlier in \cite{nn,nsh,dn,n_sh,NN_shk} and studied 
in some detail with $\zeta_a(p_a,\sigma)= p_ag_1(\sigma)=p_a(2\sigma^2)^{-1}$. 
However the arbitrarily imposed therein condition ${\cal I}(\tau)\equiv 1$ leads to 
disappearance of ${ N}_\infty\mapsto 0\leftarrowtail\aleph(0)$, 
rendering meaningless the limit (\ref{2_15})--(\ref{2_17}), (\ref{2_21}) and hidding the 
relations (\ref{2_18})--(\ref{2_21}). Unlike \cite{nn}, the derivation 
(\ref{2_28_1})--(\ref{2_29}) of the limit (\ref{2_13}) has nothing to do with the rest 
frame $\p=0$ of wave packet but explicitly requires $m>0$ (see also (\ref{D_3_01}), 
(\ref{D_3_1})). 
Thus an {\sl `independent'} limit $m\to 0$ makes sense only for the packet already with 
zero width $\sigma=0$, i.e. for a plane wave. So, for a massless particle with infinite 
Compton wave length $\lambda$ there are inevitable difficulties \cite{tir} with 
interpretation of (\ref{eq:ktp_coord}), (\ref{2_15}), (\ref{eq:packet_state_scalar}) as 
its covariant localizable states. 
Actually, the lacking of localizability for the massless states  manifests in mentioned in 
Introduction freedom of the profile of their wave packets \cite{kil}, which arises in 
(\ref{D_7})--(\ref{D_11}) due to arbitrariness of $\tau>0$ but nevertheless disappears for the 
massive case (\ref{D_3_01}), (\ref{D_3_1}). 

On the other hand, according to (\ref{eq:sc_limit_0}), (\ref{2_303}), (\ref{2_28_1}),  
the non relativistic limit $c\to\infty$ of the amplitude of the profile function 
(\ref{eq:phidef}), due to (\ref{2_3}), has exactly the form of expression (\ref{2_28_2}) 
with $e^{g_1\left(k_0-p_0\right)^2/2}=e^{g_1\left(E_{\rm k}-E_{\rm p}\right)^2/2c^2}\mapsto 1$. 
For $\epsilon=2$ \cite{nn}, whence $g_1=\sigma^{-2}$, this limit differs from the 
non relativistic profile of wave packet (\ref{eq:gauss_p}) just on the multiplier 
$(2\pi)^32mc\,(2\sigma\sqrt{\pi})^{-3/2}$ for $\sigma=\sigma_p$, whereas from 
(\ref{2_8}), (\ref{2_8_0}), it follows, that  
$\sqrt{2mc}\,\ket{0,\x}\stackunder{c\to\infty}{\longrightarrow}|\x>$. 
According to (\ref{D_6}) this recasts the wave-packet state (\ref{2_11}) exactly into 
Gaussian non relativistic one (\ref{eq:gauss_p}), (\ref{eq:gauss_x}) also with   
arbitrary initial instant of time $t_a$. 

The difference of localized states in QFT and in QM manifests itself in two 
different ways of obtaining the wave packet for ``infinitely heavy'' particle, with 
$m_ac\gg |\p_a|, |\ka|$, used, for example, in the model of Kobzarev et al. \cite{bth,kbz}.  
Its space-time and four-momentum wave functions would be proportional to 
$\propto \exp\{-im_acx^0\}\delta_3(\x-\x_a)$ and $\propto \delta(k^0-m_ac)$ 
respectively. But what about their normalization? By taking firstly the QFT-limit 
$\sigma\to\infty$, $\tau\to 0$, instead of (\ref{2_20}), (\ref{2_20_0}) we will face 
(\ref{2_10}) as localized state (\ref{2_15}) -- (\ref{2_19}) for 
$p^0_a=\sqrt{(m_ac)^2+\p^2_a}$, $k^0\mapsto k^0_a=\sqrt{(m_ac)^2+\ka^2}$, which then, 
for $m_a\to\infty$, and $p^0_a,k^0_a\to m_ac$, $\p_a=0$, recasts as: 
\bea
\fl 
F^{[\sigma=\infty]}_{p_ax_a}(x)=N_\infty \,e^{-i(p_ax_a)}\!
\int\frac{d^3{\rm k}\,e^{-i(k(x-x_a))}}{(2\pi)^3 2k^0}
\stackrel{\p_a= 0}{\stackunder{m_a\to\infty}{\longmapsto}} 
\frac{N_\infty}{2m_ac}\,e^{-im_acx^0}\,\delta_3(\x-\x_a), 
\label{4_11} \\
\fl 
\phi^\sigma(\ka,\p_a)\,\theta(k^0)\,\delta\left(k^2-(m_ac)^2\right)\equiv 
\phi^\sigma(\ka,\p_a)\,\frac{\delta(k^0-k^0_a)}{2k^0_a}
\stackrel{[\sigma=\infty]}{\stackunder{m_a\to\infty}{\longmapsto}} 
\frac{N_\infty}{2m_ac}\,\delta(k^0-m_ac). 
\label{4_12} 
\eea
On the other hand, starting from the non-relativistic limit $c\to \infty$ (\ref{D_5}) of  
$\phi^\sigma(\ka,\p_a)$, for the next limit $\sigma\to\infty$ under the 
conditions $\p_a=0$, $m_a\to\infty$, we arrive to the similar final value 
$\exp\{-im_acx^0\}\delta_3(\x-\x_a)$ but with another multiplier 
regardless the order of making the limit and Fourier transformation (\ref{2_1}). 
Thus, the normalization constants appearing here, for the first way, with $\tau=0$ 
for $\sigma=\infty$, are: 
\bea
\fl 
N_\infty=\frac{\aleph(0)}{(m_ac)^2}, \; \mbox{ and }\;\frac{1}{2m_ac},\;\mbox{ so that }
\;\frac{N_\infty}{2m_ac}=\frac{\aleph(0)}{2(m_ac)^3},\;\mbox{ but that is: }\;
\left(\frac{2\pi}{\sigma^2}\right)^{3/2}\!\!,  
\label{4_13} 
\eea
for the second contradictory way with $\sigma\to\infty$ when $\tau=\infty$ already. 
Both multipliers of the first way in (\ref{4_13}) are exactly, what were expected 
from definitions of localized states (\ref{2_15}) in QFT and (\ref{2_8}), (\ref{2_8_0}) 
in QM, where $<\x|\x_a>=\delta_3(\x-\x_a)$. 
So, the answer given by (\ref{4_11}), (\ref{4_12}) for the first way in (\ref{4_13}) 
arises as more consequent and more constructive, and it is physically more justified, 
since it directly connects the norms of localized states in QM and QFT. This illustrates 
the difference between the meaning of localized states in QM and in QFT according to above 
discussion. Note, that for both cases the initial instant $t_a$ disappeares for 
``infinitely heavy'' packet only due to limit $m_a\to\infty$ in (\ref{4_11}), (\ref{D_6}).

The used below narrow-packet approximations (\ref{2_15_41}), (\ref{3_20_0G}), 
(\ref{3_40_V}), (\ref{Pr_3_5}) imply an absence of asymptotic correction of order 
$1/\tau$ at $\tau\to\infty$ from the function ${\cal I}(\tau)\to 1$. 
This condition conforms with both limits $\tau\to\infty$ (\ref{eq:sc_limit_0}) and 
$\tau\to 0$ (\ref{2_21}) e.g. for the choice 
${\cal I}(\tau)=1+\tau^{-2}\aleph(0)/(2\pi)^2$. The further requirement of 
absence of any asymptotic corrections of higher order $\tau^{-n}$ also leaves it 
ambiguous, e.g., for $\omega_s>0$, any $\ell_s, c_s\neq 0$, as: 
\bea  
\fl 
{\cal I}(\tau)\!=\!
\prod\limits^\infty_{s=1}\!\left(\!\coth\frac{\tau^{\omega_s}}{c_s}\right)^{\ell_s}\!\!, 
\mbox{ if: }\prod\limits^\infty_{s=1}\!\left({\rm sign}(c_s)\right)^{\ell_s}\!=1,\quad  
\prod\limits^\infty_{s=1}\!(c_s)^{\ell_s}\!=\frac{\aleph(0)}{(2\pi)^2}, \quad 
\sum^\infty_{s=1}\omega_s\ell_s\!=2.  
\label{4_6} 
\eea 
The first and the second condition come from the first and second limit respectively but 
the second condition includs the first. Moreover they coincide for $\aleph(0)=(2\pi)^2$ 
if all $c_s=\pm 1$. 
Nevertheless the infinite numbers of values $\ell_s$ and $\omega_s$ remain restricted 
only by the third condition, which comes from the second limit (\ref{2_21}). The most 
``economic'' choice is: 
\bea  
{\cal I}(\tau)=
\left(\coth\frac{\tau^{\omega_1}}{c_1}\right)^{\ell_1},\;\mbox{ with: }\; 
\left(c_1\right)^{\ell_1}=\frac{\aleph(0)}{(2\pi)^2}, \;\;\;\; \omega_1\ell_1=2.  
\label{4_10} 
\eea 

The inner product of the states (\ref{eq:packet_state_scalar}), (\ref{2_19_0}), providing 
the self-adjointness of above operator $\widehat{\Xop}$ in the space of 
solutions to KG equation (\ref{eq:klein_gordon}) with the same mass $m_a=m_b$ 
\cite{tir,p_s,vrg,schw,blt,oksak,jost,s_w,stroc,Dvt}, in view of (\ref{eq:ktp_coord}) turns 
out to be naturally consistent with the inner product given by (8) of \cite{nn}, because in 
accord with (\ref{3_19_1}):  
\numparts
\bea
\fl 
\langle\{p_b,x_b,\sigma_b\}|\{p_a,x_a,\sigma_a\}\rangle =
{ N}_{\sigma a}{ N}_{\sigma b}\, e^{i(p_bx_b)-i(p_ax_a)}
\langle 0|\varphi\left(z^*_b\right)\varphi\left(z_a\right)|0\rangle 
\label{eq:scalar_dp_start_} \\
\fl 
=e^{i(p_bx_b)-i(p_ax_a)}\!\int\!\frac{d^3{\rm k}}{(2\pi)^3 2E_\ka } 
\,\phi^{\sigma_b}(\ka,\p_b)\,\phi^{\sigma_a}(\ka,\p_a)\,e^{i(k(x_a-x_b))}=
\left(F_{p_bx_b},F_{p_ax_a}\right)  
\label{2_30_0} \\
\fl 
=\!\int\! d^3{\rm x}\,F^*_{p_bx_b}(x)
(i\!\stackrel{\leftrightarrow}{\partial_x^{0}})F_{p_ax_a}(x)=
\frac{{N}_{\sigma a}{N}_{\sigma b}}{i}\,e^{i(p_bx_b)-i(p_ax_a)}\,D^-\!(z^*_b-z_a) 
\label{2_30} \\
\fl 
=\frac{{ N}_{\sigma a}{ N}_{\sigma b}}{i}\,e^{i(p_bx_b)-i(p_ax_a)}\!
\int\! d^3{\rm x}\,
D^-\!\left(z^*_b-x\right)\!\stackrel{\leftrightarrow}{\partial_x^{0}}\!
D^-\!\left(x-z_a\right).
\label{eq:scalar_dp_end}
\end{eqnarray}
\endnumparts
It is positively defined for fixed frequency type and does not depend on $x^0$ 
\cite{jost}. For $\sigma_a,\sigma_b \to 0$ independently it leads to the covariant 
orthogonality condition (\ref{2_4}) for solutions (\ref{2_04}):
\bea
\left(F_{p_bx_b},F_{q_ax_a}\right)
\longrightarrow
(f_{\rm p},f_{\q})= \braket{\p}{\q}=
(2\pi)^3 2E_{\rm p}\delta_3(\p-\q),
\label{2_30_1}
\eea
The complex 4-vector $z_a=z_a(p_a,\sigma_a)=x_a+i\zeta_a(p_a,\sigma_a)$, with time-like 
imaginary part $\zeta_a(p_a,\sigma_a)$ (\ref{z_1}) provides the correct analytic  
continuation (\ref{2_23_0}) to $V^+$ for all WFs \cite{blt,oksak,jost,s_w,stroc} in 
(\ref{2_18}), (\ref{2_19}), 
(\ref{2_30}), (\ref{eq:scalar_dp_end}) and below in (\ref{3_3})--(\ref{eq:spinor_wp_wf_end}), 
(\ref{3_07_03}). 

Absolute convergence of integral (\ref{eq:scalar_wp_wf}) due to $e^{-(k\zeta_a)}$ in the 
function (\ref{eq:phidef}) yields to estimation 
$|\psi_\sigma(\p_a,\underline{x})|\leqslant {\cal I}(\tau)$ uniformly relative to 
$x,x_a,\p_a$. 
So, non uniqueness of the scalar wave packet (\ref{eq:phidef}) ($g_2=0$) would mean an 
additional polynomial dependence on $(k-p_a)^2=2m^2_a-2(kp_a)$, i.e. on the scalar product 
$(kp_a)$ for ${N}_{\sigma}\mapsto\widehat{N}_{\sigma}$. 
This leads to formally previous expression of wave packet (\ref{2_9}), (\ref{2_10}), 
when the ``multiplier'' $\widehat{N}_{\sigma}$ is pulled out of the integrand as a 
polynomial on differential operator $\partial^2_{x_a}$, which acts on the function of 
previous wave packet: (for any function of $\zeta^2$,   
$\partial^\mu_\zeta \mapsto 2\zeta^\mu\partial_{\zeta^2}$) 
\bea
&&
\phi^\sigma(\ka,\p_a)\longmapsto
\widehat{ N}_{\sigma}\!\left(m_a,\zeta^2_a, -(k-p_a)^2\right) e^{-(k\zeta_a)}, 
\label {2_32}  \\
&&
F_{p_ax_a}(x)\longmapsto
\widehat{ N}_{\sigma}\!\left(m_a,\zeta^2_a, \partial^2_{x_a}\right)
e^{-i(p_ax_a)} \frac {1}{i} D^-_{m_a}(x-z_a),
\label {2_32_1}
\eea
and the same for the states (\ref{eq:packet_state_scalar}), (\ref{2_19_0}) and
for the scalar product (\ref{2_30}), (\ref{eq:scalar_dp_end}). 
Then, from dimensional reasoning in (\ref{2_20}), (\ref{2_20_0}) with 
$\tau^2=m^2\zeta^2_a \mapsto m^4g^2_{1}(m,\sigma)=(p_a\zeta_a)^2$ it follows, that for  
(\ref{2_20}), (\ref{2_20_0}):  
$-\partial^2_{x_a}\mapsto 2[m^2+(p_a\partial_{\zeta_a})]\mapsto 2m^2(1+\partial_\tau)$  
and 
$
m^2\widehat{ N}_{\sigma} \mapsto \widehat{\aleph}\left(\tau,\partial^2_{x_a}/m^2\right)
\mapsto \widehat{\aleph}(\tau,\partial_\tau)
$. 
The additional dependence on $\partial^2_{x_a}/m^2$ in (\ref{2_32_1}) doesn't affects the 
above plane-wave limit $\sigma\to 0$ (\ref{2_13}) and non relativistic limit 
$c\to\infty$ but, for $\Theta_a=(p_a x_a)$, leads to inadmissible contributions like 
$(p_a\partial_{x_a})\varphi(x_a)\ket{0}$, breaking down localization condition  
(\ref{2_15}) at $\sigma\to\infty$. 

Another dimensionless substitution:  
$m^2\widehat{ N}_{\sigma} \mapsto \widehat{\aleph}\left(\tau,-(k-p_a)^2/\sigma^2\right)$,
is possible in (\ref{2_32}). Contrariwise, this is innocent relative to the condition 
(\ref{2_15}), but in general breaks down the limit (\ref{2_13}), and in non relativistic 
limit leads to inevitable deformation of the Gaussian profile (\ref{eq:gauss_p}) by the 
polynomial on $(\Delta\ka)^2/\sigma^2$, and destroys its minimization properties (\ref{hein}). 
Thus, both those dimensionless combinations are excluded, and the expression 
(\ref{eq:phidef}) is the only acceptable from both physical and mathematical points  
of view \cite{Tl,g_w}. 

Nevertheless the polynomial $k$ - dependence arises below naturally for wave packet of the 
spin particle and for respective scalar product, which finally will fully exclude {\sl a 
posteriori} its arbitrary uncontrollable appearance in   (\ref{2_32}), (\ref{2_32_1}). 

\subsection{Contracted Relativistic Gaussian approximation}\label{sec:CRG_app}
\begin{figure}[htb]
\centering  
\includegraphics[width=.69\textwidth ]{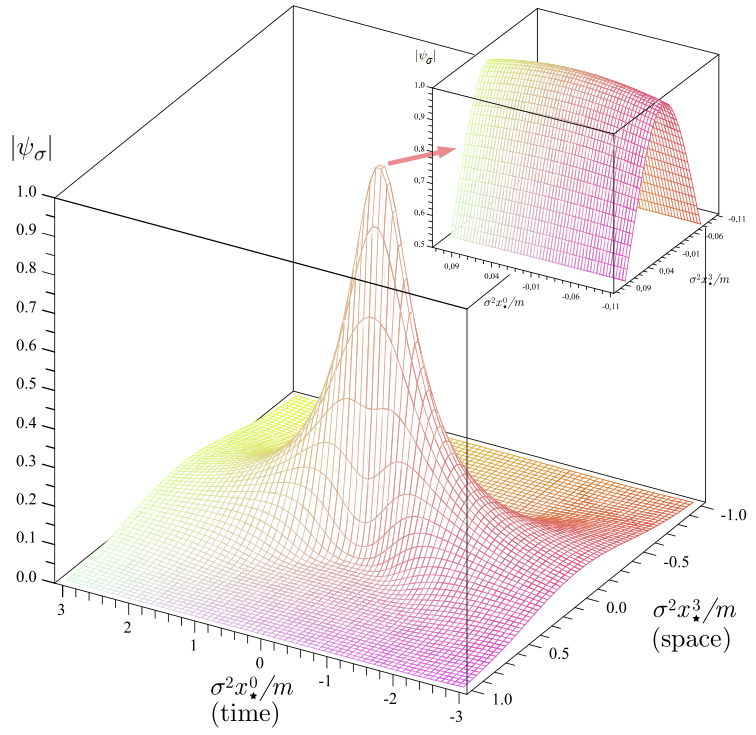}
\caption{The value of $|\psi_\sigma(\vec{0},\underline{x}_*)|$,  
(\ref{2_15_34})--(\ref{2_15_3_5}), for $\underline{x}^\beta_* =(x^0_*,0,0,x^3_*)$, 
with $\zeta_a=p_a(2\sigma^2_a)^{-1}$, ${\cal I}(\tau)=1$, as a function of 
dimensionless variables $\sigma^2 x^0_*/m$ and $\sigma^2 x^3_*/m$, for $\sigma/m=0,1$.}
\label{fig:1}
\end{figure}

A fully relativistic Gaussian approximation for the (pseudo) scalar wave packet (\ref{2_18}) 
was suggested in \cite{nn,nsh} for $g_2\equiv 0$, 
$\zeta_a(p_a,\sigma_a)=p_a(2\sigma^2_a)^{-1}$, $\tau^2=m^2\zeta^2_a\to\infty$. 
Definitions (\ref{z_1}), (\ref{2_20})--(\ref{2_21}) with the use of (\ref{A_7}), 
(\ref{2_15_36}), for: $m=m_a$, $\tau=\tau_a$, $\;\underline{x}=x_a-x$, 
\bea
\fl 
\psi_\sigma(\p_a,\underline{x})={ N}_\sigma
\langle 0|\varphi(x)\varphi\left(x_a+i\zeta_a\right)|0\rangle=
{\cal I}(\tau)\frac{h (\widehat{Z}_a(\underline{x}))}{h(\tau^2)}, \quad 
\upsilon_a=\frac{\zeta_a}{\sqrt{\zeta^2_a}},\quad \upsilon^2_a=1, 
\label{2_15_34} \\
\fl 
\widehat{Z}_a(\underline{x})=-m^2(\underline{x}+i\zeta_a)^2=
m^2\left[\zeta^2_a-2i\left(\underline{x}\zeta_a\right)-\underline{x}^2\right]=
m^2\left\{\left[\sqrt{\zeta^2_a}-i\underline{x}^0_*\right]^2+\underline{\x}^2_*\right\}, 
\label{2_15_35} \\
\fl 
\underline{x}^0_*=(\underline{x}\upsilon_a), \quad 
\underline{\x}^2_*=(\underline{x}\upsilon_a)^2-\underline{x}^2
\equiv -\underline{x}^2_\perp, \;\;\mbox{ when: }\; 
\zeta^2_a\gg (\underline{x}\zeta_a),\underline{x}^2, \quad {\cal I}(\tau)\mapsto 1,
\label{2_15_3_5} \\
\fl 
\sqrt{\widehat{Z}_a(\underline{x})}\approx m\sqrt{\zeta^2_a}
-i\,m \frac{\left(\underline{x}\zeta_a\right)}{\sqrt{\zeta^2_a} }+
\frac{m}{2(\zeta^2_a)^{3/2}}
\left[(\underline{x}\,\zeta_a)^2-\underline{x}^2\zeta^2_a\right]=
\tau-im\underline{x}^0_*+\frac{m^2}{2\tau}\underline{\x}^2_*, 
\label{2_15_37} \\
\fl 
\mbox{give: }\;                      
\psi^{CRG}_\sigma(\p_a,\underline{x})=
\exp\left\{i\,m(\underline{x}\upsilon_a)\right\}
\exp\left\{i\,\frac {3}{2}\frac{m(\underline{x}\upsilon_a)}{\tau}
-\frac{m^2}{2\tau}\left[(\underline{x}\upsilon_a)^2-\underline{x}^2\right]\right\},  
\label{2_15_41} \\
\fl 
\mbox{where: }\;
\frac{m^2}{2\tau}\left[(\underline{x}\upsilon_a)^2-\underline{x}^2\right]\equiv
\left(\underline{x}\vec{T}_a\underline{x}\right)=
T^{\beta\lambda}_a\underline{x}_\beta\underline{x}_\lambda 
\label{2_15_40_0} \\
\fl 
=\frac{m^2}{2\tau}\frac{g^2_1}{\zeta^2_a} 
\left[(\underline{x}p_a)^2-\underline{x}^2 m^2
+2\frac{g_2}{g_1}(\underline{x}p_a)(\underline{x}\widehat{w}_a)+
\frac{g^2_2}{g^2_1}\left[(\underline{x}\widehat{w}_a)^2-
\underline{x}^2\widehat{w}^2_a\right]\right]. 
\label{2_15_40} 
\eea 
The values (\ref{2_15_3_5}) define 4-vector 
$\underline{x}^\beta_*=(\underline{x}^0_*,\underline{\x}_*)$ in the rest frame of time-like 
vector $\upsilon_a$ (\ref{2_15_34}).  
The first imaginary term of last exponential (\ref{2_15_41}) was arbitrarily omitted in 
(21) of \cite{nn}. In spite of the similar to \cite{nsh} ``hidden spin'' asymmetry of the 
quadratic form (\ref{2_15_40_0}), (\ref{2_15_40}), such generalized to $g_2\neq 0$ 
CRG- approximation keeps its non negative definiteness like in \cite{nn}. 
The drawback of CRG- approximation (\ref{2_15_41}) is, that due to the Gaussian 
exponential it obviously can't be a solution to free KG equation (\ref{eq:klein_gordon}), 
i.e. can't represent an {\sl external particle} on the mass shell. 
Thus, strictly speaking, {\sl it is not a wave packet} in the usual sense, with definite 
covariant dispersion equation \cite{g_w}. 
However, if it is considered as a suitable approximation to the exact wave packet, then 
the function (\ref{2_15_41}) describes propagation with specific frequency mixing but, 
without of any spreading \cite{nn,nsh} in the ``Gaussian sense'' \cite{dn}. 
Indeed, the Figure \ref{fig:1} (see also \cite{nsh}) shows a very small spreading of 
exact wave packet near its peak\footnote{We thank V.A. Naumov for kind permission 
to use this and next pictures.}, which is analytically approximated in (\ref{2_15_41}) as 
absence of spreading.  

These properties are clarified by Fourier-image of function (\ref{2_15_41}), not 
discussed in  \cite{nn,nsh}. For the simplified case of  \cite{nn}, with 
$g_2=0$, $\upsilon_a\mapsto u_a=p_a/m$, $u^2_a=1$, it reads: 
\numparts
\bea
&&\!\!\!\!\!\!\!\!\!\!\!\!\!\!\!\!\! 
\int d^4\underline{x}\,e^{-i(k\underline{x})}\,\psi^{CRG}_{\sigma\,[1]}(\p,\underline{x})=
\int d^4\underline{x}\,e^{i((p_a-k)\underline{x})}\,
e^{-\left(\underline{x}\vec{T}_a\underline{x}\right)}  
\label{f_1} \\
&&\!\!\!\!\!\!\!\!\!\!\!\!\!\!\!\!\! 
=(2\pi)^4\delta\left((u_ak)-m\right)\left(\frac{\tau}{2\pi m^2}\right)^{3/2}
\exp\left\{\frac{\tau}{2m^2}(k^2-m^2)\right\}  
\label{f_2} \\
&&\!\!\!\!\!\!\!\!\!\!\!\!\!\!\!\!\! 
=(2\pi)^4\delta\left(k^0_*-m\right)\left(\frac{\tau}{2\pi m^2}\right)^{3/2}
\exp\left\{-\,\frac{\tau}{2m^2}\ka^2_*\right\}, 
\label{f_3}
\eea
\endnumparts
where for calculation of the integral the projective properties of tensor 
$T^{\beta\lambda}_a$ are used in the variables $\underline{x}^0_*=(u_a\underline{x})$ 
and $\underline{x}_\perp=\Pi_{u_a}\underline{x}$, with: $k_\perp=\Pi_{u_a}k$,  
$(\underline{x}_\perp k)=(\underline{x}_\perp k_\perp)=-(\underline{\x}_*\cdot\ka_*)$, 
$\underline{x}^2_\perp=(\underline{x}\Pi_{u_a}\underline{x})=-\underline{\x}^2_* <0$, 
$k^2_\perp=(k\Pi_{u_a}k)=k^2-(u_ak)^2=-\ka^2_*<0$. Indeed, since: 
\bea
\fl 
T^{\beta\lambda}_a=-\frac{m^2}{2\tau}\Pi^{\beta\lambda}_{u_a},\quad 
\Pi^{\beta\lambda}_{u_a}={\rm g}^{\beta\lambda}-u^\beta_au^\lambda_a=
\Pi^{\beta\gamma}_{u_a}\Pi^{\;\lambda}_{\gamma\,\;{u_a}}, \quad 
u_a\Pi_{u_a}=\Pi_{u_a}u_a=0, 
\label{f_4} \\
\fl 
\mbox {then, as: }\;
k^0_*=(u_ak), \quad \ka_*=\ka+\us_a\left[\frac{(\us_a\!\cdot \ka)}{u^0_a+1}-k^0\right]=
\ka-\us_a\frac{k^0+(u_ak)}{u^0_a+1}, 
\label{f_5} 
\eea 
as well as $\underline{x}^0_*$, $\underline{\x}_*$ are defined by the same Lorentz 
transformation (\ref{f_5}) to the rest frame of wave packet $\p_{a*}=0$, 
$p^\lambda_{a*}=(m,\vec{0})$. 
Expressions (\ref{f_2}), (\ref{f_3}) should be compared with the multiplied by   
$2\pi$ l.h.s. of relation (\ref{4_12}). The dispersion equation now depends on the 
reference frame as $k^0=[m+(\us_a\!\cdot \ka)]/u^0_a$. Only in the plane-wave limit 
$\tau\to\infty$ both compared expressions coincide with 
$(2\pi)^4\delta_4\left(k-p_a\right)$ in arbitrary reference frame. 

The relevance of approximation (\ref{2_15_41}) will be seen below in subsections  
\ref{sec:wf_asymp}, \ref{sec:amp_asymp}, where it will be naturally reproduced by 
asymptotic behavior at $\underline{x}=-R_{\{\}} \to\infty$ of defined therein off-shell 
and on-shell composite wave functions $\widehat{\cal F}^{(\xi)}_{\{C/D\}j}(\varrho)$ 
(\ref{C_04}), (\ref{5_20}) for neutrino creation/detection vertices, as well as for 
the oscillation amplitude (\ref{5_32}).

\section{Fermionic wave packet. Higher spins.}\label{sec:wp_ff}
The relativistic fermionic wave packet is constructed in completely similar way to
described above scalar packet. A free massive quantum Fermi-Dirac field $\psi(x)$, 
being a solotion to equation $\left[i(\gamma\partial_x)-m\right]\psi(x)=0$, has 
the form \cite{blp,p_s,vrg,tir,schw,blt,oksak,jost,s_w,stroc,Dvt,b_d,i_z}:
\bea
&&\!\!\!\!\!\!\!\!\!\!\!\!\!\!\!\!\!\!\!\!\!
\psi_\alpha(x)=\!\sum_{r=\pm 1/2} \int\! \frac{d^3{\rm k}}{(2\pi)^3 2k^0}
\left(u^{(+)}_\alpha(k,r)f_{\rm k}(x)\,b^{(+)}_{{\rm k},r}+
u^{(-)}_\alpha(k,r)f_{-\rm k}(x)\,b^{(-)\dagger}_{{\rm k},r}\right), 
\label {3_1} \\
&&\!\!\!\!\!\!\!\!\!\!\!\!\!\!\!\!\!\!\!\!\!
\mbox{for: }\;
k^0 = E_{\rm k}>0, \quad b^{(\xi)}_{{\rm k},s}|0\rangle=0,
\quad b^{(\xi)\dagger}_{{\rm k},s}|0\rangle=|(\xi);\ka,s\rangle,
\label {3_1_0}
\eea
and respectively for Dirac-conjucated field $\overline \psi(x)=\psi^\dagger(x)\gamma^0$. 
The bispinors $u^{\xi}(k, r)$ are 
solutions to the free Dirac equations $\left[(\gamma k)-\xi m\right]u^{\xi}(k, r)=0$,
with ``positive and negative'' energy according to the index $\xi = \pm$. Analogously 
(\ref{2_04}) for $k^\mu=(E_{\rm k},\ka)$, $f_{\xi\rm k}(x)=e^{-i\xi(kx)}$, 
$\overline{u}=u^\dagger\gamma^0$ they define the matrix elements as corresponding wave 
functions of initial (or final) states for fermions $u^{(+)}(k,s)$ (or 
$\overline{u}^{(+)}(k,s)$) and for antifermions $\overline{u}^{(-)}(k,s)$ (or $u^{(-)}(k,s)$) 
with mass $m$, momentum $\ka$ and spin $s=\pm 1/2$ along its quantization 4- axis 
$\widehat{s}^\mu$ \cite{blp,p_s,vrg,tir,schw,blt,oksak,jost,s_w,stroc,Dvt,b_d,i_z}:
\numparts
\bea
&&
\langle 0|\psi(x)|(+);\ka,s\rangle= u^{(+)}(k,s)f_{\rm k}(x)=
{\cal U}^{(+)}_{{\rm k},s}(x), 
\label {3_2:1} \\
&&
\langle \ka,s;(+)|\overline{\psi}(x)|0 \rangle=
\overline{u}^{(+)}(k,s)f^*_{\rm k}(x)=\overline{{\cal U}}^{(+)}_{{\rm k},s}(x),  
\label {3_2:2} \\
&&
\langle \ka,s;(-)|\psi(x)|0 \rangle= u^{(-)}(k,s)f_{-\rm k}(x)=
{\cal U}^{(-)}_{{\rm k},s}(x),
\label {3_2:3} \\
&&
\langle 0|\overline{\psi}(x)|(-);\ka,s\rangle=
\overline{u}^{(-)}(k,s)f^*_{-\rm k}(x)=\overline{{\cal U}}^{(-)}_{{\rm k},s}(x),
\label {3_2:4}  
\eea
\endnumparts
where, for any representation of $\gamma^\lambda$ and any such axis 
\cite{i_z,Nvj,Bil} $\widehat{s}^2=-1$, $(\widehat{s}k)=0$ (clf. (\ref{3_7_04})), 
with $\xi,\eta=\pm$:
\numparts
\bea
\fl 
({\cal W}\,\widehat{s})\mapsto -\,\frac 12\gamma^5(\gamma\widehat{s})\xi(\gamma k), \quad 
u^{\xi}(k,s)\overline{u}^{\xi}(k,s)=\left[(\gamma k)+\xi m\right]
\frac 12\left[{\rm I}+2s\gamma^5(\gamma \widehat{s})\right], 
\label {3_2_1_0} \\
\fl 
\overline{u}^\xi(k,s)u^\eta(k,r)=2m\xi\delta_{\xi\eta}\delta_{rs}, \quad 
u^{\xi\dagger}(k^0,\xi\ka;s)u^\eta(k^0,\eta\ka;r)=
2E_{\rm k}\delta_{rs}\delta_{\xi\eta}, 
\label {3_2_10} \\
\fl 
\left\{b^{(\xi)}_{{\rm q},r},b^{(\eta)\dagger}_{{\rm k},s}\right\}= 
\langle\q,r;(\xi)|(\eta);\ka,s\rangle=
(2\pi)^32E_{\rm k}\delta_3(\ka-\q)\delta_{rs}\delta_{\xi\eta}
= \left({\cal U}^\xi_{{\rm q},s},{\cal U}^\eta_{{\rm k},r}\right)  
\label {3_2_0} \\
\fl 
=\left({\cal U}^\eta_{{\rm k},r},{\cal U}^\xi_{{\rm q},s}\right)=
\int d^3{\rm x}\,\,{\cal U}^{\xi\dagger}_{{\rm q},s}(x)\,{\cal U}^\eta_{{\rm k},r}(x)=
\int d^3{\rm x}\,\,\overline{\cal U}^{\xi}_{{\rm q},s}(x)\,\gamma^0\,
{\cal U}^\eta_{{\rm k},r}(x). 
\label {3_2_2}
\eea
\endnumparts
Following the scalar case (\ref{eq:packet_state_scalar}), (\ref{2_19_0}), the 
wave-packet state is created by operators:
\numparts
\bea
\fl 
|(+);\{\p_a,x_a,s_a\}\rangle =\widehat{\rm N}^\xi_{\sigma a}
\overline{\psi}\left(x_a+i\zeta_a\right)|0\rangle\, 
{\cal U}^{(+)}_{{\rm p}_a,s_a}(x_a)=\widehat{\rm N}^\xi_{\sigma a}
\overline{\psi}\left(z_a\right)|0\rangle\, 
{\cal U}^{(+)}_{{\rm p}_a,s_a}(x_a),
\label{eq:spinor_wp_ket_b_fin_0} \\
\fl 
|(-);\{\p_a,x_a,s_a\}\rangle=\widehat{\rm N}^\xi_{\sigma a}
\overline{\cal U}^{(-)}_{{\rm p}_a,s_a}(x_a)
\psi\left(x_a+i\zeta_a\right)|0\rangle=\widehat{\rm N}^\xi_{\sigma a}
\overline{\cal U}^{(-)}_{{\rm p}_a,s_a}(x_a)
\psi\left(z_a\right)|0\rangle, 
\label{eq:spinor_wp_ket_b_fin_1} \\
\fl 
\mbox{with: }\widehat{\rm N}^\xi_{\sigma a}\!=\frac{\xi { N}_{\sigma a}}{2m_a},
\;\quad \Xi^{\xi}_{p_a,x_a,s_a}(x)\stackunder{\sigma\to 0}{\longrightarrow}
{\cal U}^{\xi}_{{\rm p}_a,s_a}(x),
\;\quad \overline{\Xi}^{\xi}_{p_a,x_a,s_a}(x)\stackunder{\sigma\to 0}{\longrightarrow}
\overline{\cal U}^{\xi}_{{\rm p}_a,s_a}(x), 
\label{eq:spinor_wp_ket_start} 
\eea
\endnumparts
where the wave functions of these states and the respective Dirac-conjugated ones,
according to (\ref{3_2:1})--(\ref{3_2:4}) and similarly (\ref{2_10}), (\ref{2_18}), for 
$\underline{x}=x_{a}-x$, $\zeta_a=\zeta_a(p_a,\sigma_a)$, 
$z_a=x_a+i\zeta_a$, $z^*_a=x_a-i\zeta_a$, $\xi=\pm$, with the same ${ N}_{\sigma a}$ 
as for the scalar case (\ref{2_20})--(\ref{2_21}), are defined as bispinors: 
\numparts
\bea
\fl 
\Xi^{(+)}_{p_a,x_a,s_a}(x)=\langle 0|\psi(x)|(+);\{\p_a,x_a,s_a\}\rangle=
\widehat{\rm N}^\xi_{\sigma a}
\langle 0|\psi(x)\overline{\psi}\left(x_a+i\zeta_a\right)|0\rangle\, 
{\cal U}^{(+)}_{{\rm p}_a,s_a}(x_a)  
\nonumber \\
\fl 
=(-i) \widehat{\rm N}^\xi_{\sigma a}  
S^-\left(-\underline{x}-i\zeta_a\right)\,{\cal U}^{(+)}_{{\rm p}_a,s_a}(x_a)=
(-i) \widehat{\rm N}^\xi_{\sigma a}  
S^-\left(x-z_a\right)\,{\cal U}^{(+)}_{{\rm p}_a,s_a}(x_a), \quad (\xi=+),
\label {3_3} \\
\fl 
\overline{\Xi}^{(+)}_{p_a,x_a,s_a}(x)=
\langle\{\p_a,x_a,s_a\};(+)|\overline{\psi}(x)|0\rangle=
\widehat{\rm N}^\xi_{\sigma a}\overline{\cal U}^{(+)}_{{\rm p}_a,s_a}(x_a)
\langle 0|\psi\left(x_a-i\zeta_a\right)\overline{\psi}(x)|0\rangle 
\nonumber \\
\fl 
=(-i)\widehat{\rm N}^\xi_{\sigma a}
\overline{\cal U}^{(+)}_{{\rm p}_a,s_a}(x_a)
S^-\left(\underline{x}-i\zeta_a\right)=
(-i)\widehat{\rm N}^\xi_{\sigma a}
\overline{\cal U}^{(+)}_{{\rm p}_a,s_a}(x_a)S^-\left(z^*_a-x\right), \quad (\xi=+),
\label {3_3_1} \\
\fl 
\Xi^{(-)}_{p_a,x_a,s_a,\alpha}(x)=
\langle\{\p_a,x_a,s_a\};(-)|\psi_\alpha(x)|0\rangle=
\widehat{\rm N}^\xi_{\sigma a}
\langle 0|\overline{\psi}_\beta\!\left(x_a-i\zeta_a\right)
\psi_\alpha(x)|0\rangle\,{\cal U}^{(-)}_{{\rm p}_a,s_a, \beta}(x_a) 
\nonumber \\
\fl 
=(-i)\widehat{\rm N}^\xi_{\sigma a}
S^+_{\alpha\beta}\!\left(-\underline{x}+i\zeta_a\right)
{\cal U}^{(-)}_{{\rm p}_a,s_a,\beta}(x_a)=
(-i)\widehat{\rm N}^\xi_{\sigma a}S^+_{\alpha\beta}\left(x-z^*_a\right)
{\cal U}^{(-)}_{{\rm p}_a,s_a,\beta}(x_a), \; (\xi=-),
\label {3_3_2} \\
\fl 
\overline{\Xi}^{(-)}_{p_a,x_a,s_a,\beta}(x)=
\langle 0|\overline{\psi}_\beta(x)|(-);\{\p_a,x_a,s_a\}\rangle=
\widehat{\rm N}^\xi_{\sigma a}
\overline{\cal U}^{(-)}_{{\rm p}_a,s_a,\alpha}(x_a)
\langle 0|\overline{\psi}_\beta(x)\psi_\alpha\left(x_a+i\zeta_a\!\right)|0\rangle  
\nonumber \\
\fl 
=(-i) \widehat{\rm N}^\xi_{\sigma a}
\overline{\cal U}^{(-)}_{{\rm p}_a,s_a,\alpha}(x_a)
S^+_{\alpha\beta}\left(\underline{x}+i\zeta_a\right)=
(-i) \widehat{\rm N}^\xi_{\sigma a}
\overline{\cal U}^{(-)}_{{\rm p}_a,s_a,\alpha}(x_a)
S^+_{\alpha\beta}\left(z_a-x\right)\!, \;\; (\xi=-).
\label{eq:spinor_wp_wf_end}
\eea
\endnumparts
The fermionic WFs $S^\pm(x)$ are connected mutually and with causal propagator $S^{c}(x)$,   
and with the scalar WF $D^-(x)$ (\ref{2_19}), (\ref{2_23_0}), for  
$(D^-(x))^*=D^+(x^*)=-D^-(-x^*)$ , and $D^c(x)$ (\ref{2_30_01}), by the relations 
\cite{tir,p_s,vrg,schw,blt,oksak,jost,s_w,stroc,Dvt}, containing charge conjugation matrix 
$C^\top=-C$, $C\gamma^\top_\mu C^{-1}=-\gamma_\mu$: 
\numparts
\bea
&&\!\!\!\!\!\!\!\!\!\!\!\!\!\!\!\!\!\!
\frac 1i S^{-\xi}(x-y)=\!\int\!
\frac{d^4 k}{(2\pi)^3}\,\theta(k^0)\,\delta(k^2-m^2)\, e^{-i\xi\left(k(x-y)\right)}\, 
\left[(\gamma k)+\xi m\right],
\label{2_15_51_2_} \\
&& \!\!\!\!\!\!\!\!\!\!\!\!\!\!\!\!\!\!
S^+(x)=-C S^{-\top}(-x)C^{-1}, \quad S^{c}(x)=C S^{c\top}(-x)C^{-1}, 
\label{2_15_510} \\
&& \!\!\!\!\!\!\!\!\!\!\!\!\!\!\!\!\!\!
\mbox{(see (\ref{A_1})--(\ref{A_1_2})), but with }\;
\langle 0|\psi(x)\psi(y)|0\rangle\equiv 0,   
\label{2_15_51_0} 
\eea
\endnumparts
and instead of (\ref{3_19_1}), 
for $\xi,\eta=\pm$, 
$\varrho^\mu=(\varrho^0,\vec{\rho})$ satisfy $\forall\,\varrho^0$ or space-like 
$\Sigma(\varrho)$ \cite{tir,p_s,vrg,schw,blt,oksak,jost,s_w,stroc,Dvt}: 
\bea
\fl 
\delta_{\xi\eta}\,\frac 1i S^{-\xi}(x-y)=\!
\int\!\!d^3\!\rho\,\frac 1i S^{-\xi}(x-\varrho)\,\gamma^0\,\frac 1i\,
S^{-\eta}(\varrho-y),
\;\mbox { or with }\; d^3\!\rho \gamma^0\mapsto d\Sigma_\mu(\varrho)\gamma^\mu, 
\label{eq:spinor_weightman_conv_same}
\eea
also for complex values of $x,y$. So, the similar to (\ref{3_2_0}), (\ref{3_2_2}) inner 
product for the packets (\ref{3_3})--(\ref{eq:spinor_wp_wf_end}) reads:
\numparts
\bea
\fl 
\langle\{\p_a,x_a,s_a\};(\eta)|(\xi);\{\p_c,x_c,s_c \}\rangle = 
\delta_{\xi\eta}\int d^3{\rm x}\left\{\begin{array}{c} 
\overline{\Xi}{}^{(+)}_{p_a,x_a,s_a}(x)\,\gamma^0\,\Xi^{(+)}_{p_c,x_c,s_c}(x) \\
\overline{\Xi}{}^{(-)}_{p_c,x_c,s_c}(x)\,\gamma^0\,\Xi^{(-)}_{p_a,x_a,s_a}(x)
\end{array} \right \}  
\label{eq:spinor_dp_sm_} \\
\fl 
\equiv \left(\Xi^{\,\eta}_{\{a/c\}},\,\Xi^{\xi}_{\{c/a\}}\right) = 
\delta_{\xi\eta}\widehat{\rm N}^\xi_{\sigma a}\widehat{\rm N}^\xi_{\sigma c}\! 
\left \{\!\begin{array}{c} 
\overline{{\cal U}}{}^{(+)}_{{\rm p}_a,s_a,\alpha}(x_a)\,
\langle 0| \psi_\alpha\left(z^*_a\right)\overline{\psi}_\beta\left(z_c\right)|0\rangle 
{\cal U}^{(+)}_{{\rm p}_c,s_c,\beta}(x_c)  \\
\overline{{\cal U}}{}^{(-)}_{{\rm p}_c,s_c,\alpha}(x_c)
\langle 0|\overline{\psi}_\beta\left(z^*_a\right)\psi_\alpha\left(z_c\right)|0\rangle\,
{\cal U}^{(-)}_{{\rm p}_a,s_a,\beta}(x_a)\end{array}\! \right \}  
\label{3_4_0} \\
\fl 
=\delta_{\xi\eta}\widehat{\rm N}^\xi_{\sigma a}\widehat{\rm N}^\xi_{\sigma c} 
\left\{\begin{array}{c} 
\overline{{\cal U}}{}^{(+)}_{{\rm p}_a,s_a}(x_a) \\
\overline{{\cal U}}{}^{(-)}_{{\rm p}_c,s_c}(x_c) 
\end{array}\right\}\, \frac{1}i\,
S^{-\xi}\left(\xi\left(z^*_a-z_c\right)\right)
\left\{\begin{array}{c} 
{\cal U}^{(+)}_{{\rm p}_c,s_c}(x_c) \\
{\cal U}^{(-)}_{{\rm p}_a,s_a}(x_a)
\end{array} \right \}  
\label{eq:scalar_conv} \\
\fl 
=\!\int\! d^3{\rm x} \,\overline{\Xi}{}^{\,\eta}_{\{a/c\}}(x)\,\gamma^0 \!\!
\int\! d^3{\rm y}\, \frac{\xi}i\, S^c(x-y)\,\gamma^0\, \Xi^{\xi}_{\{c/a\}}(y), 
\quad  \forall \,x^0_a,\,x^0_c, \quad \xi(x^0-y^0)>0.
\label{3_4}
\eea
\endnumparts
The expressions (\ref{2_30}), (\ref{eq:scalar_conv}) are independent of the choice of 
$x^0$, $y^0$ in (\ref{eq:scalar_dp_end}), (\ref{eq:spinor_dp_sm_}), (\ref{3_4}) 
respectively. However, according the meaning of (\ref{eq:limit_sigma_inf}) and Huygens' 
principle, for $x^0_a>x^0_c$ it may be considered as the projections (\ref{3_4}) of `final' 
packet ``$a/c$'' onto the result of causal evolution of the `initial' packet ``$c/a$'',  
according to causal propagator $S^c(x-y)$, where both these packets, due to (\ref{A_1_1}), 
(\ref{eq:spinor_weightman_conv_same}), have to select natural causal sequence of events, which 
for both $\xi=\pm$ cases is reduced to the same sequence: 
\bea 
&&
\theta(x^0_a-x^0)\theta(x^0-y^0)\theta(y^0-x^0_c)\longmapsto\theta(x^0_a-x^0_c),\quad (\xi=+),
\;\mbox{ or } 
\nonumber \\
&&
\theta(x^0-x^0_c)\theta(y^0-x^0)\theta(x^0_a-y^0)\longmapsto\theta(x^0_a-x^0_c),\quad (\xi=-),
\label{3_thet} 
\eea
if the time-ordering $\theta$-functions for specific process are formally assigned to 
the packets itself \cite{tir}. 

From (\ref{eq:scalar_conv}) for $a=c$, $\xi=\eta$, by making use of (\ref{2_30_0}), 
(\ref{3_2:1})--(\ref{3_2:4}), (\ref{3_2_1_0}), (\ref{eq:phidef})--(\ref{2_19}), (\ref{2_20}), 
(\ref{2_20_0}), with $m^2\zeta^2_a=\tau^2$, $(p_a\widehat{s}_a)=0$, for normalization of 
the Fermi packet with spin $2s_a=\pm 1$, one has: 
\bea 
\fl 
{\rm A}^2_\sigma=\langle\{\p_a,x_a,s_a\};(\xi)|(\xi);\{\p_a,x_a,s_a \}\rangle
\equiv \left(\Xi^{\xi}_{\{a\}},\,\Xi^{\xi}_{\{a\}}\right) 
\label{3_5} \\
\fl 
=\int\frac{d^3{\rm k}}{(2\pi)^3 2E_\ka} |\phi^\sigma(\ka,\p_a)|^2 
\left[\overline{{\cal U}}^{\xi}_{{\rm p}_a,s_a}(x_a)
\frac{(\gamma k)+\xi m}{(2m)^2}\,{\cal U}^{\xi}_{{\rm p}_a,s_a}(x_a)\right]  
\nonumber \\
\fl 
={ N}_\sigma^2\int\frac{d^3{\rm k}}{(2\pi)^3 2E_\ka}
e^{-2(k\zeta_a(p_a,\sigma_a))}
\mathrm{Sp}\left[\frac{(\gamma k)+\xi m}{(2m)^2}\left\{(\gamma p_a)+\xi m\right\}
\frac{{\rm I} \pm \gamma^5(\gamma\widehat{s}_a)}{2}\right]  
\nonumber \\ 
\fl 
={ N}_\sigma^2 \int\frac{d^3{\rm k}}{(2\pi)^3 2E_\ka}
e^{-2(k\zeta_a(p_a,\sigma_a))}\left[\frac 12+\frac{(kp_a)}{2 m^2}\right]  
\label{3_5_0} \\ 
\fl 
={ N}_\sigma^2\left[\frac 12-\frac{(p_a\partial_{\zeta_a})}{4 m^2}\right]\!
\frac 1i D^-_{m}\left(-2i\zeta_a(p_a,\sigma_a)\right)=
\frac{\aleph^2(\tau)}{m^2 (2\pi)^2}
\left[\frac 12-\frac{(p_a\zeta_a)}2\frac{\partial}{\partial\tau^2}\right] 
h\left(4\tau^2\right).  
\label{3_6} 
\eea                                       
For the scalar packet (\ref{2_30_0}) all the square brackets here should be omitted.
The absence of $\widehat{s}_a$ - contribution indicates an independence of 
this result also of the spin averaging and suggests again to neglect $g_2$ in (\ref{z_1}),  
whence $(p_a\zeta_a)\mapsto\tau$, and from definitions (\ref{2_20_0}), \cite{b_er} it 
follows: 
\bea
\fl 
A^2_\sigma\bigr|_{S=0}\!=\frac{\aleph^2(\tau)}{m^2 4\pi^2}h(4\tau^2)=
\frac{\aleph^2(\tau)}{m^2 4\pi^2}\frac{K_1(2\tau)}{2\tau}, \quad 
{\rm A}^2_\sigma\bigr|_{S=\frac 12}\!\mapsto\!\frac{\aleph^2(\tau)}{m^2 4\pi^2}\!
\left[\frac{K_1(2\tau)+K_2(2\tau)}{4\tau}\right]\!. 
\label{3_6_0}
\eea 
The last normalization of Fermi packet, with definite spin, conforms with the first 
normalization of scalar packet from (\ref{2_30}) for $b=a$ and coincides with it for 
$\sigma\to 0$, $\tau\to\infty$ \cite{b_er}: ${\rm A}^2_\sigma \to A^2_\sigma$.
Although such consistency will be saved also in the case of general dependence 
(\ref{2_32}), (\ref{2_32_1}), we'll emphasize, the form (\ref{eq:phidef}), 
(\ref{2_20_0})--(\ref{2_21}) of the function $\phi^\sigma(\ka,\p_a)$ is already 
uniquely defined {\sl for any spin} by uniqueness condition of analytic continuation 
\cite{jost,s_w,stroc} of state vectors (\ref{eq:packet_state_scalar}), (\ref{2_19_0}), 
(\ref{eq:spinor_wp_ket_b_fin_0}), (\ref{eq:spinor_wp_ket_b_fin_1}) and respective 
WFs \cite{blt,oksak,jost,s_w,stroc}
(clf. (\ref{2_23_0})) as a wave-packet functions (\ref{2_18}), 
(\ref{3_3})--(\ref{eq:spinor_wp_wf_end}), by the limiting 
conditions (\ref{2_13}), (\ref{2_14}) and (\ref{2_16}), (\ref{2_17}), by described 
above transformation of (\ref{2_28_2}) into (\ref{eq:gauss_p}) in nonrelativistic 
limit, and by the consistency conditions (\ref{eq:spinor_wp_ket_start}) with Fermi 
packet. 
Whereas the appearance of any polynomial $(kp_a)$ - dependence in (\ref{3_5_0}) is 
fully regulated by the spin \cite{oksak,i_z} of wave packet. 

For example, for spin $S=1$ of free real massive vector field $B^\mu(x)$ with the 
WF ${\cal D}^{\,-}_{\mu\nu}(x)$ (\ref{A_6}) and the wave function of state with 
definite momentum and polarization 
${\cal A}^{\nu}_{{\rm k},\lambda}(x)=\epsilon^\nu_{(\lambda)}(k)f_{\rm k}(x)$, 
the wave-packet state and its vector wave function are:
\bea
\fl 
|\{\p_a,x_a,\lambda_a\}\rangle ={ N}_{\sigma a}
B_\nu\left(x_a+i\zeta_a\right)|0\rangle \, 
{\cal A}^{\nu}_{{\rm p}_a,\lambda_a}(x_a)=
{ N}_{\sigma a}B_\nu\left(z_a\right)|0\rangle\,
{\cal A}^{\nu}_{{\rm p}_a,\lambda_a}(x_a), 
\label{3_07_01} \\
\fl 
{\rm F}^{\mu}_{p_a,x_a,\lambda_a}(x)=
\langle 0|B^\mu(x)|\{\p_a,x_a,\lambda_a\}\rangle= { N}_{\sigma a}
\langle 0|B^\mu(x)B_\nu\left(x_a+i\zeta_a\right)|0\rangle  
{\cal A}^{\nu}_{{\rm p}_a,\lambda_a}(x_a)  
\label{3_07_02}  \\ 
\fl 
= i{ N}_{\sigma a}\,{\rm g}^{\mu\beta}\,{\cal D}^{\,-}_{\beta\nu}\left(x-z_a\right)\,
{\cal A}^{\nu}_{{\rm p}_a,\lambda_a}(x_a), \;\mbox{ with: }\;
{\rm F}^{\mu}_{p_a,x_a,\lambda_a}(x)\stackunder{\sigma\to 0}{\longrightarrow}
{\cal A}^{\mu}_{{\rm p}_a,\lambda_a}(x).
\label{3_07_03} 
\eea
The normalization constant follows similar to (\ref{eq:spinor_dp_sm_})--(\ref{eq:scalar_conv}), 
(\ref{3_5})--(\ref{3_6}) from the inner product similar to product for the scalar packets 
(\ref{eq:scalar_dp_start_})--(\ref{eq:scalar_dp_end}) in view of (\ref{A_5}), (\ref{A_6}) 
and corresponding group property, as:
\bea
\fl 
{\rm A}^2_\sigma\bigr|_{S=1}\!=\langle\{\p_a,x_a,\lambda_a\}|\{\p_a,x_a,\lambda_a\}\rangle
\!=\!
N^2_{\sigma a}{\cal A}^{*\eta}_{{\rm p}_a,\lambda_a}(x_a) 
\langle 0|B_\eta(z^*_a)B_\beta\left(z_a\right)|0\rangle
{\cal A}^{\beta}_{{\rm p}_a,\lambda_a}(x_a)
\label{3_07_04} \\
\fl 
={\rm g}_{\mu\nu}\left({\rm F}^{\mu}_{p_a,x_a,\lambda_a},
{\rm F}^{\nu}_{p_a,x_a,\lambda_a}\right)=
iN^2_{\sigma a}{\cal A}^{*\eta}_{{\rm p}_a,\lambda_a}(x_a) 
\,{\cal D}^{\,-}_{\eta\beta}\left(z^*_a-z_a\right)\,
{\cal A}^{\beta}_{{\rm p}_a,\lambda_a}(x_a).
\nonumber
\eea
This way of construction of wave packets by means of analytic continuation in 
coordinate space differs from the one used in \cite{nn,bern} by its universality for 
any spin. On the other hand it ensures their wave functions (\ref{2_10}),  
(\ref{3_3})--(\ref{eq:spinor_wp_wf_end}), (\ref{3_07_02}) by correct tensor dimension and 
automatical satisfaction to the corresponding free KG or Dirac equations: 
\bea
\fl 
\left(\partial^2_x+m^2\right)\!F_{p_ax_a}(x)\!= 0,\;\;\, 
\left[i(\gamma\partial_x)-m\right]\!\Xi^{\xi}_{p_a,x_a,s_a}(x)\!=0= 
\overline{\Xi}^{\xi}_{p_a,x_a,s_a}(x)\!
\left[i(\gamma\!\!\stackrel{\leftarrow}{\partial}_x)+m\right]\!. 
\label{3_07} 
\eea
The remaining ambiguity of functions $g_{1,2}(m,\sigma)$, with asymptotics (\ref{2_301}), 
(\ref{2_302}), concerns only to some inessential details of dependence of 
$\tau=\tau(mc/\sigma)$, being absorbed to redefinition of invariant width $\sigma$. 
While the remaining unavoidable ambiguity in a choice of dimensionless function 
$\aleph(\tau)$ or ${\cal I}(\tau)$ in (\ref{2_20})--(\ref{2_21}) defines the general 
form of observables via averages of arbitrary operators ${\cal O}$ over packet state, 
because this ambiguity is canceled only in the ratio: 
\bea
\MVV{\cal O}\stackrel{\rm def}{==}
\frac{\langle\{\p_a,x_a,s_a\};(\xi)|{\cal O}|(\xi);\!\{\p_a,x_a,s_a \}\rangle}
{{\rm A}^2_\sigma}. 
\label{3_7} 
\eea
For example, for the mass of scalar wave packet as average of operator 
${\cal P}^2=-\partial^2_x$, with 
${\cal P}_\mu=i\partial_\mu=i(\partial_0,\vec{\nabla_\x})$, for $k^2=p^2_a=m^2_a$, 
from (\ref{2_10}), (\ref{eq:scalar_wp_wf}) explicitly, in view of 
(\ref{eq:scalar_dp_start_})--(\ref{eq:scalar_dp_end}), as well as directly from 
(\ref{3_07}), one has: 
\bea
&&\!\!\!\!\!\!\!\!\!\!\!\!\!\!\!\!\!\!\!\!\!\!\!\! 
\MVV{{\cal P}^2}\equiv
\frac{\langle\{p_a,x_a,\sigma\}|{\cal P}^2|\{p_a,x_a,\sigma\}\rangle}
{\langle\{p_a,x_a,\sigma\}|\{p_a,x_a,\sigma\}\rangle}=
\frac{\left(F_{p_ax_a},(-\partial^2_x)F_{p_ax_a}\right)}
{\left(F_{p_ax_a},F_{p_ax_a}\right)}  
\label{3_7_0} \\
&&\!\!\!\!\!\!\!\!\!\!\!\!\!\!\!\!\!\!\!\!\!\!\!\! 
\equiv 
\frac {1}{A^2_\sigma}
\int\frac{d^3{\rm k}\;k^2}{(2\pi)^3 2E_{\rm k}}\,|\phi^\sigma(\ka,\p_a)|^2=m^2_a, 
\label{3_7_01} 
\eea
and similarly for Fermi wave packets (\ref{3_3})--(\ref{eq:spinor_wp_wf_end}) with the help of 
(\ref{3_5})--(\ref{3_6}) and changing $A^2_\sigma\mapsto{\rm A}^2_\sigma$ (\ref{3_6_0}). 
Therefore any predictions for such kind averages with scalar wave packets will be the 
same as in \cite{nn} where ${\cal I}(\tau)\equiv 1$ was chosen. 

The difference may arise for averaging of matrix operators with the spin wave packets. 
For the Fermi packet by using (\ref{3_3})--(\ref{eq:spinor_wp_wf_end}) and  
(\ref{2_15_51_2_}) or (\ref{A_1}) explicitly, in view of (\ref{A_00})--(\ref{A_4}) or  
(\ref{3_2_1_0})--(\ref{3_2_2}), as well as directly from (\ref{3_07}), one has: 
\bea
&&\!\!\!\!\!\!\!\!\!\!\!\!\!\!\!\!\!\!\!\!\!\!\!\! 
\MVV{\bigl(\gamma{\cal P}\bigr)}\equiv 
\frac{\langle\{\p_a,x_a,s_a\};(\xi)|\bigl(\gamma{\cal P}\bigr)
|(\xi);\{\p_a,x_a,s_a\}\rangle}
{\langle\{\p_a,x_a,s_a\};(\xi)|(\xi);\{\p_a,x_a,s_a \}\rangle}=
\frac{\bigl(\Xi^{\xi}_{\{a\}},i(\gamma\partial_x)\Xi^{\xi}_{\{a\}}\bigr)}
{\bigl(\Xi^{\xi}_{\{a\}},\Xi^{\xi}_{\{a\}}\bigr)}   
\label{3_7_02} \\
&&\!\!\!\!\!\!\!\!\!\!\!\!\!\!\!\!\!\!\!\!\!\!\!\! 
\equiv \frac{1}{{\rm A}^2_\sigma}\int\!\frac{d^3{\rm k}}{(2\pi)^3 2E_\ka}
|\phi^\sigma(\ka,\p_a)|^2\!\left[\overline{{\cal U}}^{\xi}_{{\rm p}_a,s_a}(x_a)
\,\frac{(\gamma k)+\xi m}{(2m)^2}\,\xi(\gamma k)\,
{\cal U}^{\xi}_{{\rm p}_a,s_a}(x_a)\right]\!=m_a.
\label{3_7_03} 
\eea
According to (\ref{2_13}), (\ref{2_14}), (\ref{2_17_0}) and (\ref{2_15}), (\ref{2_16}), 
the meaning of quantum numbers of wave packet manifests only at the respective limit. 
For any operator ${\cal O}({\cal P}^\mu)$ by means of (\ref{2_20})--(\ref{eq:sc_limit_0}), 
(\ref{2_28_2})--(\ref{2_29}), (\ref{3_6_0}), for scalar packet follows:
\bea
\fl 
\MVV{{\cal O}({\cal P}^\mu)}\equiv
\frac{\left(F_{p_ax_a},{\cal O}(i\partial^\mu)F_{p_ax_a}\right)}
{\left(F_{p_ax_a},F_{p_ax_a}\right)}= \frac {N^2_\sigma}{A^2_\sigma}
\int\frac{d^3{\rm k}\,{\cal O}(k^\mu)}{(2\pi)^3\,2E_{\rm k}}\,
e^{-2\left(k\zeta_a(p_a,\sigma)\right)}
\stackunder{\sigma\to 0}{\longrightarrow}{\cal O}(p^\mu_a),
\label{3_8_00} \\
\fl 
\mbox{where: }\,
\frac {N^2_\sigma}{A^2_\sigma}e^{-2\left(k\zeta_a(p_a,\sigma)\right)}
\stackunder{\sigma\to 0}{\longrightarrow} 
2m(2\pi)^3 e^{g_1\left(E_{\rm k}-E_{\rm p}\right)^2}\!\!
\left\{\!\left(\frac{g_1}{\pi}\right)^{3/2}\!e^{-g_1(\ka-\p)^2}\!\right\}
\nonumber \\
\fl 
\stackunder{g_1\to\infty}{\longrightarrow}(2\pi)^3 2E_{\rm k}\delta_3(\ka-\p),  
\label{3_8_01} 
\eea
because this differs from (\ref{2_28_2}) only by replacement $g_1\mapsto 2g_1$, 
and the same takes place for Fermi wave packets with the formal changing 
$A^2_\sigma \mapsto {\rm A}^2_\sigma$ (\ref{3_6_0}). 
For Pauli-Lubanski operator ${\cal W}^\mu$ and any space-like vector ${\cal S}$, 
similarly (\ref{3_5})--(\ref{3_6}) by the use of (\ref{A_4_0})--(\ref{A_4_2})  
for upper sign of $2s_a=1$, with ${\cal O}=-({\cal W}{\cal S})/m_a$ and 
\bea
&&\!\!\!\!\!\!\!\!\!\!\!\!\!\!\!\!\!\!\!\!\!\!\!\! 
4({\cal W}{\cal S})\equiv 
4{\cal W}^\mu {\cal S}_\mu= -\gamma^5[\gamma^\mu,\gamma^\nu]{\cal S}_\mu{\cal P}_\nu=
-2\gamma^5\left\{(\gamma{\cal S})(\gamma{\cal P})-({\cal S}{\cal P})\right\},   
\label{3_7_04} \\
&&\!\!\!\!\!\!\!\!\!\!\!\!\!\!\!\!\!\!\!\!\!\!\!\! 
\mbox{one has explicitly: }\;
\MMV{-\, \frac{({\cal W}{\cal S})}{m_a}}=
\frac 12\MMV{\frac{\gamma^5}{m_a}\left[(\gamma{\cal S})(\gamma{\cal P})-
({\cal S}{\cal P})\right]}  
\label{3_7_05} \\
&&\!\!\!\!\!\!\!\!\!\!\!\!\!\!\!\!\!\!\!\!\!\!\!\! 
= -\,\frac 12(\widehat{s}_a{\cal S})+\frac 12  \frac {N^2_\sigma}{{\rm A}^2_\sigma}
\int\frac{d^3{\rm k}\,e^{-2\left(k\zeta_a\right)}}{(2\pi)^3\,2k^0}\left[
\frac{(k\widehat{s}_a)(k{\cal S})+(k\widehat{s}_a)(p_a{\cal S})}{2m^2_a}\right].
\label{3_7_06} 
\eea
For the pure polarized state with $\widehat{s}^2_a=-1$, $(p_a\widehat{s}_a)=0$, if    
${\cal S}\mapsto \widehat{s}_a$, then (comp. (\ref{3_2_1_0})):  
\bea 
&&\!\!\!\!\!\!\!\!\!\!\!\!\!\!\!\!\!\!\!\!\!\!\!\! 
\MMV{-\, \frac{({\cal W}{\cal S})}{m_a}} \mapsto  
\frac 12+\frac 12  \frac {N^2_\sigma}{{\rm A}^2_\sigma}
\int\frac{d^3{\rm k}\,e^{-2\left(k\zeta_a\right)}}{(2\pi)^3\,2k^0}\,
\frac{(k\widehat{s}_a)^2}{2m^2_a}\stackunder{\sigma\to 0}{\longrightarrow}\frac 12. 
\label{3_7_08}
\eea
In the plane-wave limit (\ref{3_8_01}) the last summands in (\ref{3_7_06}), 
(\ref{3_7_08}) obviously disappear as it should. 
For the states with {\sl mixed} polarization: $-1<\widehat{s}^2_a<0$. 
For the correlation with ${\cal S}$ by the rule: 
\bea
&&\!\!\!\!\!\!\!\!\!\!\!\!\!\!\!\!\!\!\!\!\!\!\!\! 
\overline{\widehat{s}^\mu_a{\cal S}^\nu}=\frac{{\rm g}^{\mu\nu}}{4}
\overline{(\widehat{s}_a{\cal S})}, \;\mbox{ one finds: }\;
\overline{ \MMV{-\, \frac{({\cal W}{\cal S})}{m_a}} } = 
-\,\frac 12 \overline{(\widehat{s}_a{\cal S})}\frac{3}{4}.
\label{3_7_07}
\eea
\section{Wave packets and  Huygens' principle}\label{sec:Disc}
The principal differences of suggested here wave packet from the one suggested in 
\cite{nn,nsh,dn,n_sh,NN_shk} and marked below as [1], manifest in spin degrees of 
freedom and thus in the Lorentz transformation and localization properties 
(\ref{2_15}) -- (\ref{2_17}), and normalization. 
According to our definitions (\ref{3_3})--(\ref{eq:spinor_wp_wf_end}), (\ref{2_15_51_2_}), 
and (\ref{z_1}), (\ref{eq:phidef}), (\ref{2_20}), (\ref{2_20_0}) but 
unlike the formulas (6), (7) of \cite{nn}, for $k_0=E_{\rm k}=\sqrt{\ka^2+m^2_a}$, 
$\underline{x}=x_{a}-x$, $\zeta_a=\zeta_a(p_a,\sigma_a)$, $\xi=+$: 
\numparts
\bea
\fl 
\Xi^{(+)}_{p_a,x_a,s_a}(x)\equiv \Xi^{(+)}_{\{a\}}(x)=
\langle 0|\psi(x)|(+);\{\p_a,x_a,s_a\}\rangle =\frac 1i \widehat{\rm N}^\xi_{\sigma a}
S^-\left(x-z_a\right)\,{\cal U}^{(+)}_{{\rm p}_a,s_a}(x_a)  
\label{4_1_2} \\
\fl 
=\widehat{\rm N}^\xi_{\sigma a}
\langle 0|\psi(x)\overline{\psi}\left(x_a+i\zeta_a\right)|0\rangle\, 
u^{(+)}(p_a,s_a)\,e^{-i(p_ax_a)}  
\label{4_1} \\
\fl 
=\!\int\!\frac{d^3{\rm k}\,\,e^{i(k\underline{x})}}{(2\pi)^3 2E_{\rm k}}\,
\phi^\sigma(\ka,\p_a)\,\frac{(\gamma k)+m_a}{2m_a}\,u^{(+)}(p_a,s_a)\,e^{-i(p_ax_a)}, 
\label{4_1_1} \\
\fl 
\mbox{but that is: } \neq \Xi^{(+)\,{[1]}}_{\{a\}}(x)=
\langle 0|\psi(x)|(+);\p_a,x_a,s_a \rangle_{[1]} 
\label{4_2_1} \\
\fl 
=\!\int\!\frac{d^3{\rm k}\,e^{i(k\underline{x})}}{(2\pi)^3 2E_{\rm k}}\, 
\phi^\sigma_{[1]}(\ka,\p_a)\,u^{(+)}(k,s_a)\,e^{-i(p_ax_a)},\,\mbox{ for}
\label{4_2} \\
\fl 
\phi^\sigma_{[1]}(\ka,\p_a)\,\mbox{ with: }\,{\cal I}^{[1]}(\tau)\equiv 1, \;\;\,
N^{[1]}_\sigma(m_a,\zeta^2_a) =\frac{(2\pi)^2}{m^2_a h(\tau^2)},\;\;\, 
g_1=\frac 1{2\sigma^2}, \;\;\,g_2 \equiv 0. 
\label{4_3}
\eea
\endnumparts
According to \cite{blp,tir,p_s,vrg,schw,blt,oksak,jost,s_w,stroc,Dvt,b_d,i_z,Nvj,Bil,Wnb}, 
if the packet realizes a state with fixed momentum $p_a$ and spin $s_a$ as irreducible 
representation of Lorentz group, its spin-quantization 4- axis $\widehat{s}_a$ in 
(\ref{3_2_1_0}) is fixed only by $(p_a\widehat{s}_a)=0$, with the fixed momentum $p_a$ of 
wave packet. This is exactly the case of (\ref{4_1_2})--(\ref{4_1_1}), containing only 
the external bispinorial plane wave 
${\cal U}^{(+)}_{{\rm p}_a,s_a}(x_a)=u^{(+)}(p_a,s_a)\,e^{-i(p_ax_a)}$. 
However this fixed spin $s_a$ can not simultaneously ``sit'' on a variable momentum 
$k^\lambda$ that in (\ref{4_2}) runs through the entire space via internal plane waves with 
bispinor $u^{(+)}(k,s_a)$, meaning that $(k\widehat{s}_a)=0$. The same is true for 
vector wave packet (\ref{3_07_02}), (\ref{3_07_03}). 

Nevertheless, the evident from the relations (\ref{3_3})--(\ref{eq:spinor_wp_wf_end}) with 
(\ref{A_1_1}), (\ref{eq:spinor_weightman_conv_same}), (\ref{3_4}) form-invariant propagation 
according to Huygens' principle for the packet (\ref{4_1_2}) -- (\ref{4_1_1}), due to the 
definitions (\ref{A_1_1}), (\ref{2_15_51_2_}), (\ref{A_1}), (\ref{3_2:1})--(\ref{3_2:4}) and 
the second orthogonality relation (\ref{3_2_10}) \cite{p_s,vrg} takes place also for the 
packet (\ref{4_2}), also as covariant solution to free equations (\ref{3_07}) (clf.  
(\ref{3_13_0_0_1})--(\ref{3_13}) below). Similar assertion, where $D^c(x)$ 
is even function (\ref{2_30_01}), takes place for the respective scalar wave packet 
(\ref{2_10}), (\ref{2_18}), $\forall\,x_a$, with $\xi(x^0-y^0)>0$, for both the initial 
and final states (cmp. \cite {feyn}, pp. 101, 102). So: 
\bea
\fl 
\Xi^{\xi}_{\{a\}}(x)=\frac{\xi}i\! 
\int\!\! d^3{\rm y} S^c(x-y)\gamma^0\Xi^{\xi}_{\{a\}}(y), \quad 
F_{p_ax_a}(x)=\frac{1}i\! 
\int\!\! d^3{\rm y}D^c(x-y)(i\!\stackrel{\leftrightarrow}{\partial^0_{y}})F_{p_ax_a}(y), 
\label{4_4} \\
\fl 
\overline{\Xi}^{\xi}_{\{a\}}(x)=\frac{\xi}i\! 
\int\!\! d^3{\rm y}\overline{\Xi}^{\xi}_{\{a\}}(y)\gamma^0 S^c(y-x), \quad 
F^*_{p_ax_a}(x)=\frac{1}i\!\int\!\! d^3{\rm y}
F^*_{p_ax_a}(y)(i\!\stackrel{\leftrightarrow}{\partial^0_{y}}) D^c(y-x), 
\label{4_5} 
\eea 
and the same with the changes $d^3{\rm y}\gamma^0\mapsto d\Sigma_\nu(y)\gamma^\nu$,  
$d^3{\rm y}\!\stackrel{\leftrightarrow}{\partial^0_{y}}\,\mapsto 
d\Sigma_\nu(y)\!\stackrel{\leftrightarrow}{\partial^\nu_{y}}$, and 
$\xi(x^0-y^0)\mapsto \xi(n_\Sigma(x-y))$ (\ref{3_13_14}), and the same for 
$\Xi^{\xi}_{\{a\}}\mapsto \Xi^{\xi\,{[1]}}_{\{a\}}$. 
Free propagation (\ref{4_4}), (\ref{4_5}) can not change the {\sl invariant width} 
$\sigma$ and the value of $\tau$ of covariant free wave packets, because such changing 
would be in contradiction with the local nature of Lorentz covariance of this 
propagation \cite {feyn}. Nevertheless, the usual Gaussian spreading shown on Figure 
\ref{fig:1} will inevitably take place at any {\sl fixed reference frame} \cite{dn}.  
So the spreading of relativistic wave packet as well as its minimal position-velocity 
uncertainty are frame-dependent feature \cite{al_h_w} (clf. \ref{ap:1_A}).

Due to dimension reduction of free Green functions \cite{g_sh} and due to conservation of 
orthogonality relation (\ref{3_2_10}) also for $m=0$, the similar picture with 
$d^3{\rm x}\mapsto d{\rm x}_\parallel$ takes place also for evolution (\ref{4_4}), 
(\ref{4_5}) of the massless wave packets, defined for fixed $\tau>0$ by (\ref{D_7}), 
(\ref{D_7_00}), (\ref{D_10}), (\ref{D_11}), (\ref{A_1}), and normalized by the condition 
(\ref{D_13}):
\numparts
\bea
\fl 
F_{p_ax_a}(x)\stackrel{m\to 0}{\stackunder{\sigma\to 0}{\longmapsto}}\,
F^{(\tau)}_{p_ax_a}(x)=e^{-i(p_ax_a)}\,\psi^\tau(\p_a,\underline{x})=
\frac{\aleph(\tau)}{(2\pi)^2}\,e^{-i(p_ax_a)}\,
h\!\left(\tau^2-2i\tau(p_a \underline{x})\right),  
\label{4_7} \\ 
\fl 
\Xi^{\xi}_{\{a\}}(x)\stackrel{m\to 0}{\stackunder{\sigma\to 0}{\longmapsto}}\,
\Xi^{\xi(\tau)}_{\{a\}}(x)=\lim_{m_a\to 0}
\frac {i(\gamma\partial_x)+m_a}{2m_a}\,F^{(\tau)}_{\xi p_a,x_a}(x)\,
u^\xi_{m_a}(p_a,s_a)  
\label{4_8} \\ 
\fl 
=\frac{\aleph(\tau)}{(2\pi)^2}\,e^{-i\xi(p_ax_a)}\,u^\xi_{0}(p_a,s_a)\,
\left[\frac {1}2-\tau \frac{\partial}{\partial Z^\xi_\tau}\right] 
h\!\left(Z^\xi_\tau\right)\biggr|_{Z^\xi_\tau=\tau^2-2i\tau\xi (p_a \underline{x})}
\label{4_8_0} \\ 
\fl 
\stackunder{\tau\to\infty}{\longmapsto}
e^{-i\xi(p_a x)}\,
\lim_{m_a\to 0}\frac {\xi (\gamma p_a)+m_a}{2m_a}u^\xi_{m_a}(p_a,s_a)=
e^{-i\xi(p_a x)}\,u^\xi_{0}(p_a,s_a)={\cal U}^\xi_{{\rm p}_a,s_a,0}(x). 
\label{4_9} 
\eea
\endnumparts
The normalization conditions (\ref{3_2_0}), (\ref{eq:spinor_dp_sm_}) of all bispinors leads 
to their finite limits with $m_a\to 0$ \cite{vrg,Nvj}. The parameter $\tau$, originated by 
(\ref{2_20}), (\ref{2_301}), (\ref{2_303}), in the rough approximation (\ref{D_14}) of 
(\ref{D_8}) may be considered as inverse dimensionless width of these wave packets on 
the light cone. From (\ref{E_1}), (\ref{E_2}) their form-invariant propagation is also 
governed by corresponding Huygens' principle:  
\bea
\fl 
F^{(\tau)}_{p_ax_a}(\overline{x})=\!\!\int\limits^\infty_{-\infty}\!\!\!
d{\rm y}_\parallel D^c_{(2)}(\overline{x}-\overline{y})\!
\stackrel{\leftrightarrow}{\partial^0_{y}}\!\! 
F^{(\tau)}_{p_ax_a}(\overline{y}), \quad 
\Xi^{\xi(\tau)}_{\{a\}}(\overline{x})\!=\frac{\xi}i\!\!\int\limits^\infty_{-\infty}\!\!\!
d{\rm y}_\parallel S^c_{(2)}(\overline{x}-\overline{y})\gamma^0\,
\Xi^{\xi(\tau)}_{\{a\}}(\overline{y}), 
\label{4_4_0} 
\eea 
where for $p^\nu_a={\rm p}(1,\n_{\rm p})$, all functions of $x$ with $m_a=0$ in   
(\ref{4_7})--(\ref{4_9}), (\ref{4_4_0}) as well as in (\ref{D_10}) -- (\ref{D_12_0}) in fact 
depend only on 2- dimensional $\overline{x}^\nu = (x^0,{\rm x}_\parallel)$ with 
${\rm x}_\parallel=(\n_{\rm p}\!\cdot\!\x)$.  

Following  (\ref{4_1_2})--(\ref{4_1_1}) it is not dificalt to construct the wave packet of 
that type also for Majorana neutrino for both massive and massless cases 
\cite{bGG,Dvor}, with the similar properties of normalization constant. 

\section{Neutrino oscillations. Intermediate wave packet treatment.}\label{sec:intrm_wp}

Spin degrees of freedom are potentially important \cite{bern} in the intermediate wave 
packet approach to the neutrino oscillations problem  \cite{gunt}. The respective 
generalization of the method of Blasone \cite{bl1,bl2,bl3} onto above mentioned Fermi wave 
packets by the use of respectively generalized calculations leads to the following expression 
for two-flavour oscillations of leptonic charge for the electronic neutrino $\nu_e$, where 
$\stackrel{\rm w}{=}$ means equality in a weak sense (with redefinition in (\ref{3_1}) 
$b^{(+)}_{{\rm k},r}=a_{{\rm k},r}$, $b^{(-)}_{{\rm k},r}=b_{{\rm k},r}$):
\bea
\fl  
Q_e(t) \stackrel{\rm w}{=} G^{-1}_\vartheta(t)Q_1 G_\vartheta(t)=
\int\frac{d^3{\rm k}}{(2\pi)^3 2E_{\ka,1}} \sum_{r=\pm 1/2}
\left[a^{e\dagger}_{{\rm k},r}(t)\,a^e_{{\rm k},r}(t)-
b^{e\dagger}_{-{\rm k},r}(t)\, b^{e}_{-{\rm k},r}(t)\right],
\label{3_8} \\
\fl  
\MVV{Q_e(t)}_e=\frac{\bra{\nu_e\{\p_1,x_1,s_1\}}Q_e(t)\ket{\nu_e\{\p_1,x_1,s_1\}}}
{{\rm A}^{2}_{\sigma e}}   
\label{3_8_10}  \\
\fl  
= \frac{{N}^2_{\sigma 1}}{{\rm A}^{2}_{\sigma 1}}
\int \frac{d^3{\rm k}\,e^{-2(k\zeta_1(p_1,\sigma_1))}}{(2\pi)^3 2E_{\ka,1} }
\left[\frac 12+\frac{(kp_1)}{2 m^2_1}\right]{\cal Q}_{\ka,e}(t), \;\,\mbox{ with:}
\label{3_9}  \\
\fl  
{\cal Q}_{\ka,e}(t)=1-\sin^2 2\vartheta
\left[|{\rm U}_\ka|^2\sin^2\left(\frac{E_{\ka,1}-E_{\ka,2}}{2}t\right)+
|{\rm V}_\ka|^2\sin^2\left(\frac{E_{\ka,1}+E_{\ka,2}}{2}t\right)\right], 
\label{3_10}  \\
\fl  
\mbox{where: }\; |{\rm U}_\ka|^2 + |{\rm V}_\ka|^2 = 1,\; \mbox { with: }\;
|{\rm U}_\ka|=\frac{|\ka|^2+(E_{\ka,1}+m_1)(E_{\ka,2}+m_2)}
{2\sqrt{E_{\ka,1}E_{\ka,2}(E_{\ka,1}+m_1)(E_{\ka,2}+m_2)}}, 
\nonumber \\
\fl  
|{\rm V}_\ka|=\frac{(E_{\ka,2}+m_2)-(E_{\ka,1}+m_1)}
{2\sqrt{E_{\ka,1}E_{\ka,2}(E_{\ka,1}+m_1)(E_{\ka,2}+m_2)}}\,|\ka|, 
\;\mbox { for: }\;  m_1<m_2,\;\;\mbox{ because:}  
\nonumber \\
\fl  
{\rm A}^{2}_{\sigma e}=
\langle \nu_e\{\p_1, x_1, s_1\}\ket{\nu_e\{\p_1,x_1, s_1\}} = 
\langle\nu_1\{\p_1, x_1, s_1\}\ket{\nu_1\{\p_1,x_1, s_1\}}={\rm A}^{2}_{\sigma 1},
\;\mbox { for:}
\label{3_11}  \\
\fl  
a^{e\dagger}_{{\rm k},r}(x^0)\stackrel{\rm w}{=}
G^{-1}_\vartheta(x^0)a^{1\dagger}_{{\rm k},r} G_\vartheta(x^0), 
\;\mbox { and: }\;  
\ket{\nu_e\{\p_1,x_1, s_1\}}=G^{-1}_\vartheta(x^0_1)\ket{\nu_1\{\p_1,x_1, s_1\}},
\label{3_11_0}
\eea
is the wave-packet state with definite electronic flavour $e$, defined according to 
\cite{bl2} by the same {\sl improper transformation} \cite{bl1,bl2,bl3} (\ref{3_8}), 
(\ref{3_11_0}) of the above packet states (\ref{eq:spinor_wp_ket_b_fin_0}), 
(\ref{eq:spinor_wp_ket_b_fin_1}) for initial 
massive field (\ref{3_1}), (\ref{3_1_0}) $\psi(x_1)=\nu_{1}(x_{1})$ with mass $m_1$
at the {\sl same instant of time} $x^0=x^0_1$, and normalized by the same constant 
(\ref{3_6}). 
The conserved charge $Q_1$ \cite{i_z,bl3} is defined by expression (\ref{3_8}) for 
$\vartheta=0$ via operators $a^1_{{\rm k},r}$, $b^1_{{\rm k},r}$ of the field (\ref{3_1}). 
Notice the time $t$ of operator $Q(t)$ (\ref{3_8}) has nothing to do with the local 
packet time $x^0_a = x^0_1$ in (\ref{3_8_10}), (\ref{3_11_0}), is automatically canceled 
in   (\ref{3_9}). Finally for each spin $s_1$, from (\ref{3_9}), (\ref{3_10}) one has: 
\bea
&&\!\!\!\!\!\!\!\!\!\!\!\!\!\!\!\!\!\!\!\!\!\!\!\! 
\MVV{Q_e(t)}_e=1+
\frac{{N}^2_{\sigma 1}}{{\rm A}^{2}_{\sigma 1}}
\int \frac{d^3{\rm k}\,e^{-2(k\zeta_1(p_1,\sigma_1))}}{(2\pi)^3 2E_{\ka,1} }
\left[\frac 12+\frac{(kp_1)}{2 m^2_1}\right]\left[{\cal Q}_{\ka,e}(t)-1\right],
\label{3_12} 
\eea 
that due to (\ref{3_8_01}) surely reproduces the plane-wave limit 
$\MVV{Q_e(t)}_e\stackunder{\sigma\to 0}{\longrightarrow}{\cal Q}_{\p_a,e}(t)$  
with the other cases of \cite{bl1,bl2} and after time averaging of (\ref{3_10}) 
yieldes to the well known result: $\overline{\MVV{Q_e(t)}_e}=1-(1/2)\sin^2 2\vartheta$.  
The mixing angle $\vartheta$ defines the mixing matrix $U$ in (\ref{3_12_00}) below, where 
the case of two-flavour oscillation is chosen only for simplicity.   

It should be stressed, since the first relation in (\ref{3_11_0}) takes place 
only in a weak sense \cite{bl1,bl2,bl3}, according to the second relation (\ref{3_11_0}) 
it generates wave packets of flavour states for $\{\ell,j\}=\{e,1\}$ or $\{\mu,2\}$, 
with $\zeta_{aj}(p_{aj},\sigma_j)=p_{aj} g_1(m_j,\sigma_j)+
\widehat{w}_{aj} g_2(m_j,\sigma_j)$, 
$p^0_{aj}=E_{\p_a,j}=+\sqrt{\p^2_a+m^2_j}$ only over respective time dependent flavour vacuum 
$\ket{0(x^0)}_{e,\mu}$ by making use of the {\sl improper transformation} 
$G_\vartheta(x^0)$ of operators and states at the same instant $x^0_a$ with 
$a^{\#}=a,a^\dagger$ and $b^{\#}=b,b^\dagger$:
\bea
\fl  
G_\vartheta(x^0)=\exp\left[\vartheta \!\int \! d^3x\left(\nu_1^\dagger(x)\nu_2(x)-
\nu_2^\dagger(x)\nu_1(x)\right)\right], \quad \;\;
U=\left( \begin{array}{cc} \cos \vartheta & \sin \vartheta \\
- \sin \vartheta & \cos \vartheta \\ \end{array} \right), 
\label{3_12_00} \\
\fl  
\left(a^{\ell\#}_{{\rm k},r}(x^0),\, b^{\ell\#}_{{\rm k},r}(x^0)\right)\stackrel{\rm w}{=}
G^{-1}_\vartheta(x^0)\left(a^{j\#}_{{\rm k},r},\,b^{j\#}_{{\rm k},r}\right) 
G_\vartheta(x^0), \quad 
\ket{0(x^0)}_{e,\mu} = G^{-1}_\vartheta(x^0)\ket{0}_{1,2}, 
\label{3_12_0}
\eea
with the help of definitions: 
\bea
\fl  
\left\{\!\!
\begin{array}{c}
\overline{\nu}^{(+)}_{\ell}[x_a;\,\zeta_{a j}]\ket{0(x^0_a)}_{e,\mu}=
G^{-1}_\vartheta(x_a^0)\overline{\nu}_{j}(x_a+i\zeta_{a j})\ket{0}_{1,2},  \\
\nu^{(-)}_{\ell}[x_a;\,\zeta_{a j}]\ket{0(x^0_a)}_{e,\mu}=
G^{-1}_\vartheta(x_a^0)\nu_{j}(x_a+i\zeta_{a j})\ket{0}_{1,2}, 
\end{array}\right. \;\; \mbox{ where:} 
\label{3_12_01} \\
\fl  
\left.\!\!
\begin{array}{c}
\overline{\nu}^{(+)}_{\ell}[x_a;\, \zeta_{a j}] \\
\nu^{(-)}_{\ell}[x_a;\, \zeta_{a j}]
\end{array}\!\!\right\}=\!\!\sum_{r=\pm 1/2}\int\!
\frac{d^3{\rm k}\,e^{i(kx_a)-(k\zeta_{aj})}}{(2\pi)^3\,2E_{\ka,j}}\left\{\!\!
\begin{array}{c} 
\overline{u}^{(+)}_j(k,r)\,a^{\ell\dagger}_{{\rm k},r}(x^0_a) \\
u^{(-)}_j(k,r)\,b^{\ell \dagger}_{{\rm k},r}(x^0_a) \end{array}\!\! \right\}, \quad 
k^0=E_{\ka,j}. 
\label{3_12_02} 
\eea
Here the (\ref{3_12_02}) only for $\zeta_{aj}=0$ recoveres a ``negative 
frequency parts'' of local field operators with definite flavour $\nu_\ell(x_a)$ as a mix 
of massive fields {\sl at the same point} $x_a$ \cite{bl1}: 
\bea
&&\!\!\!\!\!\!\!\!\!\!\!\!\!\!\!\!\!\!
\overline{\nu}_\ell(x_a)=\overline{\nu}^{(+)}_{\ell}[x_a;0]+
\overline{\nu}^{(-)}_{\ell}[x_a;0]\stackrel{\rm w}{=}
G^{-1}_\vartheta(x_a^0)\overline{\nu}_{j}(x_a)G_\vartheta(x_a^0)= 
\sum\limits_{n}U^*_{\ell n}\overline{\nu}_{n}(x_a), 
\label{3_12_05} \\
&&\!\!\!\!\!\!\!\!\!\!\!\!\!\!\!\!\!\!
\nu_\ell(x_a)\stackrel{\rm w}{=}
G^{-1}_\vartheta(x_a^0)\nu_{j}(x_a)G_\vartheta(x_a^0)=
\sum\limits_{n}U_{\ell n}\nu_{n}(x_a),\;\;\;j,n=1,2.   
\label{3_12_03}
\eea
However the conventional mixing relations for the {\sl states} follow from there only by 
neglecting the difference (\ref{3_12_0}) between the flavour and massive vacua 
\cite{bl1,bl2}: $\ket{0(0)}_{e,\mu}\mapsto \ket{0}_{1,2}$ thus ignoring the weak sense of 
(\ref{3_12_05}), (\ref{3_12_03}), and strictly speaking, only for the states with 
definite momentum $|\ka\rangle$, implying the plane-wave limit (\ref{2_14}), (\ref{2_29}) 
of states (\ref{eq:spinor_wp_ket_b_fin_0}), (\ref{eq:spinor_wp_ket_b_fin_1}), 
(\ref{3_11_0}), (\ref{3_12_01}), 
(\ref{3_12_02}), when $\sigma_j\to 0$, $\zeta^\lambda_{aj}\to\infty$. 
Corresponding mixing relations for the states of wave packets with definite mass 
$m_j$ require Lorentz invariant profile functions $\phi^\sigma(\ka,\p_a)$ to be independent 
of $j$ \cite{bern,bl3}, which, as explained in Introduction, is impossible for the 
interpolating wave-packets states (\ref{eq:phidef}), (\ref{3_3})--(\ref{eq:spinor_wp_wf_end}) and gives 
the another reason, why the conventional mixing relations for wave-packet states 
\cite{bern,gunt}: $|\nu_\ell\{\p_a,x_a,s_a\}\rangle\stackrel{?\,?}{\mapsto}
\sum_{n}U^*_{\ell n}|\nu_{n}\{\p_a,x_a,s_a\}\rangle$, 
are inconsistent with general principles of QFT \cite{bl1} already for two-flavour case.    
However, at the same time, for $\{\ell,j\}=\{e,1\},\; \{\mu,2\}$: 
\bea
\fl  
|\nu_\ell\{\p_a,x_a,s_a\}\rangle = \widehat{\rm N}^\xi_{\sigma j}
\overline{\nu}^{(+)}_{\ell}[x_a;\,\zeta_{a j}]\ket{0(x^0_a)}_{e,\mu}=
G^{-1}_\vartheta(x_a^0)\widehat{\rm N}^\xi_{\sigma j}
\overline{\nu}_{j}(x_a+i\zeta_{a j})\ket{0}_{1,2}.  
\label{3_12_04} 
\eea
Eventually only (\ref{3_11_0}), (\ref{3_12_01}), (\ref{3_12_02}), (\ref{3_12_04}), 
with the expressions (\ref{eq:spinor_wp_ket_b_fin_0}), (\ref{eq:spinor_wp_ket_b_fin_1}) 
for $\psi(z_a)\mapsto \nu_{j}(z_{a j})$, are meaningful for the states with flavour 
wave packet in the approach with intermediate wave packets. 

\section{Diagrammatic treatment of oscillations. Composite wave functions.}
\label{sec:diagr_tr}
\begin{figure}[htb]  
\centering  
\includegraphics[height=.44\textheight]{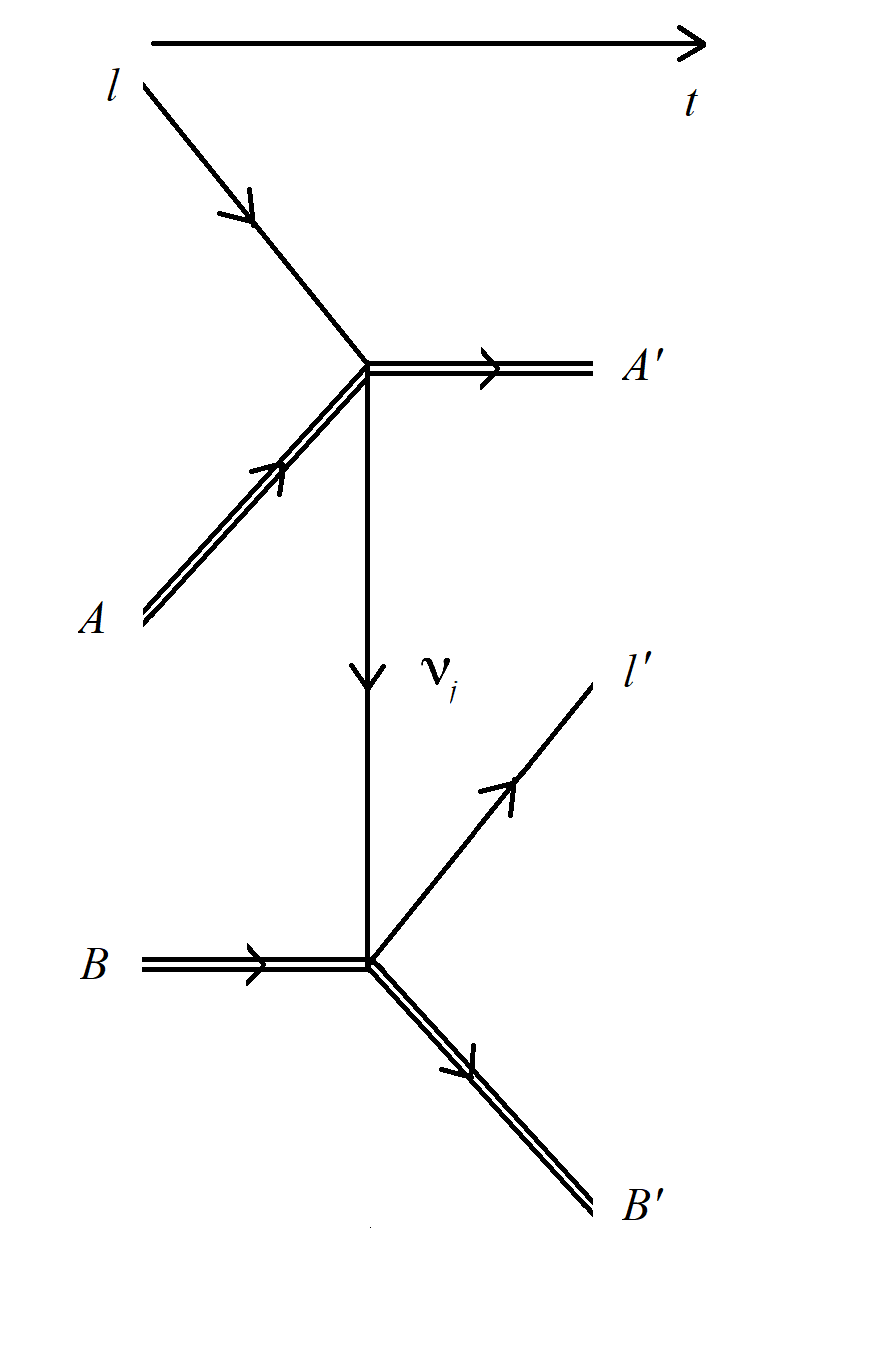}
\vskip -1cm
\caption{Tree diagram ``little donkey'' of scattering of initial particles $\ell,A$ 
in neutrino source on the initial particle $B$ in neutrino detector.}
\label{fig:2}
\end{figure}
The method of macroscopic Feynman diagrams 
\cite{nn,nsh,dn,n_sh,NN_shk,kt_1,rich,bth,ahm,Dl_1,Nis,kbz,GS,NN_anp} 
describes oscillation as a {\sl scattering} of initial particles in neutrino source 
(e.g. $\pi^\pm$ or $\ell,A$ on Figure \ref{fig:2}) on the initial particles (e.g. $B$) in 
neutrino detector and, unlike the previous approach, uses the wave packets only for 
the external particles relatively to the intermediate ``t-channel'' neutrino with definite 
mass $m_j$. 
Thus, the weak sense of the (\ref{3_12_05}), (\ref{3_12_03}) as well as difference 
between the vacua (\ref{3_12_0}) are not manifest because these flavour fields have not 
appear at all.  
This method also emphasizes the suggested form of wave packet (\ref{2_10}), (\ref{2_18}), 
(\ref{3_3})--(\ref{eq:spinor_wp_wf_end}) and (\ref{3_07_03}). Indeed, according to 
\cite{bth,ahm} the wave packet of intermediate neutrino of previous approach arises in the 
expression of amplitude of any such tree macro diagram like on Figure \ref{fig:2}:  
\bea
\fl  
{\cal A}^j_{DC}=\!\!\int\! d^4{x}\,\Psi^*_{DF}(x)\Psi_{DI}(x)\widetilde{M}_{jD}
\!\int\! d^4{y}\,\frac{1}i S^c_j(x-y)\,\widetilde{M}_{jC}\Psi^*_{CF}(y)\Psi_{CI}(y)
=\sum_{\xi=\pm} {\cal A}^{j(\xi)}_{DC}, 
\label{3_13_1}
\eea 
by making use of the `pole integration', which for $|x^0-y^0|\gg1/m_j$ actually is 
equivalent to replacement of the causal propagator $S^c_j(x-y)$ onto its relevant 
on-shell frequency part $S^{-\xi}_j(x-y)$ of its Lorentz covariant time-ordered 
decomposition (\ref{A_1_1}), (\ref{E_01}) into the WFs (\ref{2_15_51_2_}), 
(\ref{A_1}) (or (\ref{2_19}) for scalar case) with $\xi=+$ for neutrino, $\xi=-$ for 
antineutrino:
\bea
\fl  
{\cal A}^{j(\xi)}_{DC}=\!\!\int\!\! d^4{x}\Psi^*_{DF}(x)\Psi_{DI}(x)
\widetilde{M}_{jD}\!\!\int\!\! d^4{y}\theta(\xi(x^0-y^0))\frac{\xi}iS^{-\xi}_j(x-y) 
\widetilde{M}_{jC}\Psi^*_{CF}(y)\Psi_{CI}(y).   
\label{3_13_11}
\eea
This differs from  (\ref{3_4}) with the same replacement, due to   
(\ref{eq:spinor_weightman_conv_same}) exactly applicable therein, only by changing of 
three-dimensional integrals onto the four-dimensional ones and by replacement of the 
`source', being factor with only one wave packet of {\sl free intermediate} 
neutrino $\gamma^0\Xi^{\xi}_{\{\!c/a\}j}\!(y)$, onto the source as a product of 
respective vertex $\widetilde{M}_{jC}$ or $\widetilde{M}_{jD}$ with such wave packets 
(or products of them) $\Psi_{CI}(y)$ or $\Psi_{DI}(x)$ for {\sl incoming} and with 
$\Psi^*_{CF}(y)$ or 
$\Psi^*_{DF}(x)$ for {\sl outgoing} particles \cite{nn,rich,bth,ahm} at respective 
neutrino {\sl creation} 4-point $y$ or neutrino {\sl detection} 4-point $x$, but vice 
versa $x\leftrightharpoons y$ for the antineutrino with simultaneous interchange of 
indexes $D\leftrightharpoons C$ \cite{p_s,vrg}. 
This conforms also with discussion of Huygens' principle after   (\ref{3_4}),  
because in agreement with \cite{feyn} and   (6.49), (6.50) of \cite{b_d}, from 
(\ref{A_1_1}) and the solution to Cauchy problem (also with 
$d^3{\rm y}\gamma^0\mapsto d\Sigma_\mu(y)\gamma^\mu$): 
\numparts
\bea
&&\!\!\!\!\!\!\!\!\!\!\!\!\!\!\!\!\!\!
\Xi^{\xi}_{\{c/a\}j}(x)\,\delta_{\xi\eta}=\frac{1}i\!
\int\! d^3{\rm y}\,S^{-\eta}_j(x-y)\,\gamma^0\,\Xi^{\xi}_{\{c/a\}j}(y), 
\;\mbox{ for any }\; x^0,y^0,
\label{3_13_0_0_1} \\
&&\!\!\!\!\!\!\!\!\!\!\!\!\!\!\!\!\!\!
\mbox{it follows: }\; \theta(\xi(x^0-y^0))\Xi^{\xi}_{\{c/a\}j}(x)= 
\frac{\xi}i\! \int d^3{\rm y}\,S^c_j(x-y)\,\gamma^0\,\Xi^{\xi}_{\{c/a\}j}(y)  
\label{3_13_0} \\
&&\!\!\!\!\!\!\!\!\!\!\!\!\!\!\!\!\!\!
=\theta(\xi(x^0-y^0))\frac{1}i\!
\int\! d^3{\rm y}\,S^{-\xi}_j(x-y)\,\gamma^0\,\Xi^{\xi}_{\{c/a\}j}(y). 
\label{3_13}
\eea
\endnumparts
So, the group property (\ref{eq:spinor_weightman_conv_same}) of WF enable to recast 
amplitude (\ref{3_13_11}) into the form of Lorentz invariant scalar product similar to 
(\ref{eq:spinor_dp_sm_}), for any 
$\xi y^0<\xi\varrho^0<\xi x^0$ or infinite space-like hypersurface $\Sigma(\varrho)$, 
$\varrho^\mu=(\varrho^0,\vec{\rho})$ with element $d\Sigma_\mu(\varrho)$, 
$d^4\varrho=d\varrho^\mu d\Sigma_\mu(\varrho)$, and with indexes $\{C/D\}$, that 
are not correlated at all with indexes $\xi=\pm$ for the {\sl off-shell composite wave 
functions} $\widehat{\Upsilon}^{\xi}_{\{C/D\}j}(\varrho)$: 
\bea 
\fl  
{\cal A}^{j(\xi)}_{DC}\delta_{\xi\eta}=\xi\!\int\!d^3\!\rho\,
\widehat{\overline{\Upsilon}}{}^{\eta}_{\{D/C\}j}(\varrho)\,\gamma^0\,
\widehat{\Upsilon}^{\xi}_{\{C/D\}j}(\varrho)= \xi\!\int\!d\Sigma_\mu(\rho)\,
\widehat{\overline{\Upsilon}}{}^{\eta}_{\{D/C\}j}(\varrho)\,\gamma^\mu\,
\widehat{\Upsilon}^{\xi}_{\{C/D\}j}(\varrho), 
\label{3_13_12} \\ 
\fl  
\widehat{\Upsilon}^{\xi}_{\{C/D\}j}(\varrho)=\!
\int\! d^4{y}\,\theta(\xi(\varrho^0-y^0))\frac{1}i S^{-\xi}_j(\varrho-y)
\widetilde{M}_{j\{C/D\}}\Psi^*_{\{C/D\}F}(y)\Psi_{\{C/D\}I}(y),
\label{3_14_0} \\ 
\fl  
\widehat{\overline{\Upsilon}}{}^{\xi}_{\{D/C\}j}(\varrho)=\!
\!\int\! d^4{x}\,\Psi^*_{\{D/C\}F}(x)\Psi_{\{D/C\}I}(x)\widetilde{M}_{j\{D/C\}}\,
\theta(\xi(x^0-\varrho^0))\frac{1}i S^{-\xi}_j(x-\varrho), 
\label{3_14} \\ 
\fl  
\mbox{where: }\;\theta(\xi(\varrho^0-y^0))\mapsto \theta(\xi(n_\Sigma(\varrho-y))),\;
\mbox{ for: }\; d^3\!\rho\mapsto d\Sigma^\mu(\rho)=n^\mu_\Sigma d\Sigma,\quad 
n^2_\Sigma=1, 
\label{3_13_14} \\
\fl  
\mbox{with: }\theta(t)\theta(-t)=0,\;\;\theta^2(t)=\theta(t),\;\;\,
\theta(\xi(x^0-\varrho^0))\theta(\xi(\varrho^0-y^0)) \Leftrightarrow 
\theta(\xi(x^0-y^0)),   
\label{3_13_13} 
\eea
at least for $\xi\varrho^0=\xi\left(x^0+y^0\right)/2 \in \left(\xi y^0,\xi x^0\right)$, 
and thus for all admissable $\varrho^0$ due to Lorentz covariance of 
(\ref{3_13_12})--(\ref{3_13_13}) and similarly (\ref{3_thet}). 
An approximate transformation of the product of external wave packets into the scalar 
counterpart (\ref{C_04}) (below) of composite wave functions (\ref{3_14_0}), 
(\ref{3_14}), was shown in \cite{ahm}. 
The Lorentz covariance of definitions (\ref{3_13_12})--(\ref{3_13_13}) with 
{\sl arbitrary} $\varrho^0$ and time-like $n^\mu_\Sigma$ immediately resolves both 
problems with causality and with covariant equal time prescription, these mentioned in 
paragraph 5.2 of \cite{bth}. The exact transition (\ref{3_13_12}) to scalar product 
for the off shell fermionic composite wave functions eventually means that the 
transformation of amplitude (\ref{3_13_1}) into the {\sl on shell} scalar product similar 
to (\ref{eq:spinor_dp_sm_}) implies some kind of the on shell reductions: 
\bea
\fl  
{\cal A}^j_{DC}\longmapsto {\cal A}^{j(\xi)}_{DC}, \quad 
\widehat{\Upsilon}^{\xi}_{\{C/D\}j}(\varrho)\longmapsto \Xi^{\xi}_{\{C/D\}j}(\varrho), 
\quad \widehat{\overline{\Upsilon}}{}^{\xi}_{\{D/C\}j}(\varrho)\longmapsto 
\overline{\Xi}{}^{\xi}_{\{D/C\}j}(\varrho),
\label{3_15} 
\eea
into the on-shell wave functions with $\xi=+$ for neutrino and $\xi=-$ for antineutrino, 
that now become correlated with indexes $\{C/D\}$. Effectively this looks like the simple 
reduction of the sources for some $\xi\widetilde{y}^0<\xi\varrho^0<\xi\widetilde{x}^0$, 
at first step as: 
\bea
\fl 
\theta(\xi(\varrho^0-y^0))\widetilde{M}_{j\{C/D\}}\Psi^*_{\{C/D\}F}(y)
\Psi_{\{C/D\}I}(y)\mapsto \theta(\xi(\varrho^0-y^0))
\delta(y^0-\widetilde{y}^0)\gamma^0\,\Xi^{\xi}_{\{C/D\}j}(y), 
\nonumber \\
\fl 
\theta(\xi(x^0-\varrho^0))\Psi^*_{\{D/C\}F}(x)\Psi_{\{D/C\}I}(x)
\widetilde{M}_{j\{D/C\}}\mapsto \theta(\xi(x^0-\varrho^0))
\delta(x^0-\widetilde{x}^0)\,\overline{\Xi}{}^{\xi}_{\{D/C\}j}(x)\,\gamma^0.   
\nonumber
\eea
In accordance with above discussions of Huygens' principle and (\ref{3_4}), this    
means the wave functions (\ref{3_14_0}), (\ref{3_14}) become again solutions to free 
Dirac equation (\ref{3_07}) only after omitting the time-ordering $\theta$-functions:   
$\widehat{\Upsilon}\mapsto\Upsilon$. 
This is well justified for amplitudes (\ref{3_13_1}), (\ref{3_13_11}) only when the 
sources in the l.h.s. of their reductions are well localized in time near some fixed 
points: $\widetilde{y}^0\sim Y^0_{\{C\}}$, $\widetilde{x}^0\sim X^0_{\{D\}}$, with 
$\xi Y^0_{\{C/D\}}\ll\xi\varrho^0\ll\xi X^0_{\{D/C\}}$,  
$\xi \mapsto [\pm] \Leftrightarrow [C/D]$, for each macrodiagram (\ref{3_13_1}). 
The approximate reduction (\ref{3_15}), reproducing also the next ones for the 
sources, also without the temporal $\theta$- factors, was shown in \cite{gunt,bth} with 
non relativistic Gaussian profile (\ref{eq:gauss_p}) of $\phi^\sigma(\ka,\p_a)$ at 
$\x_a=0$ for all external wave packets $\Psi_{\{C/D\}I,F}(x)$ (as is shown above, follows 
from (\ref{eq:phidef}) only at $c\to\infty$). 
By remaining exact solutions for free relativistic equations (\ref{3_07}), they 
propagate with the inevitable frequency mixing and usual Gaussian spreading 
\cite {b_d,i_z}. 
\subsection{Composite wave function for pion decay vertex}\label{sec:comp_wf}
\begin{figure}[htb]
\centering  
\includegraphics[height=.42\textheight]{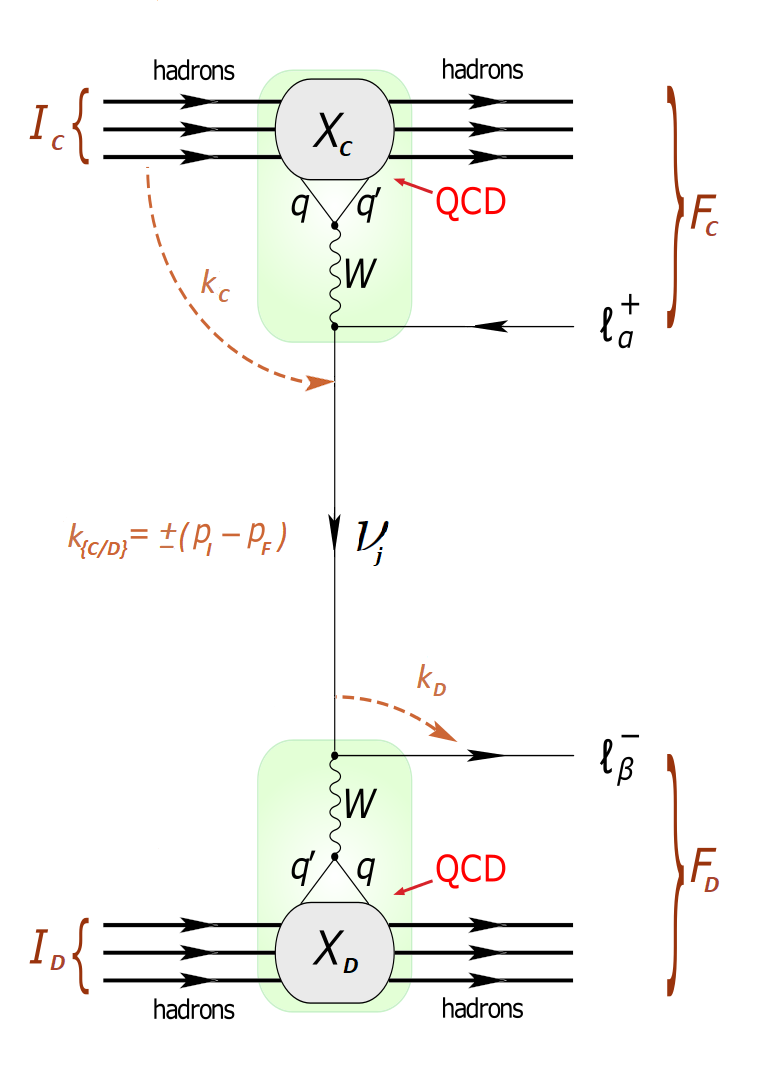}
\caption{Macroscopic Feynman diagram. Definition of points and momentums.}
\label{fig:3}
\end{figure}

For decay vertices $\pi^+\to\mu^+\,\nu_\mu$ and $\pi^-\to\mu^-\,\overline{\nu}_\mu$ 
(as well as for the cases $\pi^+\to e^+\,\nu_e$ or $\pi^-\to e^-\,\overline{\nu}_e$) 
the wave functions (\ref{3_14_0}), (\ref{3_14}) of creation process at the point $\{C\}$ 
imply:  
\bea 
\fl  
\Psi^*_{\{C\}F}(y)\Psi_{\{C\}I}(y)\!\mapsto\!\Xi^{(-)}_{p_\mu,Y_\mu,s_\mu}(y)
F^{\pi^+}_{p_\pi Y_\pi}(y),\,\mbox{ or }\,\Psi^*_{\{C\}F}(x)\Psi_{\{C\}I}(x)\!\mapsto\! 
\overline{\Xi}{}^{(+)}_{p_\mu,Y_\mu,s_\mu}(x)F^{\pi^-}_{p_\pi Y_\pi}(x),
\nonumber \\
\fl  
\widehat{\Upsilon}{}^{(+)}_{\{C\}j}(\varrho)= 
\frac{1}i\!\int\! d^4{y}\,\theta(\varrho^0-y^0)\,S^{-}_{j}(\varrho-y)\widetilde{M}_{jC}
\Xi^{(-)}_{p_\mu,Y_\mu,s_\mu}(y)F^{\pi^+}_{p_\pi Y_\pi}(y),\;
\mbox{ for }\;\nu_{\mu(e)} \;\,(\nu_j),
\label{3_18} \\
\fl  
\widehat{\overline{\Upsilon}}{}^{(-)}_{\{C\}j}(\varrho)=
\frac{1}i\!\int\! d^4{x}F^{\pi^-}_{p_\pi Y_\pi}(x) 
\overline{\Xi}{}^{(+)}_{p_\mu,Y_\mu,s_\mu}(x)\widetilde{M}_{jC}
\,S^{+}_{j}(x-\varrho)\,\theta(\varrho^0-x^0), \; \mbox{ for }\;\overline{\nu}_{\mu(e)}
\;\,(\overline{\nu}_j), 
\label{3_19}
\eea
where, according to definitions of scalar wave packets (\ref{2_10}), 
(\ref{eq:phidef})--(\ref{2_19}) with their scalar products (\ref{2_30}), 
(\ref{eq:scalar_dp_end}), (\ref{3_19_1}) and in view of the absence of Dirac sea 
\cite{schw,blt,oksak,jost,s_w,stroc,Dvt,b_d,i_z} for the pseudo-scalar charged pions 
$\pi^\mp$, with thus undirected external lines one has: 
\bea
\fl  
e^{i(p_\pi Y_\pi)}F^{\pi^\mp}_{p_\pi Y_\pi}(y)=
(\mp i)N_{\sigma\pi}D^{\mp}_{m_\pi}(\pm(y-Y_\pi-i\zeta_\pi))=
-i N_{\sigma\pi}D^-_{m_\pi}(y-Y_\pi-i\zeta_\pi). 
\label{3_19_0} 
\eea
Then, by dropping the temporal factors $\theta(t)$ for $y$- independent matrix 
$\widetilde{M}_{jC}$, these candidates into the free (anti-) neutrino wave packets recast 
into the following solutions of free Dirac equations (\ref{3_07}), denoted below as 
$\widehat{\Upsilon}^{\xi}_{\{C\}j}(\varrho)\longmapsto\Upsilon^{\xi}_{\{C\}j}(\varrho)$: 
\bea 
\fl  
\left.\!\! \begin{array}{c}
\Upsilon^{(+)}_{\{C\}j}(\varrho) \\
\overline{\Upsilon}^{(-)}_{\{C\}j}(\varrho)
\end{array}\!\right\}\! = e^{-i\Phi_{\{C\}}} 
\left\{\!\begin{array}{c}
\left[i(\gamma\partial_\varrho)+m_j\right]
\widetilde{M}_{jC}\left[m_\mu-i(\gamma\partial_{Y_\mu})\right]\!u^{(-)}(p_\mu,s_\mu) \\
\overline{u}^{(+)}(p_\mu,s_\mu)\!\left[m_\mu+i(\gamma\partial_{Y_\mu})\right]
\widetilde{M}_{jC}\left[i(\gamma\partial_\varrho)-m_j\right]
\end{array}\!\right\}\!\frac {{\cal G}^{(+)}_{\{C\}j}(\varrho)}{2 m_\mu},
\label{3_20_0} \\
\fl  
\mbox {where: }\;\Phi_{\{C\}}= (p_\pi Y_\pi)-(p_\mu Y_\mu), 
\label{3_20_3} \\
\fl  
{\cal G}^{(+)}_{\{C\}j}(\varrho)=i N_{\sigma\mu}N_{\sigma\pi}
\!\int\! d^4{y}\,D^{-}_{m_j}(\varrho-y) 
D^{-}_{m_\mu}(Y_\mu-i\zeta_\mu-y) D^{-}_{m_\pi}(y-Y_\pi-i\zeta_\pi), 
\label{3_20} 
\eea
that is explicitly calculated by Fourier transformation, with $Z_\pi=Y_\pi+i\zeta_\pi$, 
$Z_\mu=Y_\mu+i\zeta_\mu$: 
\bea
\fl  
{\cal G}^{(+)}_{\{C\}j}(\varrho)=
2\pi \!\int\!\frac{d^4 q}{(2\pi)^4}\,e^{-i(q\varrho)}\,
\theta(q^0)\delta(q^2-m^2_j)\widehat{V}_{\{C\}}(q),\;\;
\mbox{ where, }\;\forall\,q:\;
\label{3_20_1} \\
\fl  
\widehat{V}_{\{C\}}(q)=- N_{\sigma\mu}N_{\sigma\pi} 
\!\int\!d^4{y}\,e^{i(qy)}\, D^{-}_{m_\mu}(Z^*_\mu-y) D^{-}_{m_\pi}(y-Z_\pi),   
\;\mbox{ or more generally:} 
\label{3_20_2} \\
\fl  
\widehat{V}_{[C/D]}(q)=\!\int\! d^4{y}\,e^{[\pm]i(qy)}
\psi^*_{\sigma_{bF}}(\p_{bF},Y_{bF[C/D]}-y)\,\psi_{\sigma_{aI}}(\p_{aI},Y_{aI[C/D]}-y),
\label{3_21}
\eea 
(cmp. (\ref{C_05})) are the overlap functions \cite{nn,bth,ahm} of creation/detection 
processes. 
For the independent plane-wave limit (\ref{eq:limit_sigma_0}) of packet functions (\ref{2_18}) 
with $\sigma_\pi,\sigma_\mu\to 0$ they keep energy-momentum conservation for 4-vector 
$k=p_\pi-p_\mu\equiv k_{\{C\}}$, Figure \ref{fig:3}, giving: 
\bea
\fl  
\widehat{V}_{\{C\}}(q) \longmapsto 
(2\pi)^4\,e^{i\Phi_{\{C\}}}\,\delta_4(q-k),\;\;\forall\,q, \;\; \mbox{ with the function:}
\label{3_22} \\
\fl  
{\cal G}^{(+)}_{\{C\}j}(\varrho)\longmapsto 2\pi\,e^{i\Phi_{\{C\}}}
e^{-i(k\varrho)}\,\theta(k^0)\,\delta(k^2-m^2_j), \;
\mbox{ and the wave functions:}
\label{3_31} \\
\fl  
\left.\!\! \begin{array}{c}
\Upsilon^{(+)}_{\{C\}j}(\varrho) \\
\overline{\Upsilon}^{(-)}_{\{C\}j}(\varrho)
\end{array}\right\}\longmapsto 2\pi\, e^{-i(k\varrho)} 
\sum_{s=\pm 1/2} \left\{ \begin{array}{c}
u^{(+)}_j(k,s)\left(\!\overline{u}^{(+)}_j(k,s)\widetilde{M}_{jC}
u^{(-)}(p_\mu,s_\mu)\!\right) \\ 
\left(\!\overline{u}^{(+)}(p_\mu,s_\mu)\widetilde{M}_{jC}u^{(-)}_j(k,s)\!\right)
\overline{u}^{(-)}_j(k,s)
\end{array}\right\}
\nonumber \\
\fl  
\times \,\theta(k^0)\,\delta(k^2-m^2_j),\;\mbox{ where, if the matrix }\; 
\widetilde{M}_{j\{C\}} \propto ({\rm I}\pm\gamma^5), 
\label{3_23}  
\eea
as for a charged-current case, 
then it can eliminate the sum selecting here only one value of helicity $s$ for the 
massless (anti-) neutrino ${u}^\xi_{m_j=0}(k,s)$ \cite{nn,bth,ahm}. Thus the plane-wave limit 
for external packets automatically leads to plane wave for internal packet on the respective 
mass shell. For $\Delta=\Delta(q)$ the explicit value (\ref{B_14_0})--(\ref{B_15}) reads: 
\bea
\fl  
\widehat{V}_{\{C\}}(q)=\frac{N_{\sigma\mu}N_{\sigma\pi}}{8\pi}\,
\exp\left\{i(qZ_\eta)\right\}\,\frac{\theta(\Delta)}{W}
\biggl\{\exp\left(\chi_\Delta W\right)-\theta(q^2)
\exp\left(-\varepsilon(q^0)\chi_\Delta W\right)\biggr\}, 
\label{3_24} \\
\fl  
\chi_\Delta=\frac{\Delta^{1/2}(q)}{2q^2},\quad \chi_0=\frac{m^2_\pi-m^2_\mu}{2q^2}, \quad 
\eta_{\case{2}{1}}=
\frac 12\pm \chi_0, \quad 
Z_\eta=\eta_2Z_\pi+\eta_1Z^*_\mu = Y_\eta+i\zeta_\eta,
\label{3_25} \\
\fl  
\left\{\!\!
\begin{array}{c} Y_\eta\!=\eta_2Y_\pi\!+\eta_1Y_\mu \\ 
\zeta_\eta\!=\eta_2\zeta_\pi\!-\eta_1\zeta_\mu \end{array}\!\!\right\}\!, \;\;\,
Z_-\!=Z_\pi\!-Z^*_\mu = Y_-\!+i\zeta_+,\;\;\,
\left\{\!\!\begin{array}{c} Y_\pm \!=Y_\pi\pm Y_\mu \\ 
\zeta_\pm \!=\zeta_\pi\pm \zeta_\mu \end{array} \!\!\right\}\!,\;\;\,
W\!={\mathrm w}^{1/2}\!,  
\label{3_26} \\
\fl  
\Delta(q)=\left[(m_\pi+m_\mu)^2-q^2\right]\left[(m_\pi-m_\mu)^2-q^2\right]>0,
\,\mbox{ for }\, q^2\sim m^2_j\ll m^2_\mu<m^2_\pi. 
\label{3_27} 
\eea
$\chi_0\simeq 7/2$, $\chi_\Delta\to 0$, for $\sqrt{q^2}\to m_\pi-m_\mu$, but     
$\chi_0\simeq 10^{17}$, $\chi_0-\chi_\Delta\simeq 1,79$ for $\sqrt{q^2}=0,2\,eV$. 
Calculations of (\ref{3_24}) with $W(\mathrm w)=\left[q^2 Z^2_- -(qZ_-)^2\right]^{1/2}$, 
for $q^2\leqslant (m_\pi-m_\mu)^2$, are given in  \ref{ap:C}. In  \ref{ap:D} 
it is shown how the plane-wave limit (\ref{3_22}) of   (\ref{3_24}) is realized without any 
addition constraints onto the universal functions ${\cal I}(\tau_a)$, 
$g_{1}(m_a,\sigma_a)$, $a=\pi, \mu$.

Note, that unlike the common belief \cite{bth,ahm} the overlap functions (\ref{3_21}), 
(\ref{3_24}) explicitly depend on the centers of initial and final wave packets, and thus 
from the effective points $X_{\{D/C\}}$, $Y_{\{C/D\}}$ of (anti-) neutrino 
creation/detection, that in turn, define the base time $T=\xi(X^0_{\{D/C\}}-Y^0_{\{C/D\}})$ 
and the 
base line $\vec{\rm L}=\xi(\vec{\rm X}_{\{D/C\}}-\vec{\rm Y}_{\{C/D\}})$ of oscillations 
\cite{nn,nsh,dn,n_sh,NN_shk}. Due to opposite signs of $y$ for initial and final 
wave packets the properties of   (\ref{3_20_2}) are essentially different from the ones of 
usual two-particle phase volume \cite{blt,b_k} (clf.  \ref{ap:C}). Similarly \cite{nn} 
its extreme properties define the points $X_{\{D/C\}}$, $Y_{\{C/D\}}$ (clf.  
\ref{ap:D}). 
This becomes transparent for the approximate expression (\ref{3_40}), which in fact  
represents the effective off-shell Gaussian-like profile for narrow wave packet of 
intermediate (anti-) neutrino, arising for external packets are close to plane waves. 
From this approximation it is clear, that with the substitution into (\ref{3_20_0})  
(see (\ref{3_37})--(\ref{3_37_A})) the (\ref{3_20_3}) and: 
\bea
\fl  
{\cal G}^{(+)}_{\{C\}j}(\varrho)= 
e^{i\Phi(k))}{\cal F}^{(+)}_{\{C\}j}(\varrho),\;\mbox{ for }\;
\Phi(k)=\Phi_{\{C\}}(k)=(p_\pi Y_\pi)-(p_\mu Y_\mu)\equiv \Phi_{\{C\}}, 
\label{3_20_G} 
\eea
up to the small corrections, that by symmetry are reduced to orders of 
$O(q-k)\mapsto O\left((q-k)^2\right)\sim O(g^{-1}_{1a})$, one can replace 
the (\ref{3_20_0})--(\ref{3_21}) to the following simple ones:
\bea 
\fl  
\left. \begin{array}{c}
\Upsilon^{(+)}_{\{C\}j}(\varrho) \\
\overline{\Upsilon}^{(-)}_{\{C\}j}(\varrho)
\end{array}\right\} \longmapsto 
\left\{\begin{array}{c} \left[m_j+i(\gamma\partial_\varrho)\right]
\widetilde{M}_{jC}u^{(-)}(p_\mu,s_\mu) \\
\overline{u}^{(+)}(p_\mu,s_\mu)\widetilde{M}_{jC}
\left[i(\gamma\partial_\varrho)-m_j\right] \end{array}\right\} 
{\cal F}^{(+)}_{\{C\}j}(\varrho), \;\;\mbox{where:}
\label{3_20_0G} \\
\fl  
{\cal F}^{(+)}_{\{C\}j}(\varrho)=
\!\!\int\!\frac{d^4 q}{(2\pi)^4}\,e^{-i(q\varrho)}\,
2\pi\,\theta(q^0)\delta(q^2-m^2_j)\widehat{\cal V}_{\{C\}}(q)\equiv
\!\int\!\frac{d^4 q}{(2\pi)^4}\,e^{-i(q\varrho)}\,{\cal H}^{(+)}_{\{C\}j}(q),
\label{3_20_1_G} \\
\fl  
\mbox{and }\;\forall\,q:\;
\widehat{V}_{\{C\}}(q)\equiv e^{i(p_\pi Y_\pi)-i(p_\mu Y_\mu)}\widehat{\cal V}_{\{C\}}(q),
\;\mbox{ or }\; 
\widehat{\cal V}_{\{C\}}(q)\equiv e^{-i\Phi_{\{C\}}(k))}\widehat{V}_{\{C\}}(q), 
\label{3_21_G} 
\eea
is ``reduced'' overlap function, given approximately by (\ref{3_40}) for $q^2\gtrless 0$ as: 
\bea
\fl  
\widehat{\cal V}_{\{C\}}(q)\approx  
(2\pi)^4\,\theta(\Delta)\,\widehat{\delta}_{\{C\}}(q-k)
e^{i\left((q-k)Y_{\{C\}}\right)} e^{\Sigma_k(Y_-)},\quad 
\widehat{\delta}_{\{C\}}(\kappa)\!=\!
\left[\frac{|\vec{\cal T}|}{\pi^{4}}\right]^{1/2}\!\!\!
e^{-\left(\kappa\vec{\cal T}\kappa\right)}, 
\label{3_40_V} \\
\fl  
\mbox{with: }\;\kappa=q-k,\;\;\, k\equiv k_{\{C\}},\;\;\,
\vec{\cal T}\equiv \vec{\cal T}_{\{C\}},\quad 
{\cal Y}^\lambda_-=\left(\Pi_{P_{\Pi}}\Pi_k Y_-\right)^\lambda=
\left(\Pi_{P_{\Pi}}Y_{\Pi_k -}\right)^\lambda,
\label{3_40_Y} \\
\fl  
\Pi_{P_{\Pi}}\Pi_k=\Pi_k\Pi_{P_{\Pi}}, \;\;\,  
\Sigma_k(Y_-)\equiv \Sigma^{\{C\}}_k(Y_-)
=\frac{1}{2g_+}\!\left[Y^2_{\Pi_k -}-\frac{(P_{\Pi_k}Y_{\Pi_k -})^2}
{P^2_{\Pi_k}}\right]\!=\frac{{\cal Y}^2_-}{2g_+}<0,
\label{3_40_S} 
\eea
because $(k{\cal Y}_-)=(P_{\Pi_k}{\cal Y}_-)=0$, and for $P=p_\pi+p_\mu$, either 
$k$ or $P_{\Pi_k}=\Pi_k P$ is time-like. The function $\widehat{\delta}_{\{C\}}(\kappa)$ 
in (\ref{3_40_V}) for symmetrical positively defined tensor 
${\cal T}^{\beta\lambda}_{\{C\}}$ (\ref{3_39}) similarly \cite{nn} keeps the approximate 
energy-momentum conservation according to (\ref{3_41}). The effective impact point 
$Y_{\{C\}}$ (\ref{3_37_A}) in (\ref{3_40_V}) depends on the widths 
$\sigma_{\pi},\sigma_\mu$ only via dimensionless combination $|\varsigma|\leqslant 1$ for 
${\rm O}^{\lambda\beta}_{(\pi)}+{\rm O}^{\lambda\beta}_{(\mu)}=2{\rm g}^{\lambda\beta}$, 
${\rm O}^{\lambda\beta}_{(a)}(k,P)={\rm O}^{\lambda\beta}_{(a)}(P,k)$, with:   
\bea
&&\!\!\!\!\!\!\!\!\!\!\!\!\!\!\!\!\!\!\!\!\!\!\!
2Y^\lambda_{\{C\}}=Y^\lambda_+ +\partial^\lambda_q(Q(q)Y_-)\bigr|_{q=k}=
\sum_{a=\pi,\mu}{\rm O}^{\lambda}_{(a)\,\beta}Y^\beta_{a}, \qquad 
\varsigma=\frac{g_\pi-g_\mu}{g_\pi+g_\mu}\equiv \frac{g_-}{g_+}, 
\label{3_40_S_1} \\
&&\!\!\!\!\!\!\!\!\!\!\!\!\!\!\!\!\!\!\!\!\!\!\!
{\rm O}^{\lambda\beta}_{\left(\case{\pi}{\mu}\right)}(k,P)= 
{\rm g}^{\lambda\beta}(1\mp\varsigma) \pm 
\frac{(k-\varsigma P)^\lambda (\Pi_{k}P)^\beta k^2 +
(P -\varsigma k)^\lambda(\Pi_{P}k)^\beta P^2}{(kP)^2-k^2P^2}.
\label{3_40_S_2} 
\eea
This definition of point $Y_{\{C\}}$ of neutrino creation may be compared with (\ref{Pr_3_1}) 
below, that is the generalization of what was given in \cite{nn}. 
In view of (\ref{3_40_S}) the geometric suppression factor $\exp\left\{\Sigma_k(Y_-)\right\}$,  
depends here only on the relative separation of initial and final wave packets 
$Y_-\equiv Y^{\{C\}}_- =Y_\pi-Y_\mu$, but not of their distances from effective point 
$Y_{\{C\}}$. Only its transverse to the $k$ and $P_{\Pi_k}$ space-like two-dimension part 
${\cal Y}_-$ (\ref{3_40_Y}) is significant for suppression. 
In spite of exactly the same structure of obtained approximation (\ref{3_40_V}) as in 
\cite{nn}, with exactly the same physical meaning of corresponding values 
$\widehat{\delta}_{\{C\}}(\kappa)$, $\Sigma_k(Y^{\{C\}}_-)$, $Y_{\{C\}}$, $k_{\{C\}}$, 
$\vec{\cal T}_{\{C\}}$, their explicit expressions (\ref{3_40_Y}) -- (\ref{3_40_S_2}) 
may be essentially different from respective $\widetilde{\delta}_{C}(K)$, 
$\Sigma_C$, $Y_{C}$, $\widetilde{k}_{C}$, $\vec{\Re}^{-1}_{C}/4$ given therein. 

Indeed, substitution of (\ref{2_15_41}), (\ref{2_15_40_0}) to (\ref{3_21}) with arbitrary 
number of packets in initial $\{I\}$ and final $\{F\}$ states, 
$Y^\beta_a\mapsto x^{C\beta}_{a_C}$, $a_C\in\{I \oplus F\}_C$, by means of Gaussian 
integration over Minkowski space \cite{kt_3} (cmp. \cite{nn,NN_anp}), for 
$R^\lambda=X^\lambda_D-Y^\lambda_C=(T,\vec{\rm L})$ gives: 
\bea
\fl  
\widehat{\cal V}_{\{C\}}(q) \approx 
\widehat{\cal V}^{CRG}_{\{C\}}(q)=(2\pi)^4\,
\widetilde{\delta}_C(K)\,\exp\left\{\Sigma_C\right\}\,
\exp\left\{i\left(K{Y}_C\right)\right\},  \quad 
K=q-\widetilde{k}_C, 
\label{Pr_3_5} \\
\fl  
\mbox{with: }\; 
\widetilde{\delta}_C(K)=\frac{|\vec{\Re}^{-1}_{C}|^{1/2}}{(4\pi)^2}
\exp\left[-\frac 1{4}\left(K\vec{\Re}^{-1}_C K\right)\right]
\stackunder{\Re\to 0+}{\longmapsto}\delta_4(K), \;\mbox{ and: }\;
K\stackunder{C\mapsto D}{\longmapsto}-K, 
\label{Pr_3_4} 
\eea
\bea
\fl  
\widetilde{k}_{C}=\!\sum^{I_C}_{a_{iC}=1}\widetilde{p}^C_{a_{iC}}-
\sum^{F_C}_{a_{fC}=1}\widetilde{p}^C_{a_{fC}},  
\quad \;
\widetilde{k}_{D}=\!\sum^{F_D}_{a_{fD}=1}\widetilde{p}^D_{a_{fD}}-
\sum^{I_D}_{a_{iD}=1}\widetilde{p}^D_{a_{iD}}, \quad \;
\vec{\Re}_C=\sum_{a_C}\vec{T}_{\!\!C a_C}, 
\label{Pr_3_3} \\
\fl  
\vec{T}_{\!\!C a_C} \leftarrowtail 
T^{\lambda\beta}_{a}=
\frac{m^2_a}{2\tau_a}
\left[\upsilon^\lambda_a\upsilon^\beta_a-{\rm g}^{\lambda\beta}\right], 
\quad \widetilde{p}_{a}=m_a\upsilon_{a}\left(1+\frac 3{2\tau_a}\right), \quad 
\upsilon_{a}\stackunder{g_2\to 0}{\longmapsto}u_{a}=\frac{p_a}{m_a},
\label{Pr_3_3_2} \\
\fl  
\Sigma_{C}= 
- \sum_{a_C}\!\!\left((x^C_{a_C}-Y_{C})\vec{T}_{\!\!C a_C}(x^C_{a_C}-Y_{C})\right), 
\,\mbox{ or: }\,C\mapsto D,\;\; Y_C=X_C\mapsto X_D=Y_D, 
\label{Pr_3_5_1}  \\
\fl  
Y_{C}=\vec{\Re}^{-1}_C \sum_{a_C}\vec{T}_{\!\!C a_C}x^C_{a_C},\;\mbox{ i.e. }\;
\sum_{a_C}\vec{T}_{\!\!C a_C}(x^C_{a_C}-Y_{C})=0, \quad 
\sum_{a_C}\partial_{x^C_{a_C}}^\lambda \Sigma_{C}=0,
\label{Pr_3_1}  \\ 
\fl  
\sum_{a_{C/D}}\left(\mp \partial_{x^{C/D}_{a_{C/D}}}^\lambda\right)
\left(\mp Y^\beta_{C/D}\right)=
\frac 12\left(\sum_{a_D}\partial_{x^D_{a_D}}^\lambda -
\sum_{a_C}\partial_{x^C_{a_C}}^\lambda \right)(X^\beta_{D}-Y^\beta_{C})=
\partial_{R}^\lambda R^\beta={\rm g}^{\lambda\beta}.
\label{Pr_3_2} 
\eea
Unlike the quadratic form with tensor $\vec{\cal T}_{\{C\}}$ 
(\ref{3_41}) -- (\ref{3_44_1}), the forms with tensors $\vec{\Re}^{-1}_{C,D}$ 
(\ref{Pr_3_3}) for two or more external packets are positively defined merely almost 
everywhere \cite{nn}. 
Moreover, founding only on different leading terms for powers of $x$ and $x^2$ in 
(\ref{2_15_41}), one can't neglect with the same accuracy of order 
$O(\tau^{-1}_a)=O(g^{-1}_{1a})$ the difference between $\widetilde{p}_{a}$ and 
$p_{a}=m_au_a$ in (\ref{Pr_3_3_2}) even for $g_2=0$ \cite{nn} and so one can't neglect the 
difference between $\widetilde{k}_{C/D}$ (\ref{Pr_3_3}) and $k_{C/D}$. 
Nevertheless, for two-packet case, the expressions (\ref{3_40_S}) and (\ref{Pr_3_5_1}) for  
geometric suppression factor $\Sigma_C$ coincide at specific reference frame, given in 
(81) of \cite{NN_anp}, but (\ref{3_40_Y}), (\ref{3_40_S}) automatically determine this factor 
in relatavistically invariant form (cmp. (184) of \cite{NN_20}). 
Moreover, at least two wave packets per vertex are necessary in order to define and to save 
the effective impact point $Y_{\{C\}}$ (\ref{3_40_S_1}), (\ref{3_40_S_2}) in reduced overlap 
function (\ref{3_40_V}). 

Similarly (\ref{3_20_0}) the Feynman amplitude (\ref{3_13_1}) admits an exact operator 
factorization of its spin structure in view of (\ref{3_07}), (\ref{A_1}), (\ref{A_1_2}) by 
means of translation invariance of the representations 
(\ref{3_3})--(\ref{eq:spinor_wp_wf_end}) and the function $\psi_\sigma(\p_a,\underline{x})$ 
(\ref{2_18}) in overlap function (\ref{3_21}), with integration by parts over $x$ and/or $y$, 
for example ($\eta^\prime \mapsto \eta_{a_D}$, $\eta \mapsto \eta_{a_C}$), as:  
\bea
\fl  
{\cal A}^j_{DC}=\!\prod_{a_D\in\{I \oplus F\}_D}\!\!
\overline{\cal U}^{(\eta^\prime)}_{p^D_{a_D}}(x^D_{a_D})\!
\left(\frac{1}{2m_{a_D} }
\left[m_{a_D}\!+i\left(\!\gamma\partial_{x^D_{a_D}}\!\right)\right]\!
\widetilde{M}_{jD}\left[m_j+i(\gamma\widehat{\partial})\right]\!\widetilde{M}_{jC} \right. 
\label{3_60_0}  \\
\fl  
\times\!\! \prod_{a_C\in\{I\oplus F\}_C}\!\! \left.
\left[m_{a_C}\!-i\left(\gamma\partial_{x^C_{a_C}}\!\right)\right]\frac{1}{2m_{a_C}}
\!\int\!\!\frac{d^4 q}{(2\pi)^4}\widehat{V}_{\{D\}}(q)
\frac{(-i)}{m^2_j-q^{2}-i0}\,\widehat{V}_{\{C\}}(q)\right)
{\cal U}^{(\eta)}_{p^{C}_{a_C}}(x^{C}_{a_C}).
\nonumber
\eea
All the differentiations act here on $\widehat{V}_{\{D/C\}}(q)$ only; 
$\widehat{\partial}$ is the any one of differential operators defined in (\ref{Pr_3_2}); 
${\cal U}^{(\eta)}_{p^{C}_{a_C}}(x^{C}_{a_C})$ are bispinorial plane waves 
(\ref{3_2:1})--(\ref{3_2:4}) of wave packets (\ref{3_3})--(\ref{eq:spinor_wp_wf_end}) of 
external particles at the creation/detection points $C/D$. 
Similarly (\ref{3_20_0G}) the off-shell narrow-packets approximation of (\ref{3_24}) as 
(\ref{3_21_G})--(\ref{3_40_S_2}), or in general, as (\ref{Pr_3_5})--(\ref{Pr_3_1}), then 
gives for oscillation amplitude (\ref{3_13_1}), similarly \cite{nn}: 
\bea
\fl  
{\cal A}^j_{DC}\approx \overline{\vec{u}}^{\{\eta^\prime\}}(\{p^{D}_{f,i}\})\frac 1i
\widetilde{M}_{jD}\!\!\int\!\!\frac{d^4 q}{(2\pi)^4}
\widehat{\cal V}_{\{D\}}(q)\,\frac{(\gamma q)+m_j}{m^2_j-q^{2}-i0}\,
\widehat{\cal V}_{\{C\}}(q)
\widetilde{M}_{jC}\,\vec{u}^{\{\eta\}}(\{p^{C}_{f,i}\}), 
\label{Pr_3_6} 
\eea
where $\vec{u}^{\{\eta\}}(\{p^{C}_{f,i}\})$ is a short notation for product of 
bispinors arising in the products of ${\cal U}^{(\eta)}_{p^{C}_{a_C}}(x^{C}_{a_C})$ 
in (\ref{3_60_0}). The final result of factorization for those approximations depends on the 
actual pole integration procedure of propagator, selecting neutrino/antineutrino pole with 
definite $\xi=\pm$ and, similarly (\ref{3_23}), for 
$k_{\{C\}}\mapsto k_C\approx k_D \leftarrowtail k_{\{D\}}$, in general case reads as:
\bea
\fl  
{\cal A}^j_{DC}\mapsto 
\overline{\vec{u}}^{\{\eta^\prime\}}(\{p^{D/C}_{f,i}\})\widetilde{M}_{j\{D/C\}}\!\!\!
\sum_{s=\pm 1/2}\!\! u^{\xi}_j(k_C,s)\overline{u}^{\xi}_j(k_C,s)\widetilde{M}_{j\{C/D\}}
\vec{u}^{\{\eta\}}(\{p^{C/D}_{f,i}\})\!\left[\widetilde{\cal A}^{j(\xi)}_{DC}\right]\!. 
\label{3_60_1} 
\eea
A possibility of this reduction depends also on subsidiary kinematical conditions 
\cite{Nis} in  both vertices of the macrodiagram (\ref{3_13_1}). Here the amplitudes 
$\widetilde{\cal A}^{j(\xi)}_{DC}$, like (\ref{3_13_11}) for (\ref{3_13_1}), represent 
the respective fractions (\ref{C_02}), (\ref{3_54}) below of the given below scalar 
Feynman amplitude (\ref{3_56_0}), (\ref{C_01}) corresponding to (\ref{3_60_0}), 
(\ref{Pr_3_6}) as: 
\bea
\widetilde{\cal A}^{j}_{DC}=\frac 1i
\int\frac{d^4q}{(2\pi)^4}\,\widehat{\cal V}_{\{D\}}(q)\, \frac{1}{m^2_j-q^2-i0}\,
\widehat{\cal V}_{\{C\}}(q)=\sum_{\xi=\pm} \widetilde{\cal A}^{j(\xi)}_{DC}.  
\label{3_56_0}
\eea
The $\left[\widetilde{\cal A}^{j(\xi)}_{DC}\right]$ denotes the respective on-shell value 
(\ref{C_06}) below, arises as corresponding pole contribution of propagator here or in 
(\ref{3_54}) below. At least for charged-current case (\ref{3_23}) the factor 
befor $\left[\widetilde{\cal A}^{j(\xi)}_{DC}\right]$ in (\ref{3_60_1}) is simplified for 
$m_j\to 0$ with factorization into the factors independent of $j$, for 
$s=\mp\frac 12$ refered as Left/Right helicity\footnote{Here $\overline{M}$ means 
Dirac conjugation.} and $\xi=\pm$: 
\bea
\fl {\cal A}^j_{DC}\!\mapsto\! \overline{M}^\xi_{\{D/C\}}M^\xi_{\{C/D\}}\!\!
\left[\widetilde{\cal A}^{j(\xi)}_{DC}\right]\!, \quad  
M^\xi_{\{C/D\}}\!=
\overline{u}^{\xi}_j\!\left(k_{C},-\case{\xi}{2}\right)\!\widetilde{M}_{j\{C/D\}}
\vec{u}^{\{\eta\}}(\{p^{C/D}_{f,i}\})\biggr|_{m_j=0}\!\!.
\label{3_60_2} 
\eea
Since the last factorizations take place only for the case of on-shell (anti-) neutrino, 
the transition amplitude like on Figure \ref{fig:2} from flavour $\ell=\alpha(\xi)$ to 
flavour $\ell^\prime=\beta(\xi)$, in view of (\ref{3_12_05}), (\ref{3_12_03}), for 
corresponding choice of matrix $\widehat{U}^{(\xi=+,-)}\mapsto\{U,U^*\}$ recasts as: 
\bea
\fl 
{\cal A}^{\beta\alpha}_{DC}=
\sum_j\widehat{U}^{(\xi)}_{\beta j}{\cal A}^j_{DC}\widehat{U}^{(\xi)*}_{\alpha j}
\longmapsto\overline{M}^\xi_{\{D/C\}}M^\xi_{\{C/D\}}
\sum_j\widehat{U}^{(\xi)}_{\beta j}\left[\widetilde{\cal A}^{j(\xi)}_{DC}\right]
\widehat{U}^{(\xi)*}_{\alpha j}. 
\label{3_60_3} 
\eea
Now we will focus on the further properties of composite wave functions simplified 
so up to their scalar versions independently defined below. 
\subsection{Scalar composite wave function} 
In order to outline the differences between the spinor (\ref{3_13_1}) and scalar 
(\ref{3_56_0}) cases, we briefly give here a scalar version of formalism of composite 
wave function $\widehat{\cal F}^{(\xi)}_{[C/D]j}(\varrho)$ following the above 
fermionic case (\ref{3_13_1})--(\ref{3_13_13}). We start with Feynman amplitude 
(\ref{3_56_0}) in configuration space (without matrix factors 
$\widetilde{M}_{j\{D/C\}}$): 
\bea
\fl 
\widetilde{\cal A}^{j}_{DC}\equiv \!\!
\int\!d^4 x \,\Psi^*_{DF}(x)\Psi_{DI}(x)\!\int\!d^4 y\,
\frac {1}i D^c_{m_j}(x-y)\,\Psi^*_{CF}(y)\Psi_{CI}(y)= 
\sum_{\xi=\pm} \widetilde{\cal A}^{j(\xi)}_{DC},
\label{C_01} \\
\fl 
\widetilde{\cal A}^{j(\xi)}_{DC}=\!\!
\int\!d^4 x\,\Psi^*_{DF}(x)\Psi_{DI}(x)\!\int\!d^4 y \,
\theta(\xi(x^0-y^0))\,\frac {\xi}i\,D^{-\xi}_{m_j}(x-y)\,\Psi^*_{CF}(y)\Psi_{CI}(y),  
\label{C_02}
\eea
defining then the off-shell scalar composite wave functions as: 
\bea
\fl 
\widehat{\cal F}^{(\xi)}_{[C/D]j}(\varrho)=\!\int\!d^4 y \,
\theta\left([\pm]\xi(\varrho^0-y^0)\right) \frac {\xi}i\,
D^{-\xi}_{m_j}\left([\pm](\varrho-y)\right) \Psi^*_{[C/D]F}(y) \Psi_{[C/D]I}(y), 
\label{C_04} \\
\fl 
\delta_{\eta\xi}\widetilde{\cal A}^{j(\xi)}_{DC}=
\xi\!\!\int\!d^3\!\rho \,
\widehat{\cal F}^{(\eta)}_{\{D\}j}(\varrho)\,
[i\!\stackrel{\leftrightarrow}{\partial_\varrho^{0}}]\,
\widehat{\cal F}^{(\xi)}_{\{C\}j}(\varrho)\Longleftrightarrow 
\xi\!\!\int \!d^3\!\rho  
\,\widehat{\cal F}^{(\eta)}_{\{D\}j}(\varrho)\,
(i\!\stackrel{\leftrightarrow}{\partial^{0}_\varrho})\,
\widehat{\cal F}^{(\xi)}_{\{C\}j}(\varrho).  
\label{C_03} 
\eea
The first two equations follow here from the decomposition (\ref{E_01}). The second two  
equations reflects the scalar group property (\ref{3_19_1}), where the (\ref{C_04}),   
for $[C/D]$ correlated with $[\pm]$, combines both fermionic relations (\ref{3_14_0}), 
(\ref{3_14}), since the scalar propagator is undirected. 

In (\ref{C_03}) the first differential operation 
$[i\!\stackrel{\leftrightarrow}{\partial_\varrho^{0}}]$, 
unlike the second usual one $(i\!\stackrel{\leftrightarrow}{\partial^{0}_\varrho})$, 
does not feel the temporal $\theta$- functions from (\ref{C_04}). 
Their difference repeats the difference between Dyson's and Wick's prescriptions for 
chronological product in (\ref{3_13_13}). This difference is proportional to 
$\xi\theta(\xi(x^0-y^0))[\delta(\varrho^0-y^0)+\delta(\varrho^0-x^0)]$ and becomes 
negligible for the `sources' well localized in time near any points differ from 
$\varrho^0$, that was discussed after (\ref{3_15}). So, the use of wave packets for 
external states smoothes the difference between these two types of chronological ordering. 
Moreover this observation allowes to connect directly the off-shell spinor composite wave 
functions with their scalar counterparts by the same relations (\ref{3_20_0}), 
(\ref{3_20_G}), and thus (\ref{3_20_0G}), as for the on-shell ones. 
As the two spinor functions $\widehat{\Upsilon}^{(+)}_{\{C/D\}j}(\varrho)$ (\ref{3_18}) 
and $\widehat{\overline{\Upsilon}}{}^{(-)}_{\{C/D\}j}(\varrho)$ (\ref{3_19}) 
are connected with the same scalar function $\widehat{\cal F}^{(+)}_{[C/D]j}(\varrho)$, as 
well the two spinor ones $\widehat{\Upsilon}^{(-)}_{\{C/D\}j}(\varrho)$ (\ref{3_14_0}) and 
$\widehat{\overline{\Upsilon}}{}^{(+)}_{\{C/D\}j}(\varrho)$ (\ref{3_14})  
are connected with the same scalar function $\widehat{\cal F}^{(-)}_{[C/D]j}(\varrho)$. 

Assuming e.g. only two external wave packets (\ref{3_21}) for detection point $\{D\}$ also 
\cite{bth,ahm}, one can connect immediately the scalar counterpart of (\ref{3_18}), 
(\ref{3_19}), written now for the wave function 
$\widehat{\cal F}^{(+)}_{\{C\}j}(\varrho)$ (\ref{C_04}) instead of 
${\cal F}^{(+)}_{\{C\}j}(\varrho)$ (\ref{3_20_G}), (\ref{3_20_1_G}), with the function 
$\widehat{\cal G}^{(+)}_{\{C\}j}(\varrho)$ instead of ${\cal G}^{(+)}_{\{C\}j}(\varrho)$ 
(\ref{3_20}), (\ref{3_20_1}), by replacing $\pi\mapsto I$, $\mu\mapsto F$, for $[\pm]$ 
correlated with $[C/D]$ as in (\ref{3_21}), (\ref{C_04}). In accord with (\ref{3_14_0}), 
(\ref{3_14}) that is: 
\bea
\fl  
\widehat{\cal G}^{(+)}_{[C/D]j}(\varrho)= 
i N_{\sigma F}N_{\sigma I}\!\!\int\!\!d^4y\,\theta \left([\pm](\varrho^0-y^0)\right)
D^{-}_{m_j}\left([\pm](\varrho-y)\right) D^{-}_{m_F}(Z^*_F\!-y) D^{-}_{m_I}(y-Z_I).  
\nonumber 
\eea
Thus, from (\ref{3_56_0}), (\ref{C_01}), due to (\ref{E_00}), (\ref{E_3}), one has the 
reduced overlap function $\widehat{\cal V}_{[C/D]}(q)$, which for the products of 
arbitrary number of initial and final wave packets $\Psi^*_{[C/D]F}(y) \Psi_{[C/D]I}(y)$ 
allowes also to restrict the amplitude (\ref{C_02}) onto the mass shell as a 
corresponding pole contribution by dropping the $\theta(t)$- function in (\ref{C_02}), 
(\ref{C_04}).  For 
\bea
\fl  
\widehat{\cal V}_{[C/D]}(q)=\!\int\!d^4{y} \,e^{[\pm] i(qy)}\,
\Psi^*_{[C/D]F}(y) \Psi_{[C/D]I}(y),\; \mbox { with the off-shell} 
\label{C_05} \\
\fl  
 \mbox {wave function (\ref{C_04}) as: }\; 
\widehat{\cal F}^{(\xi)}_{[C/D]j}(\varrho)=
\!\int\!\!\frac{d^4 q}{(2\pi)^4}\,e^{[\mp]i\xi(q\varrho)}\,
\widehat{\cal H}^{(\xi)}_{[C/D]j}(q), 
\label{3_49} \\ 
\fl  
\mbox{and its invers: }\;  
\widehat{\cal H}^{(\xi)}_{[C/D]j}(q)=
\!\int\!d^4\varrho \,e^{[\pm]i\xi(q\varrho)}\,\widehat{\cal F}^{(\xi)}_{[C/D]j}(\varrho)=
\frac{(-i)\,\widehat{\cal V}_{[C/D]}(\xi q)}{2E_{{\rm q}j}(E_{{\rm q}j}-q^0-i0)}
\label{3_52}  \\ 
\fl  
\equiv \frac{{\cal H}^{(\xi)}_{[C/D]j}(q)}{2}+
\frac{\widehat{\cal V}_{[C/D]}(\xi q)}{2E_{{\rm q}j}}\,
{\rm PP}\frac{(-i)}{(E_{{\rm q}j}-q^0)},\;\;
\mbox{ reduces to the on-shell one as:}
\label{3_53} \\ 
\fl  
\widehat{\cal F}^{(\xi)}_{[C/D]j}(\varrho)\stackrel{\theta(t)\mapsto 1}{\longmapsto}
{\cal F}^{(\xi)}_{[C/D]j}(\varrho)=
\!\int\!\frac{d^4 q}{(2\pi)^3}\,e^{[\mp]i\xi(q\varrho)}\,\theta(q^0)\,
\delta(q^2-m^2_j)\,\widehat{\cal V}_{[C/D]}(\xi q), 
\label{3_53_0} \\ 
\fl  
\mbox {the amplitude (\ref{C_02}), (\ref{C_03}): }\; 
\widetilde{\cal A}^{j(\xi)}_{DC}=
\frac 1i \!\!\int\!\frac{d^4 q}{(2\pi)^4} 
\frac{\widehat{\cal V}_{\{D\}}(\xi q)\widehat{\cal V}_{\{C\}}(\xi q)}
{2E_{{\rm q}j}(E_{{\rm q}j}-q^0-i0)},\;\mbox{ transcribes}  
\label{3_54} \\  
\fl  
\mbox{as: }\;
\widetilde{\cal A}^{j(\xi)}_{DC}\Longleftrightarrow 
\!\int\!\frac{d^3q}{(2\pi)^3}\!\int\!\frac{dp^0}{2\pi}\!\int\!\frac{dq^0}{2\pi}
\,e^{i\xi\varrho^0(p^0-q^0)}(p^0+q^0)
\widehat{\cal H}^{(\xi)}_{\{D\}j}(p^0,\q)
\widehat{\cal H}^{(\xi)}_{\{C\}j}(q^0,\q), 
\label{3_55} \\  
\fl  
\mbox {but reduces as: }\;
\widetilde{\cal A}^{j(\xi)}_{DC} 
\stackrel{\theta(t)\mapsto 1}{\longmapsto}
\!\left[\widetilde{\cal A}^{j(\xi)}_{DC}\right]\!=
\!\int\!\frac{d^4q}{(2\pi)^3}\,\widehat{\cal V}_{\{D\}}(\xi q)\,
\theta(q^0)\delta(q^2-m^2_j)\widehat{\cal V}_{\{C\}}(\xi q).   
\label{C_06} 
\eea
(Here PP stays for Cauchy principal value.)
Indeed, in addition to the first pure on-shell contribution (\ref{3_20_1_G}), (\ref{3_53_0}) 
with (\ref{3_21_G}), (\ref{3_24}) or (\ref{C_05}), at $q^0=E_{{\rm q}j}=\sqrt{\q^2+m^2_j}$,  
the expression (\ref{3_53}) contains also the second off-shell contribution with the 
function (\ref{C_05}) or (\ref{3_21_G}) for $q^2\neq m^2_j$. 
On the other hand, the off-shell version of scalar product, given by the first equality 
(\ref{C_03}), corresponds exactly to the off-shell fractions (\ref{C_02}), (\ref{3_54}) 
of scalar amplitude (\ref{3_56_0}). 
The equivalence of ``Wick's and Dyson's prescription'' leads to the 
second relation (\ref{C_03}), which coincides with the on-shell scalar product of free 
packets (\ref{2_30}) and recasts these fractions to the form (\ref{3_55}). 
Its $\varrho^0$- independence implies contribution only from $p^0=q^0$. First of all 
that is the pole contribution of (\ref{3_52}) in the lower half-planes of $q^0, p^0$ 
with the residues containing both pieces of (\ref{3_53}). 
Convergence of both these integrals is provided by overlap function  
$\widehat{\cal V}_{[C/D]}(q)$ (\ref{C_05}) kind of (\ref{3_40_V}), (\ref{Pr_3_5}) in the 
general expressions (\ref{3_52}), (\ref{3_55}). 
This pole contribution is fully equivalent to simple on-shell replacement 
$\widehat{\cal H}^{(+)}_{[C/D]j}(q)\mapsto{\cal H}^{(+)}_{[C/D]j}(q)$, 
i.e. $\widehat{\cal F}^{(+)}_{[C/D]j}(\varrho)\mapsto {\cal F}^{(+)}_{[C/D]j}(\varrho)$, 
ensuing from (\ref{C_04}), (\ref{3_53_0}) with omitted temporal factor $\theta(x^0-y^0)$, 
that in turn is fully equivalent to the residue of amplitude (\ref{3_56_0}) in the lower 
half-plane ${\rm Im}\,q^2\leqslant 0$ for the case of  `pole integration' 
\cite{bth,ahm,fujii} in accord with (\ref{C_06}), as (\ref{3_56_0}) $\mapsto$ (\ref{3_54}) 
$\Leftrightarrow$ (\ref{3_55}) 
\bea
\fl 
\longmapsto 
\!\int\!\frac{d^4q}{(2\pi)^3}\,\widehat{\cal V}_{\{D\}}(q)\,
\theta(q^0)\,\delta(q^2-m^2_j)\,\widehat{\cal V}_{\{C\}}(q)=
\!\int\!d^3\!\rho\, {\cal F}^{(+)}_{\{D\}j}(\varrho)
(i\!\stackrel{\leftrightarrow}{\partial^{0}_\varrho})
{\cal F}^{(+)}_{\{C\}j}(\varrho)
\label{3_56} \\
\fl   
=\left[\widetilde{\cal A}^{j(+)}_{DC}\right]
=\!\int\!d^4 x\!\int\!d^4 y\,\Psi^*_{DF}(x)\Psi_{DI}(x) \frac 1i D^-_{m_j}(x-y)
\Psi^*_{CF}(y)\Psi_{CI}(y).  
\label{3_57} 
\eea
This explanation of equivalence of the `pole integration' for the Feynman amplitude 
(\ref{3_56_0}) to the similar pole approximation for the off-shell composite wave 
functions (\ref{3_49}), (\ref{3_52}), that transform them into respective free on-shell 
ones (\ref{3_20_1}), (\ref{3_20_G}), (\ref{3_20_1_G}), (\ref{3_53_0}), surely 
takes place also for the spinor case (\ref{3_14_0}), (\ref{3_14}): 
$\widehat{\Upsilon}_{\{C\}j}(\varrho)\mapsto {\Upsilon}_{\{C\}j}(\varrho)$, as  
(\ref{3_18}), (\ref{3_19}) $\mapsto$ (\ref{3_20_0}). 
Moreover the self-consistency of both these pole approximations conforms with 
Lorentz invariance, leading automatically to independence of $\varrho^0$ for amplitude 
(\ref{C_03}), (\ref{3_55}). This simplifies also further asymptotic analysis for 
off-shell composite wave functions (\ref{C_04}), (\ref{3_49}), (\ref{3_52}) with fixed 
$\varrho^\lambda$. 
For $q^\lambda=(q^0,\q)$, $\q={\rm q}\vec{\omega}$, ${\cal K}_j(q^0)=(q^2_0-m^2_j)^{1/2}$, 
$E_{{\rm q}j}\equiv E_j({\rm q})=(\q^2+m^2_j)^{1/2}$, with the main branches of square 
roots, for $[C/D]\leftrightarrow \pm$, where 
$\mp i(q\varrho)=\mp iq_0\varrho_0\pm i(\q\cdot\vec{\rho})$, 
$\rho=|\vec{\rho}|$, $\vec{\rho}=\rho\,\n_\rho$, that is: 
\bea
\fl  
\widehat{\cal F}^{(+)}_{[C/D]j}(\varrho)=
\!\int\!\!\frac{d^4 q}{(2\pi)^4}\frac{(-i)\,e^{\mp i(q\varrho)}
\widehat{\cal V}_{[C/D]}(q)}{2E_{{\rm q}j}(E_{{\rm q}j}-q^0-i0)}
=\!\int\!\!\frac{d^4 q}{(2\pi)^4i}\frac{e^{\mp i(q\varrho)}\;
\widehat{\cal V}_{[C/D]}(q)}{(m^2_j-q^2-i0)}\,
\frac 12\left[1+\frac{q_0}{E_{{\rm q}j}}\right]\!. 
\label{5_00} 
\eea
Separating the rapidly oscillating exponential functions of $q$ in (\ref{3_40_V}) 
or in (\ref{Pr_3_5}), for $k_{\{\}}=k_{\{C/D\}}$, $Y_{\{\}}=Y_{\{C/D\}}$, 
$(\varrho-Y_{\{\}})^\lambda=R^\lambda_{\{\}}=(R^0,\vec{\rm R})=R^\lambda$,    
$\vec{\rm R} = {\rm R}\n \equiv {\rm R}_{\{\}}\n_{\{\}}$, we rewrite (\ref{5_00}) 
and then calculate its asymptotic behavior as:  
\bea
\fl  
\widehat{\cal V}_{[C/D]}(q)=e^{\pm i(qY_{\{\}})}\,\widehat{\Psi}_{[C/D]}(q), 
\;\;\mbox{ where:}
\label{5_00_01} \\ 
\fl  
\widehat{\Psi}_{[C/D]}(q)\approx 
(2\pi)^4\,\theta(\Delta)\,e^{\mp i(k_{\{\}}Y_{\{\}})}
\,\widehat{\delta}_{[C/D]}(q-k_{\{\}})\,e^{\Sigma_k(Y^{\{\}}_-)}, \;\;
\mbox{ whence:}
\label{5_001} \\
\fl  
\widehat{\cal F}^{(+)}_{[C/D]j}(\varrho)=
\!\int\!\!\frac{d^4 q}{(2\pi)^4i}\frac{e^{\mp i(qR)}\,
\widehat{\Psi}_{[C/D]}(q)}{2E_{{\rm q}j}(E_{{\rm q}j}-q^0-i0)}=
\!\int\!\!\frac{d^4 q}{(2\pi)^4i}\frac{e^{\mp i(qR)}\widehat{\Psi}_{[C/D]}(q)}
{(\q^2-{\cal K}^2_j-i0)}\frac 12\!\left[1+\frac{q_0}{E_{{\rm q}j}}\right]\!, 
\label{5_002} \\
\fl  
\stackunder{{\rm R}\to\infty}{\longmapsto}\int\!\frac{dq_0\,e^{\mp iq_0R_0}}
{2i(2\pi)^2}\,\theta(q_0)\,\theta({\cal K}^2_j)\,\frac{e^{i{\cal K}_j{\rm R}}}{{\rm R}}\,
\widehat{\Psi}_{[C/D]}(q_0;\pm\n{\cal K}_j),
\;\mbox{ for }\;
\theta(q_0)=\frac 12\left[1+\frac{q_0}{|q_0|}\right]\!, 
\label{5_01} \\
\fl  
\mbox{or, that is: }\;\,
\stackrel{\pm R_0=T}{\stackunder{T\to\infty}{\longmapsto}}
\int\!\frac{d^3{\rm q}}{(2\pi)^3}\,\frac{e^{-iTE_{{\rm q}j}}}{2E_{{\rm q}j}}\,
e^{\pm i(\q\cdot\vec{\rm R})}\widehat{\Psi}_{[C/D]}(E_{{\rm q}j};\q)=
{\cal F}^{(+)}_{[C/D]j}(\varrho),
\label{5_0_02} \\
\fl  
\mbox{and then: }\;\,
\stackrel{{\rm R}\to\infty}{\stackunder{T\to\infty}{\longmapsto}}
\!\int\limits^\infty_{m_j}\!\frac{dq_0\;e^{-iq_0T}}{2i(2\pi)^2}\,
\frac{e^{i{\cal K}_j{\rm R}}}{{\rm R}}\,
\widehat{\Psi}_{[C/D]}(q_0;\pm\n{\cal K}_j). 
\label{5_02} 
\eea
The asymptotic expression (\ref{5_01}) follows at ${\rm R}\to\infty$ 
$(|\vec{\rm Y}_{\{\}}|\to\infty)$ from the second equality (\ref{5_002}) by using the 
Grimus-Stockinger theorem \cite{GS}. The first $\theta$- function comes here from the 
square brackets, while the second $\theta$- function comes from the conditions of theorem. 
The asymptotic expression (\ref{5_0_02}) appeares from the first equality (\ref{5_002}) 
due to Jacob-Sachs theorem at $\pm R^0=T\to\infty$ \cite{bth}. Then its final form 
(\ref{5_02}) is obtained at ${\rm R}\to\infty$ by using the well known asymptotics of 
plane wave (\ref{B_as}), integration over solid angle in 
$d^3{\rm q}={\rm q}^2d{\rm q}d\Omega(\vec{\omega})$ and change of variable 
$E_j({\cal K}_j(q^0))=q^0$. The above theorems are adduced in 
(\ref{E_5})--(\ref{E_6}).
Contribution of converging spherical wave (\ref{B_as}) is omitted in (\ref{5_02}) 
due to faster oscillations of the total exponential for this term and since it furnishes 
the spatial vector $\q$ of $q^\lambda=(q^0,\q)$ with wrong direction $\mp\n_{\{\}}$. 
The momentum $q$ is globally defined for amplitudes of macrodiagram (\ref{3_60_0}), 
(\ref{3_56_0}), (\ref{3_54}), (\ref{3_55}) and for composite wave functions (\ref{3_21}), 
(\ref{Pr_3_5}), (\ref{C_05})--(\ref{3_53_0}) of both $\{C\}$ and $\{D\}$ vertices. 
For the $(+)$ case the momentum $q$ leaks from point $Y_{\{C\}}$ to $Y_{\{D\}}$, when    
$-\infty \leftarrow Y^0_{\{C\}}\ll \varrho^0\ll Y^0_{\{D\}}\to+\infty$. 
Coincidence of expressions (\ref{5_01}) and (\ref{5_02}) means, that 
${\rm R}_{\{\}}\to+\infty$ for $\pm\vec{\rm R}_{\{C/D\}}=
\pm (\vec{\rho}-\vec{\rm Y}_{\{C/D\}})=
\pm {\rm R}_{\{C/D\}}\n_{\{C/D\}}\simeq {\rm R}_{\{\}} \n_{\{C\}}$, already implies 
$\pm R^0_{\{C/D\}}=T\to+\infty$. On the one hand, the expression (\ref{5_0_02})  
is nothing but on-shell composite wave function 
(\ref{3_20_1_G}), (\ref{3_53_0}). So, the further difference between the off-shell 
$\widehat{\cal F}^{(+)}_{[C/D]j}(\varrho)$ and the on-shell composite wave functions 
${\cal F}^{(+)}_{\{C/D\}j}(\varrho)$ fully disappears in this asymptotical regime. 
On the other hand, the $j$- dependence of asymptotical integrand for these composite wave 
functions may be changed by using above simple change of variable: 
\bea
\fl  
\!\int\limits^\infty_{0}\!\!\frac{{\rm q} d{\rm q}}{(2\pi)^2i}\,
\frac{e^{-iTE_{{\rm q}j}}}{2E_{{\rm q}j}}
\frac{e^{i{\rm q}{\rm R}}}{{\rm R}}\,
\widehat{\Psi}_{[C/D]}(E_{{\rm q}j};\pm\n{\rm q})=
\int\limits^\infty_{m_j}\!\!\frac{dq_0\;e^{-iq_0T}}{2i(2\pi)^2}\,
\frac{e^{i{\cal K}_j{\rm R}}}{{\rm R}}\,
\widehat{\Psi}_{[C/D]}(q_0;\pm\n{\cal K}_j)  
\label{5_03} \\
\fl  
=m_j\!\!\int\limits^\infty_{0}\!\!\frac{d\beta \sinh\beta\, e^{-iT m_j\cosh\beta}}
{2i(2\pi)^2}\,\frac{e^{i{\rm R} m_j\sinh\beta}}{{\rm R}}\,
\widehat{\Psi}_{[C/D]}(m_j\cosh\beta;\pm\n_{\{C/D\}} m_j\sinh\beta),  
\label{5_04} 
\eea 
where $\n_{\{C\}}\mapsto \n_\rho$ only if coordinate system origin is placed to the 
point $\vec{\rm Y}_{\{C\}}$. 
These expressions show that the often discussed difference between both ``equal energy'' 
and ``equal momentum'' scenarios \cite{bth,ahm,Dl_1,Nis,kbz,GS} may be considered 
on the same footing and in fact is very conditionally. An accounting in 
(\ref{5_02}) the above omitted contribution of converging spherical wave (\ref{B_as}) 
replaces the first expression (\ref{5_03}) to the following one (see \cite{LL} \S 130): 
\bea
&&\!\!\!\!\!\!\!\!\!\!\!\!\!\!\!\!\!\!\!\!
\widehat{\cal F}^{(+)}_{[C/D]j}(\varrho) 
\stackrel{{\rm R}\to\infty}{\stackunder{T\to\infty}{\longmapsto}}
\!\int\limits^\infty_{-\infty}\!\!\frac{{\rm q} d{\rm q}}{(2\pi)^2i}\,
\frac{e^{-iTE_{{\rm q}j}}}{2E_{{\rm q}j}}
\frac{e^{i{\rm q}{\rm R}}}{{\rm R}}\,
\widehat{\Psi}_{[C/D]}(E_{{\rm q}j};\pm {\rm q}\n_{\{C/D\}}),
\label{5_06} 
\eea
where one can again put ${\rm q}=m_j\sinh\beta$. Expression (\ref{5_03}) implies only 
contribution of positive saddle point for the further approximation of function 
$\widehat{\Psi}_{[C/D]}(q^0;\pm\n{\rm q})$, with $q^0_j, {\rm q}_j>0$ near 
${\rm q}\simeq {\rm k}_{\{\}}$, $q^0\simeq k^0_{\{\}}$ with additional assumption about 
its ultrarelativistic position ${\rm q}_j,q^0_j\gg m_j$.
Unlike (\ref{5_03}), the dominant contribution of saddle point for   (\ref{5_06}) 
is not restricted by these assumptions and may be estimated here without them. 
Nevertheless we will adopt these assumptions below in order to compare different ways 
of estimations and to see the correspondence with previous works. 
According to the obtained above by different ways the same structure (\ref{3_40_V}), 
(\ref{Pr_3_5}) of reduced overlap function with the narrow wave packets of external particles, 
we adopt the approximation (\ref{5_00_01}), (\ref{5_001}) for general case.
Below in this subsection we consider corresponding saddle-point estimations for the wave 
function and amplitude following the spirit of the works \cite{nn,nsh,dn,n_sh,NN_shk,NN_anp}.  

To this end we unify and simplify our notations. We omit subscript $\{\}$ where it is 
possible and define 4-vectors and 4-tensors: 
$\ell^\lambda=\ell^\lambda_{\{\}}=(1,\vec{\rm l})$, or 
$\overline{\ell}^\lambda=(1,\vec{l})$, $\ell^2=\overline{\ell}^2=0$, 
$\vec{\rm l}^2=\vec{l}^2=1$, for $\vec{\rm l}\equiv\vec{\rm l}_{\{\}}=\pm\n_{\{\}}$, or    
for $\vec{\rm X}_D-\vec{\rm Y}_C=\vec{\rm L}=L\vec{l}$; with 
$H^\lambda_{\{\}}=(k\vec{\cal T})^\lambda_{\{\}}=(H^0,\vec{\rm H})$,  
$U^\lambda_{\{\}}=(\ell\vec{\cal T})^\lambda_{\{\}}=(U^0,\vec{\rm U})$, or  
$\overline{H}^\lambda=H^\lambda_{D}+H^\lambda_{C}$,
$\overline{\vec{\cal T}}\equiv\vec{\cal T}_{DC}=\vec{\cal T}_{D}+\vec{\cal T}_{C}$, 
$\overline{U}^\lambda=(\overline{\ell}\overline{\vec{\cal T}})^\lambda$;  
${\cal Q}^\lambda={\rm Q}\ell^\lambda$, or 
$\overline{\cal Q}^\lambda=\overline{\rm Q}\overline{\ell}^\lambda$;
${\cal Q}^2=\overline{\cal Q}^2=0$, ${\rm Q}_{\{\}}=(H\ell)/(U\ell)$, or 
$\overline{\rm Q}=(\overline{H}\overline{\ell})/(\overline{U}\overline{\ell})$. 
Besides this we define:  
$q^\lambda_j=(q^0_j,{\rm q}_j\vec{\rm l})_{\{\}}={\cal Q}^\lambda+w^\lambda_j$, or  
$\overline{q}^\lambda_j=(\overline{q}^0_j,\overline{\rm q}_j\vec{l})=
\overline{\cal Q}^\lambda+\overline{w}^\lambda_j$, $q^2_j=\overline{q}^2_j=m^2_j$. 
Up to next to the leading order $O(m^4_j)$, these definitions mean that 
$(\ell w_j)=m^2_j/(2{\rm Q})$, but 
$(\overline{\ell}\overline{w}_j)=m^2_j/(2\overline{\rm Q})$. 
The following expansions for 
$q^0=q^0_j+(q^0-q^0_j)$, ${\rm q}={\rm q}_j+({\rm q}-{\rm q}_j)$ are also useful: 
\bea
\fl  
{\cal K}_j(q^0)\approx 
{\rm q}_j+\frac 1{{\rm v}_j}(q^0-q^0_j)-\frac{m^2_j}{2{\rm q}^3_j}(q^0-q^0_j)^2, \quad  
{\rm q}_j={\cal K}_j(q^0_j), \quad 
{\rm v}_j=\frac{{\rm q}_j}{q^0_j}, \quad 
\overline{\rm v}_j=\frac{\overline{\rm q}_j}{\overline{q}^0_j},
\label{5_08}  \\
\fl  
E_j({\rm q})\approx 
q^0_j+{\rm v}_j({\rm q}-{\rm q}_j)+\frac{m^2_j}{2(q^0_j)^3}({\rm q}-{\rm q}_j)^2, 
\quad \; q^0_j=E_j({\rm q}_j), \;\mbox{ and so on}. 
\label{5_07} 
\eea

\subsubsection{Asymptotic of composite wave function.}\label{sec:wf_asymp}

For $q^\lambda=(q^0,{\rm q}\vec{\rm l}_{\{\}})$ the quadratic form of 
$\widehat{\delta}_{\{\}}(q-k_{\{\}})$ in   (\ref{3_40_V}), (\ref{5_001}), (\ref{5_03}) 
reads: 
\bea
\fl  
((q-k)\vec{\cal T}(q-k))_{\{\}}=(q\vec{\cal T}q)-2(qH)+(k\vec{\cal T}k)\equiv 
{\rm f}_{\{\}}(q^0,{\rm q}), \;\mbox{ where:}
\label{5_09} \\
\fl  
{\rm f}_{\{\}}(q^0,{\rm q})=q^2_0{\cal T}^{00}+
{\rm q}^2(\vec{\rm l}\widehat{\vec{\rm T}}\vec{\rm l})-
2q^0{\rm q}(\underline{\vec{\rm T}}\cdot\vec{\rm l})-
2[q^0H^0-{\rm q}(\vec{\rm H}\cdot\vec{\rm l})]+{\rm f}_{\{\}}(0,0), 
\label{5_10} \\
\fl  
\vec{\cal T}=
\!\left(\!\!\begin{array}{cc}
{\cal T}^{00} & \underline{\vec{\rm T}}^{\top}  \\
\underline{\vec{\rm T}}  & \widehat{\vec{\rm T}}   
\end{array}\!\! \right) \equiv {\cal T}^{\beta\lambda}= 
\!\left(\!\!\begin{array}{cc}
{\cal T}^{00} & {\cal T}^{0r}  \\
{\cal T}^{s0}  & {\cal T}^{sr}   
\end{array}\!\! \right),\;\; s,r=1,2,3; \quad 
{\rm f}_{\{\}}(0,0)\equiv (k\vec{\cal T}k),   
\label{5_11} \\
\fl  
\mbox{whence: }\, {\rm F}_j({\rm q})={\rm f}_{\{\}}(E_j({\rm q}),{\rm q}), 
\;\mbox{ while }\; \widetilde{\rm F}_j(q^0)={\rm f}_{\{\}}(q^0,{\cal K}_j(q^0)), 
\label{5_12} 
\eea
respectively for the first and second expression (\ref{5_03}). The extremum condition 
of zero first derivatives of these functions: ${\rm F}^\prime_j({\rm q})=0$ and 
$\widetilde{\rm F}^\prime_j(q^0)=0$, gives the same equation for saddle point 
$q^\lambda_j$, similar to \cite{nn,NN_anp}, but it gives the different expressions for  
second derivatives and different complex ``effective widths'' ${\cal D}_j$ of Gaussins 
integrals arised by the use of expansions\footnote{That are {\sl not} the expansions 
on power of $m_j$.} (\ref{5_08}), (\ref{5_07}) for powers of exponentials of 
(\ref{5_03}), (\ref{5_06}): 
\bea
\fl  
q^0\equiv\frac{m_j}{\sqrt{1-{\rm v}^2}}=\frac{{\rm v}H^0-(\vec{\rm H}\cdot\vec{\rm l})}
{{\rm v}(U\ell)-(1-{\rm v})^2(\underline{\vec{\rm T}}\cdot\vec{\rm l})},\quad \; 
{\rm v}\equiv \frac {\rm q}{q^0}=\frac{dq^0}{d{\rm q}}, \quad {\rm v}\mapsto {\rm v}_j, 
\quad {\rm q}\mapsto {\rm q}_j, 
\label{5_13} \\
\fl  
\frac 1{{\cal D}^2_j}\equiv 
{\rm F}^{\prime\prime}_j({\rm q}_j)+iT \frac{m^2_j}{(q^0_j)^3},\;\;\,
\frac 1{{\rm v}^2_j\widetilde{\cal D}^2_j}\equiv 
\widetilde{\rm F}^{\prime\prime}_j(q^0_j)+i{\rm R}\frac{m^2_j}{{\rm q}^3_j}, \;\;\,
\frac 1{\widetilde{\cal D}^2_j}={\rm F}^{\prime\prime}_j({\rm q}_j)+
i\frac{{\rm R}\; m^2_j}{{\rm v}_j(q^0_j)^3}\biggr|_{{\rm F}^\prime_j=0}\!,   
\label{5_14} \\
\fl  
\int\limits^\infty_{-\infty}\!\!\frac{{\rm q} d{\rm q}}{2E_j({\rm q})}\,
\frac{e^{i{\rm R}{\rm q}-iTE_j({\rm q})}}{{\rm R}}\,e^{-{\rm F}_j({\rm q})}\approx 
\frac{{\rm v}_j}{2{\rm R}}e^{i{\rm R}{\rm q}_j-iTq^0_j}
\,e^{-{\rm F}_j({\rm q}_j)}\,\sqrt{2\pi{\cal D}^2_j}\,  
e^{-({\cal D}^2_j/2)({\rm R}-T{\rm v}_j)^2}. 
\label{5_15} 
\eea
The second representation (\ref{5_03}) leads to the same expression exactly with  
replacement ${\cal D}^2_j\mapsto \widetilde{\cal D}^2_j$. Since 
${\rm q}={\rm v} q^0$ but $dq^0={\rm v} d{\rm q}$, both complex effective widths 
(\ref{5_14}) will coincide only on the neutrino classical trajectory after substitution 
of ${\rm R}\mapsto T{\rm v}_j$. Taking into account for above defined 
$R^\lambda_{\{\}}$ and $q^\lambda_{j\{\}}$ the Lorentz invariance of: 
\bea
\fl  
i\left[{\rm q}_j{\rm R}-q^0_j T\right]_{\{\}}=-i(q_j R)_{\{\}}, \quad 
(q^0_j{\rm R}-{\rm q}_j T)^2_{\{\}}=\left[(q_j R)^2-m^2_j R^2\right]_{\{\}},
\label{5_16} \\
\fl  
{\rm F}_j({\rm q}_j)=\widetilde{\rm F}_j(q^0_j)={\rm f}_{\{\}}(q^0_j,{\rm q}_j)=
((q_j-k)\vec{\cal T}(q_j-k))_{\{\}}, \;\mbox { with:}
\label{5_17} \\
\fl  
\Omega_{j\{\}}(R)=i(q_j R)_{\{\}}+
\frac{{\cal D}^2_j}{2(q^0_j)^2}\left[(q_j R)^2-m^2_j R^2\right]_{\{\}}=
i(q_j R)+\frac{m^2_j}{2\tau_j}\left[(u_j R)^2-R^2\right], 
\label{5_18} \\
\fl  
\mbox{for: }\;u_j=\frac{q_j}{m_j}, \quad \tau_j=\frac{(q^0_j)^2}{{\cal D}^2_j}=
(q^0_j)^2{\rm F}^{\prime\prime}_j({\rm q}_j)+iT \frac{m^2_j}{q^0_j},\;\mbox{ or with }\;
T\longrightarrow \frac{\rm R}{{\rm v}_j}, 
\label{5_19}
\eea
one obtains an ``almost'' Lorentz covariant asymptotics 
(or with $\Sigma_k(Y^{\{\}}_-)\mapsto\Sigma_{\{C/D\}}$): 
\bea
\fl  
\widehat{\cal F}^{(+)}_{[C/D]j}(\varrho)\, 
\stackrel{{\rm R}\to\infty}{\stackunder{T\to\infty}{\longmapsto}}\,
(-i)(2\pi)^{5/2}\,e^{\mp i(k_{\{\}}Y_{\{\}})}\,e^{\Sigma_k(Y^{\{\}}_-)}\,
\widehat{\delta}_{\{\}}(q_j-k_{\{\}})\,\frac{{\rm v}_j}{2{\rm R}}\,{\cal D}_j \,
e^{-\Omega_{j\{\}}(R) }. 
\label{5_20}
\eea
Equation (\ref{5_20}) confirms the intuitive reasons leading to similar asymptotic form,  
qualitatively intrduced in \cite{nn,nsh} for intermediate neutrino wave function. 
Here the function and its asymptotic is strictly defined in QFT framework by parameters 
of external wave packets only at one vertex $\{C\}$ or $\{D\}$ as it should, and has 
necessary representation variable $\varrho$ as the off-shell one (\ref{3_14_0}), 
(\ref{3_18}), (\ref{3_20_0}), 
(\ref{3_20_G}) with (\ref{C_04})--(\ref{3_52}), (\ref{5_00})--(\ref{5_002}), or as the 
on-shell one with (\ref{3_20_1_G}) $\equiv$ (\ref{5_0_02}).  

It is worth to note that the expressions (\ref{5_08})--(\ref{5_20}) 
fulfil without any assumption about value of $m_j$, but leads to ambiguity (\ref{5_14}), 
(\ref{5_19}) of imaginary part of effective width (\ref{5_14}), which besides this, 
does not have consistent physical interpretation \cite{nn} of Gaussian oscillations thus 
appearing. 
Assumption of small $m_j$ allowes to avoid this ambiguity and leads to unique fully 
invariant expression for effective width. Indeed, from   (\ref{5_09}) -- (\ref{5_14}), 
(\ref{3_39}) -- (\ref{3_43}), 
for $(k\vec{\cal T}k)=g_+k^2{\rm T}_0/4\mapsto{\rm K}^2(\ell{\cal T}\ell)$, 
with $q^\lambda_j={\cal Q}^\lambda+w^\lambda_j$, $(U\ell)\equiv(\ell{\cal T}\ell)>0$, 
$U^2<0$, $H^2<0$ \cite{kt_3}, up to the next to the leading order $O(m^4_j)$ it 
follows, that for:  
\bea
\fl  
{\rm F}_j({\rm q})\approx 
{\rm f}_{\{\}}(0,0)-{\rm Q}(H\ell)+(U\ell)\left({\rm q}-{\rm Q}\right)^2+
m^2_j\left[U^0-\frac{H^0}{\rm q}\right], \quad 
{\rm r}_j\equiv \frac{m^2_j}{2{\rm Q}^2}\ll 1, 
\label{5_21} \\
\fl  
\mbox{one has: }\; 
{\rm v}_j\approx 1-{\rm r}_j, 
\quad w^\lambda_j\approx
-\frac{{\rm r}_j}{(U\ell)}\left((\vec{\rm H}\cdot\vec{\rm l}),H^0\vec{\rm l}\right),
\quad (Hw_j)=0, \quad \tau_j \simeq \tau_{\{\}},
\label{5_22} \\
\fl  
\frac 1{{\cal D}^2_j}\approx 
2(U\ell)\left\{1-\frac{{\rm r}_j}{(H\ell)}[2H^0-iT_{\{\}}]\right\}\simeq 2(U\ell)>0, 
\quad \tau_{\{\}}=2{\rm Q}(H\ell)\equiv 2\frac{(H\ell)^2}{(U\ell)}, 
\label{5_23} 
\eea
where only the main contribution is saved. This also allowes to avoid some inconsistency 
of the used saddle-point calculation \cite{nn}, because   (\ref{5_14}) implies an 
``exact'' saddle point ${\rm q}_j$, which in fact is defined by zero derivative of entire 
power of exponential of integrand in (\ref{5_15}): 
${\rm F}^\prime_j({\rm q})+i(T{\rm v}({\rm q})-{\rm R})=0$. Whence ${\rm q}_j$, 
acquires an inadmissible imaginary part, which is generated by value 
$i(T{\rm v}({\rm q})-{\rm R})$. 
To extract the $j$- dependence, the following relation, ensuing with the same accuracy 
from (\ref{5_17}), (\ref{5_21}, (\ref{5_22}), is observed: 
\bea
\fl 
{\rm F}_{j\{\}}({\rm q}_j)=((q_j-k)\vec{\cal T}(q_j-k))_{\{\}}=
(({\cal Q}-k)\vec{\cal T}({\cal Q}-k))_{\{\}}+\Theta_{j\{\}}, \;\mbox { where:}
\label{5_24} \\
\fl 
\Theta_{j\{\}}\approx m^2_j\left[U^0-\frac{H^0}{\rm Q}\right]=
\frac{m^2_j}{(H\ell)}\left[H^0(\vec{\rm U}\cdot\vec{\rm l})-
U^0(\vec{\rm H}\cdot\vec{\rm l})\right], \;\mbox { whence:}
\label{5_25} \\
\fl 
\widehat{\cal F}^{(+)}_{[C/D]j}(\varrho)\, 
\stackrel{{\rm R}\to\infty}{\stackunder{T\to\infty}{\longmapsto}}\,
(-i)(2\pi)^{5/2}\,e^{\mp i(k_{\{\}}Y_{\{\}})}\,e^{\Sigma_k(Y^{\{\}}_-)}\,
\widehat{\delta}_{\{\}}({\cal Q}-k_{\{\}})\,\frac{{\rm v}_j}{2{\rm R}}\,{\cal D}_j \,
e^{-\Omega_{j\{\}}(R)-\Theta_{j\{\}} }. 
\label{5_26} 
\eea
Similarly \cite{nn} the value of $\Theta_{j\{\}}$ is negligible due to smeared 
delta-function of approximate energy-momentum conservation, keeping 
$k^\lambda\simeq {\rm K}\ell^\lambda$, $(H\ell)\simeq {\rm K}(U\ell)$, 
when $\Theta_{j\{\}}=0$.  

This result should be compared with evolution of neutrinos emitted by 
classical sources, given by (11.19) of \cite{Dvor}. We observe that the 
spherical wave $e^{-i(q_j R)}/{\rm R}$ given therein, now is modulated by the same 
damping Gaussian factor (\ref{5_15}), (\ref{5_20}), that appears in CRG narrow-packet 
approximation for one-packet state (\ref{2_15_41}) with $\underline{x} \mapsto -R_{\{\}}$: 
\bea
\exp\left\{-\frac{{\cal D}^2_j}{2}({\rm R}-T{\rm v}_j)^2_{\{\}}\right\}=
\exp\left\{-\frac{m^2_j}{2\tau_j}\left[(u_j R)^2-R^2\right]_{\{\}}\right\}, 
\label {5_26_0}
\eea 
and thus define via $\tau_j$ (\ref{5_23}) the corresponding ``invariant width''  for this 
effective intermediate neutrino wave-packet. This factor is equal to one only on the 
neutrino classical trajectory for    
${\rm R}=T{\rm v}_j$ with ${\rm R}_{\{C/D\}}=|\vec{\rho}-\vec{\rm Y}_{\{C/D\}}|$, 
$T_{\{C/D\}}=\pm(\varrho^0-Y^0_{\{C/D\}})$.

\subsubsection{Asymptotic of oscillation amplitude.}\label{sec:amp_asymp} 

Now let us show how the definitions of scalar products (\ref{C_03}), (\ref{3_54}), 
(\ref{3_55}) for the off shell and (\ref{C_06}), (\ref{3_57}) for the on shell 
composite wave functions reproduce the asymptotic of oscillation amplitude \cite{nn}. 
Unfortunately the time derivative $\partial_\varrho^{0}$ and $d^3\rho$- 
integration in (\ref{C_03}) is not uniform with respect to asymptotical 
expansions over $T$ and ${\rm R}$, and substitution of the results (\ref{5_20}), 
(\ref{5_26}) into (\ref{C_03}) is not correct. 
Starting from substitution of (\ref{3_52}), (\ref{5_00_01}) into (\ref{3_55}), in view of 
(\ref{5_001}), we perform at first the integration over $p^0$ at the limit $Y^0_D\to+\infty$, 
i.e. $T_D\to+\infty$ with the help of Jacob-Sachs (JS) theorem (\ref{E_6}). 
Thus we arrive to very similar to (\ref{5_002}) $d^4q$- integral, which may be estimated by 
the same two ways (\ref{5_01}), (\ref{5_02}) as above. The use for example again of JS theorem 
for $q^0$- integration at $Y^0_C\to -\infty$, i.e. $T_C\to+\infty$, with $T=T_D+T_C$, 
$R^\lambda=(T,\vec{\rm L})$, $\vec{\rm L}=L\vec{l}$, 
$\overline{\ell}^\lambda=(1,\vec{l})$, gives: 
\bea
&&\!\!\!\!\!\!\!\!\!\!\!\!\!\!\!\!\!\!\!\!
\widetilde{\cal A}^{j(+)}_{DC}\;
\stackrel{(Y^0_C\to-\infty)}{\stackunder{(Y^0_D\to +\infty)}{\longmapsto}} 
\int\!\!\frac{d^3{\rm q}}{(2\pi)^3}\,\frac{e^{-iE_{{\rm q}j}T}}{2E_{{\rm q}j}}\,
e^{i(\q\cdot\vec{\rm L})}\widehat{\Psi}_{[D]}(E_{{\rm q}j};\q)
\widehat{\Psi}_{[C]}(E_{{\rm q}j};\q)=\left[\widetilde{\cal A}^{j(+)}_{DC}\right], 
\label{5_27} 
\eea
which is equal to direct substitution of $\widehat{\cal V}_{[C/D]}(q)$ (\ref{5_00_01}) 
into the on-shell expressions (\ref{C_06}), (\ref{3_56}). Similarly (\ref{5_03}), 
(\ref{5_06}), the use again of the plane-wave asymptotic (\ref{B_as}) leads to the same 
integral for $\left[\widetilde{\cal A}^{j(+)}_{DC}\right]$ as well as for the amplitude: 
\bea
&&\!\!\!\!\!\!\!\!\!\!\!\!\!\!\!\!\!\!\!\!
\widetilde{\cal A}^{j(+)}_{DC}\;
\stackrel{L\to\infty}{\stackunder{T\to\infty}{\longmapsto}}\,
\frac 1i \!\int\limits^\infty_{-\infty\,[0]}\!\!\frac{{\rm q} d{\rm q}}{(2\pi)^2}\,
\frac{e^{-iTE_{j}({\rm q})} }{2E_{j}({\rm q})}
\frac{e^{i{\rm q}L } }{L}\,
\widehat{\Psi}_{[D]}(E_{j}({\rm q});{\rm q}\vec{l})
\widehat{\Psi}_{[C]}(E_{j}({\rm q});{\rm q}\vec{l}),
\label{5_28} 
\eea
whose saddle-point estimation with substitution of (\ref{5_001}) repeats all the above 
steps with simple replacing of all defining above values onto the same but with the bar:   
$H\mapsto\overline{H}$, $U\mapsto\overline{U}$, ${\cal Q}\mapsto\overline{\cal Q}$, 
${\rm Q}\mapsto\overline{\rm Q}$, $\vec{\cal T}\mapsto\overline{\vec{\cal T}}$, 
$q^\lambda_j\mapsto\overline{q}^\lambda_j$, $w_j\mapsto\overline{w}_j$, 
$\vec{\rm l}\mapsto\vec{l}$. Thus, with $q^\lambda=(q^0,{\rm q}\vec{l})$ one has for 
quadratic form  
$((q-k_C)\vec{\cal T}_C(q-k_C))+((q-k_D)\vec{\cal T}_D(q-k_D)\equiv  
\overline{\rm f}(q^0,{\rm q})$: 
\bea
\fl  
\overline{\rm F}_j({\rm q})=\overline{\rm f}(E_j({\rm q}),{\rm q})=
{\rm F}_{jC}({\rm q})+{\rm F}_{jD}({\rm q}), \quad 
\overline{\rm F}^\prime_j({\rm q})=0,\;\mbox{ so: }\; 
{\rm q}\mapsto\overline{\rm q}_j, \; \mbox{ with}
\label{5_29} \\
\fl  
q^0\equiv\frac{m_j}{\sqrt{1-\overline{\rm v}^2} }=
\frac{\overline{\rm v}\overline{H}^0-(\overline{\vec{\rm H}}\cdot\vec{l})}
{\overline{\rm v}(\overline{U}\overline{\ell})-(1-\overline{\rm v})^2
(\underline{\overline{\vec{\rm T}}}\cdot\vec{l})},\quad  
\overline{\rm v}\equiv \frac {{\rm q}}{q^0}\mapsto\overline{\rm v}_j,
\quad \overline{u}_j=\frac{\overline{q}_j}{m_j}, 
\label{5_30} \\
\fl  
\frac 1{\overline{\cal D}^2_j}\equiv 
\overline{\rm F}^{\prime\prime}_j(\overline{\rm q}_j)+
iT \frac{m^2_j}{(\overline{q}^0_j)^3},  \quad 
\overline{\tau}_j=\frac{(\overline{q}^0_j)^2}{\overline{\cal D}^2_j},\quad 
\overline{\Omega}_{j}(R)=i(\overline{q}_j R)+
\frac{m^2_j}{2\overline{\tau}_j}\left[(\overline{u}_j R)^2-R^2\right]\,:  
\label{5_31} \\ 
\fl  
\widetilde{\cal A}^{j(+)}_{DC}\,
\stackrel{L\to\infty}{\stackunder{T\to\infty}{\longmapsto}}\,
\frac{(2\pi)^6}{i}\,e^{i\Theta_{DC}}\,e^{\overline{\Sigma}_{DC}}\,
\widehat{\delta}_{\{D\}}(\overline{q}_j-k_D)\,
\widehat{\delta}_{\{C\}}(\overline{q}_j-k_C)\,\sqrt{2\pi}\,
\frac{\overline{\rm v}_j}{2L}\,\overline{\cal D}_j\,e^{-\overline{\Omega}_j(R)}, 
\label{5_32} \\
\fl  
\mbox{where: }\,\Theta_{DC}=(k_DX_D)-(k_CY_C), \quad 
\overline{\Sigma}_{DC}=\Sigma_{k_D}(X^D_-)+\Sigma_{k_C}(Y^C_-)\mapsto 
\Sigma_{D}+\Sigma_{C},   
\label{5_33}
\eea
for the case (\ref{Pr_3_5_1}).
This result is exactly the scalar part of (39) in \cite{nn}. Of course it may be 
simplified as above for small $m_j$ to the form given in \cite{NN_anp} for 
$\overline{q}^\lambda_j=(\overline{q}^0_j,\overline{\rm q}_j\vec{l})=
\overline{\cal Q}^\lambda+\overline{w}^\lambda_j$: 
\bea
\fl  
\widetilde{\cal A}^{j(+)}_{DC}\,
\stackrel{L\to\infty}{\stackunder{T\to\infty}{\longmapsto}}
\frac{(2\pi)^6}{i} e^{i\Theta_{DC}}\,e^{\overline{\Sigma}_{DC}}
\widehat{\delta}_{\{D\}}(\overline{\cal Q}-k_D)
\widehat{\delta}_{\{C\}}(\overline{\cal Q}-k_C)\sqrt{2\pi}\,
\frac{\overline{\rm v}_j}{2L}\,\overline{\cal D}_j\,e^{-\overline{\Omega}_j(R)}
e^{-\overline{\Theta}_j}, 
\label{5_34} \\
\fl  
\mbox{with: }\,\overline {\rm r}_j\equiv \frac{m^2_j}{2\overline{\rm Q}^2}\ll 1,\quad    
\overline{\rm v}_j\approx 1-\overline{\rm r}_j, \quad 
\overline{w}^\lambda_j\approx
-\frac{\overline{\rm r}_j}{(\overline{U}\overline{\ell})}
\left((\overline{\vec{\rm H}}\cdot\vec{l}),\overline{H}^0\vec{l}\right),\;\; 
(\overline{H}\overline{w}_j)=0, 
\label{5_35} \\
\fl  
\frac 1{\overline{{\cal D}}^2_j}\approx 2(\overline{U}\overline{\ell})\!
\left\{1-\frac{\overline{\rm r}_j}{(\overline{H}\overline{\ell})}
[2\overline{H}^0-iT]\right\}\!\simeq 
2(\overline{U}\overline{\ell})\equiv 
2(\overline{\ell}\overline{\vec{\cal T}}\overline{\ell})>0, \quad 
\overline{\tau}_j\simeq\overline{\tau}=
2\overline{\rm Q}(\overline{H}\overline{\ell}), 
\label{5_36} \\
\fl  
\mbox{or: }\;
\overline{\tau}=2\frac{(\overline{H}\overline{\ell})^2}{(\overline{U}\overline{\ell})}>0, 
\quad \; \overline{\Theta}_{j}\approx 
m^2_j\left[\overline{U}^0-\frac{\overline{H}^0}{\overline{\rm Q}}\right]=
\frac{m^2_j}{(\overline{H}\overline{\ell})}
\left[\overline{H}^0(\overline{\vec{\rm U}}\cdot\vec{l})-
\overline{U}^0(\overline{\vec{\rm H}}\cdot\vec{l})\right]. 
\label{5_37} 
\eea
The value $\overline{\Theta}_{j}$  disappears now for 
$k^\lambda_C=k^\lambda_D\simeq {\rm K}\overline{\ell}^\lambda$, \cite{NN_anp}. 
Moreover, for this case: $H^\lambda\simeq {\rm K}U^\lambda$ and so on, and 
${\rm Q}\simeq\overline{\rm Q}\simeq{\rm K}$. Nevertheless, 
$\vec{\rm l}_{\{\}}\neq\vec{l}$ and $w_j\neq\overline{w}_j$. 

The assumption of small $m_j$ here repairs again the self-consistency of our calculations 
and, similarly (\ref{2_15_41}), (\ref{5_26_0}), leads to the invariant width   
(\ref{5_23}) of the approximate non-spreading CRG- like wave function of intermediate 
neutrino and to the invariant width (\ref{5_37}) for oscillation amplitude as in \cite{nn}. 
Since the imaginary contribution to the widths depends on the used integration variable, 
it should be neglected as artefact of the using estimation method. It also implies 
inadmissible complex value of saddle point $\overline{q}^\lambda_j$, depending on base 
time $T$ and base line $L$. 
The additional smallness of $\overline{\rm r}_j(\overline{H}\overline{\ell})^{-1}$ in 
(\ref{5_36}) also indicates such neglect. 

It is worth to note, that exactly the same mechanism \cite{LL,Mess,Tl,g_w} 
of Gaussian integration, (\ref{5_14})--(\ref{5_20}) or (\ref{5_31}), (\ref{5_32}), 
leads to usual inevitable spreading in time of free non relativistic Gaussian wave packet 
(\ref{eq:gauss_x}) discussed in Section \ref{sec:wp_QM}. 
The usually discussed spreading of relativistic wave packet \cite {dn,b_d,i_z,bern} 
also is based on the using of {\sl Gaussian approximation} and on the calculation of 
quantities directly inspired by Gaussian distribution as in \ref{ap:1_A}. 

On the other hand, the shown above form-invariance (\ref{4_4}), (\ref{4_5}) of covariant 
on shell one-packet wave function, similarly (\ref{3_13_0_0_1})--(\ref{3_13}) assures its 
propagation without change of its relativistically invariant width $\sigma$ or 
$g_{1,2}(m,\sigma)$.  
This in turn manifests in CRG- approximation of subsection \ref{sec:CRG_app}, as the non 
spreading at all the packet wave function (see figure \ref{fig:1}). 
We view this distinction as a manifestation of the different role of time and the different 
meaning of localization in QM and in QFT, as discussed in section \ref{sec:wp_sf_QFT}. 

Evidently, the performing in (\ref{5_00}) at first the plane-wave limits 
$\sigma_{\pi,\mu}\to 0$, (\ref{3_41}), (\ref{3_22}) of the reduced overlap function 
(\ref{3_21_G}), (\ref{3_40_V}), (\ref{5_00_01}), (\ref{5_001}) will fully eliminate its 
dependence on effective impact points $Y_{\{C/D\}}$ for $\kappa=q-k=0$, and the above 
asymptotic limits become impossible to take at all. 
Moreover, now unlike $q^\lambda$ in (\ref{5_04}), (\ref{5_06}), 4-vector 
$k^\lambda=k^\lambda_{\{C/D\}}$ will be in principle off the mass shell:  
\bea
\fl  
\widehat{\cal V}_{[C/D]}(q)\bigr|_{{\rm p.w.}}=
(2\pi)^4\,\delta_4(q-k), \quad \;\;
\widehat{\cal F}^{(+)}_{[C/D]j}(\varrho)\bigr|_{{\rm p.w.}}=
\frac{(-i)\, e^{\mp i(k\varrho)}}{2E_{{\rm k}j}(E_{{\rm k}j}-k^0-i0)}.
\label{5_05} 
\eea 
This also illuminates the above main difficulty of the problem under 
consideration: the asymptotical expansion of 
$\widehat{\cal F}^{(+)}_{[C/D]j}(\varrho)$ over parameters $T,{\rm R}$ is not 
uniform with respect to the limit of parameters $\sigma_a,m_j$ and with respect 
to the time derivative $\partial_\varrho^{0}$ and $d^3\rho$- integration in   
(\ref{C_03}). 
That means also that more uniform representation of composite wave-packet state may be useful. 
The representation of such kind for on-shell composite wave function is obtained and 
analyzed below in \ref{sec:one_p_rep}. 


\section{Conclusions}\label{sec:concl}
Due to the time relativity only the transition amplitude (\ref{eq:limit_sigma_inf}) 
from point $\x_a$ to point $\x$ during the time interval $T=x^0-x^0_a>0$ is meaningful in 
relativistic QFT, unlike the non relativistic quantum-mechanical probability amplitude 
of finding the particle localized at any point $\x$ at any instant of time $t$. 
This time relativity is reflected by respective formulation of Huygens' principle 
\cite{b_d}. 

It is shown here, how the consistent use of QFT constraints and minimization properties   
\cite{schw,blt,oksak,jost,s_w,stroc,Dvt} determines the general form of relativistic covariant 
interpolating wave packet for single-particle states of free massive fields with 
arbitrary spin. These states are simply expressed via corresponded field operators. 
The analytic continuation of Wightman functions for these fields in complex Minkowski 
coordinate space specifies the universal way to wave-packet construction for massive 
particles with arbitrary spin and thus elucidates profound physical meaning of that analytic 
continuation and the meaning of remained arbitrariness of corresponding state normalization.
This interpolating wave packet contains covariant particle (antiparticle) states 
\cite{tir} only with positive (negative) energy (frequency) sign without their mixing and  
propagates without change of its relativisticaly invariant width. It has non-relativistic 
Gaussian wave packet (\ref{eq:gauss_p}), as a precise non-relativistic 
limit (\ref{2_28_2}), (\ref{D_6}), independent of the remained ambiguity of its normalization 
(\ref{2_20}), (\ref{2_20_0}), (\ref{4_6}). 
So, the conventional belief about inevitable mixing of states with opposite energy 
sign, starting to propagate inside the pure non-relativistic Gaussian wave packet 
\cite{b_d,i_z,bern}, has the only limited sense. A similar assertion concerns 
relativistic CRG- approximation \cite{nn,nsh,dn,n_sh,NN_shk} for the wave 
packet, which by definition describes the off-shell evolution (\ref{f_1})--(\ref{f_3}).  
Thus, unlike the non relativistic quantum mechanics and optics, the relativistic QFT 
admits the massive wave-packets interpolating in above covariant sense only in the form 
of (\ref{eq:phidef}), (\ref{2_18}) and (\ref{eq:spinor_wp_ket_b_fin_0}), 
(\ref{eq:spinor_wp_ket_b_fin_1}), (\ref{3_3})--(\ref{eq:spinor_wp_wf_end}), 
(\ref{3_07_02})--(\ref{3_07_03}), respectively. 
This wave packet has also a finite limit of $m\to 0$, $\sigma\to 0$ simultaneously, 
with fixed $\tau\to (mc/\sigma)^\epsilon$, $\epsilon>0$, given by (\ref{4_7}), 
(\ref{4_8_0}). 
It separates naturally the light-cone degrees of freedom and elucidates the origin of 
possible arbitrariness of wave packet for the massless case. 

Application of the interpolating wave packets to neutrino oscillation problem reveals 
their natural appearance in amplitude of macroscopic Feynman diagram by making use of any
kind of `pole integration' \cite{bth,ahm,fujii}, which arises naturally in any 
space-time asymptotic regime. It arises also as a natural equivalence of on-shell 
reduction of the off-shell composite wave function (\ref{3_15}), by omitting the time-ordering 
$\theta$-functions: $\widehat{\Upsilon}\mapsto\Upsilon$. 
The notion of composite wave function allowes to strictly transform the amplitude of 
Feynman diagram into the form of scalar product, similar to one that is used in intermediate 
wave-packet picture of neutrino oscillations \cite{bern,gunt,bl1,bl2,bl3}. Thus, this 
notion gives effective language for detailed description of oscillations, naturally 
connecting both these pictures. The `pole integration' procedure for oscillation 
amplitude is shown to be fully equivalent to the pole approximation for composite wave 
functions defined by certain part of neutrino propagator and external interpolating wave 
packets of any one vertex of macrodiagram.    

On the one hand, unlike the non relativistic Gaussian packets and the CRG- approximation 
\cite{nn,nsh}, formula (\ref{3_4}) demonstrates, that the usual causal evolution 
(\ref{3_13_0_0_1})--(\ref{3_13}) of any free interpolating relativistic wave packet 
(cmp. \cite{feyn}) does not mix the states of particles with different energy sign 
(with antiparticles). 
On the other hand, as elucidated here, exactly for those wave packets (\ref{eq:phidef}), 
(\ref{2_18}), (\ref{3_3})--(\ref{eq:spinor_wp_wf_end}) the causal ordering, naturally 
arising for the macroscopic scattering processe (\ref{3_13_1}) similarly (\ref{3_4}) and 
according to Huygens' principle, leads to the natural transformation 
(\ref{3_13_1})--(\ref{3_15}) of integrated causal neutrino propagator $S^c_j(x-y)$ into 
the composite off-shell wave function (\ref{3_18}), (\ref{3_19}), and then into the 
composite on-shell wave function (\ref{3_20_0})--(\ref{3_20_1}), as linear superposition 
(\ref{3_29}) of {\sl free interpolating wave-packet} states with the same 
(neutrino) mass $m_j$ but with different widths and centers. The expression (\ref{3_29})
may be recursivelly generalized to arbitrary number of external wave packets, and it may 
be fully localized for $V^+\ni\zeta_a\to 0$ as in (\ref{3_29_00}). 
Non accidentally both the causal and Wightman functions are just the differently defined 
boundary values in complex Minkowski space (\ref{2_23_0}), (\ref{A_7_0}) and (\ref{2_30_01}), 
(\ref{A_7_1}) or (\ref{A_1}) and (\ref{A_1_2}) respectively for scalar or spinor case, 
actually of the same invariant analytic function $h(Z)$ \cite{oksak}, defined by 
(\ref{2_20_0}), (\ref{A_7}). 

The advantage of the presented formalism, unlike many previous works  
\cite{bern,gunt,rich,bth,ahm,Dl_1,Nis,kbz,GS,bGG,Dvor,bl1,bl2,bl3,fujii}, is that the 
effective impact points $Y_C/X_D$ of neutrino creation/detection with the base line $L$ 
and base time $T$ between them here are not introduced ``by hand''.  
The covariant and strict QFT ground of the formalism presented in sections 
\ref{sec:wp_sf_QFT}, \ref{sec:wp_ff}, 
\ref{sec:diagr_tr} gives the answer on the main part of the problem under consideration: 
how these effective points of neutrino creation/detection do arise. 
Moreover, their dependence on the momentums of two external particles also is found 
exactly. 

The given here exact calculations of overlap functions and of respective universal 
asymtotic behavior of composite wave functions (\ref{5_26}) may be successfully 
repeated for the vertices with three external particles of a ``little donkey'' diagram Figure 
\ref{fig:2} for the process like $(\ell+{\rm A})\oplus{\rm B}\overset{\nu_j}{\longrightarrow}
{\rm A}^\prime\oplus({\rm B}^\prime+\ell^\prime)$ \cite{Dl_1,Nis}. To this end one 
can use the wave packets only for incoming particles $I_{C\oplus D}\in\ell$,A,B 
and for one of outgoing ${\ell}^\prime$, but using plane waves for the one 
of outgoing unregistred particles $F_{C\oplus D}\in{\rm A}^\prime,{\rm B}^\prime$,  
in both $\{C\}$ and $\{D\}$ vertices. 
That calculations show, that the effective impact points of neutrino 
creation/detection $Y_C,\,X_D$ are given by the expression similar to (\ref{3_40_S_1}), 
(\ref{3_40_S_2}) and remain independent of the momentum of plane waves in accordance with 
the intuitive picture. 
They confirm that at least one wave packet per vertex is needed to get a finite coherence 
length, but at least two wave packets per vertex are necessary to get the definite base line 
$L$, base time $T$, and the geometric suppression factors $\exp\{\Sigma_{C/D}\}$. 
The recent results of \cite{n_sh,NN_shk, kt_1} demonstrate the possibility of  
direct experimental manifestation of the wave packets of external particles also for the 
small-distance effects, such as reactor antineutrino anomaly \cite{anom}. 
Calculation of oscillation probabilities for amplitudes (\ref{5_34}) is goes along 
standard way \cite{NN_20}. 

\ack
Authors thank V. Naumov and D. Naumov for usefull discussions, A. N. Vall, I. F. Ginzburg, 
E. Kh. Akhmedov, N. V. Ilyin, E. G. Aman for important comments.

\appendix 
\section{Minimization properties of wave packet}\label{ap:1_A}
Here in order to advocate the minimization properties of the function (\ref{eq:phidef}) we 
briefly generalize some results of \cite{al_h_w}. For any hermitian position operator $\x$ in 
momentum representation and velocity operator $\vec{\rm v}=\vec{\nabla}_{\rm p}E(\p)$ with 
an arbitrary spectrum E(\p), the wave packet with minimal uncertainty should saturate 
position-velocity uncertainty relation. This relation is determined by the minimum value of 
positive definite quadratic form for operator with complex parameters: 
\bea
\fl 
\langle\left(\vec{A}^\dagger\cdot\vec{A}\right)\rangle \geqslant 0,\;\mbox{ for }\;
\vec{A}=-i\x+\alpha \vec{\rm v}-\vec{\beta}, \;\mbox{ with }\;
\alpha=\alpha_r+i\alpha_i,\quad \vec{\beta}=\vec{\beta}_r+i\vec{\beta}_i,  
\label{1_A_0} \\ 
\fl 
\mbox{wherefrom for }\; 
\frac{\partial}{\partial\alpha_{r,i}}\langle\left(\vec{A}^\dagger\cdot\vec{A}\right)\rangle=0, 
\quad \vec{\nabla}_{\beta_{r,i}}\langle\left(\vec{A}^\dagger\cdot\vec{A}\right)\rangle=0, 
\;\mbox{ one has:}
\label{1_A_1} \\ 
\fl 
\langle\left(\vec{A}^\dagger\cdot\vec{A}\right)\rangle \geqslant 
\langle (\Delta\x)^2\rangle-|\alpha|^2\langle(\Delta\vec{\rm v})^2\rangle \geqslant 0,
\;\mbox{ for }\; \vec{\beta}_r=\alpha_r\langle\vec{\rm v}\rangle,\quad 
\vec{\beta}_i=\alpha_i\langle\vec{\rm v}\rangle-\langle\x\rangle, 
\label{1_A_2} \\ 
\fl 
\alpha_r=\frac{\langle\vec{\nabla}^2_{\rm p}E(\p)\rangle} 
{2\langle (\Delta\vec{\rm v})^2\rangle},\quad 
\alpha_i=\frac{\widetilde{C}(\x,\vec{\rm v})}{2\langle (\Delta\vec{\rm v})^2\rangle}, \quad 
\widetilde{C}(\x,\vec{\rm v})=
\langle\left(\x\cdot\vec{\rm v}\right)+\left(\vec{\rm v}\cdot\x\right)\rangle
-2\left(\langle\x\rangle\cdot\langle\vec{\rm v}\rangle\right). 
\label{1_A_3} 
\eea
Here besides the standard definitions  
$\langle (\Delta\x)^2\rangle=\langle\x^2\rangle-\langle\x\rangle^2$, 
$\langle(\Delta\vec{\rm v})^2\rangle =\langle\vec{\rm v}^2\rangle-\langle\vec{\rm v}\rangle^2$, 
we introduce $\alpha_i$ and use general form of position operator with some function 
$\Theta({\rm p})$, which makes it self-adjoint with respect to given scalar product of chosen 
Hilbert space:    
\begin{equation}
\x=i\Theta^{-1}({\rm p})\vec{\nabla}_{\rm p}\Theta({\rm p})=
i\vec{\nabla}_{\rm p}+i\vec{\nabla}_{\rm p}\ln\Theta({\rm p}).
\label{1_A_4}  
\end{equation}
The position-velocity uncertainty relation takes the form 
\bea
4 \langle (\Delta\x)^2\rangle \langle(\Delta\vec{\rm v})^2\rangle \geqslant 
\langle\vec{\nabla}^2_{\rm p}E(\p)\rangle^2+\widetilde{C}^2(\x,\vec{\rm v}). 
\label{1_A_5} 
\eea
Even for relativistic spectrum $E(\p)\mapsto E_{\rm p}$ it could not be rewritten in Lorentz- 
invariant form even for conventional case with omitted covariation 
$\widetilde{C}(\x,\vec{\rm v})$. 
So it is not suprising that corresponding wave packet with minimal uncertainty generally  
also is not Lorentz-invariant \cite{al_h_w}. This wave packet as a coherent state for 
operator (\ref{1_A_0}), (\ref{1_A_4}) with arbitrary parameters $\alpha,\vec{\beta}$ 
satisfies to equation with the following solution, where $N$ is normalization constant: 
\bea
\fl 
\vec{A}\Psi(\p)=0, \quad  \Psi(\ka)=
\frac{N}{\Theta({\rm k})}\exp\left\{-\alpha E(\ka)+(\vec{\beta\cdot\ka})\right\}\longmapsto 
\frac{N}{\Theta({\rm k})} e^{i(kx_a)}e^{-(k\zeta_a)}, 
\label{1_A_6} \\
\fl 
\mbox{with }\; k^0=E_{\rm k},\;\mbox{ and }\; 
\zeta^\mu_a=(\alpha_r,\vec{\beta}_r)=\alpha_r(1,\langle\vec{\rm v}\rangle), \quad 
\zeta^2_a=\alpha^2_r(1-\langle\vec{\rm v}\rangle^2)\geqslant 0, 
\label{1_A_7}  \\
\fl 
x^\mu_a=-(\alpha_i,\vec{\beta}_i)=(0,\langle\x\rangle)-\alpha_i(1,\langle\vec{\rm v}\rangle),   
\label{1_A_8}
\eea
as 4-vectors with respect to Lorentz transformations \cite{al_h_w}. So, the wave function of 
wave packet with minimal uncertainty consists of two factors. The first one is the 
Lorentz-invariant exponential function $N e^{i(kx_a)}e^{-(k\zeta_a)}$ with amplitude 
(\ref{eq:phidef}), (\ref{D_3}), which due to (\ref{2_11}) up to inessential phase 
$e^{-i(p_ax_a)}$ correctly determines the Lorentz-covariant wave packet states 
(\ref{eq:packet_state_scalar}) or (\ref{eq:spinor_wp_ket_b_fin_0}), (\ref{3_07_01}). 
The second one is defined only by self-adjointness of position operator. For non relativistic 
case we have $\Theta({\rm k})=1$ with quadratic spectrum and come to usual Gaussian wave 
packet (\ref{eq:gauss_p}). For the case of relativistic position operator $\widehat{\Xop}$ in 
Section \ref{sec:wp_sf_QFT}, 
we have the above Lorentz-invariant exponential factor as in \cite{al_h_w} and 
$\Theta^{-1}({\rm k})=\sqrt{E_{\rm k}}$ as in (\ref{2_8_0}), \cite{blt}. 

\section{Various limiting cases of covariant wave packet.}\label{ap:A}
The another way to check the (\ref{2_29}) uses the relations for 
$|\ka-\p|\ll E_{\rm p}=\sqrt{\p^2+m^2}$: 
\bea
\fl 
E_{\rm k}\approx E_{\rm p}+\left(\vs_{\rm p}\cdot(\ka-\p)\right)+\frac{1}{2E_{\rm p}}
(\ka-\p)^j\left(\delta^{jl}-{\rm v}^j_{\rm p}{\rm v}^l_{\rm p}\right)(\ka-\p)^l, \;
\mbox{ with: }\; \vs_{\rm p}=\frac{\p}{E_{\rm p}}, 
\label{D_1} \\ 
\fl 
(kp_a)=E_{\rm k}E_{\rm p}-(\ka\cdot\p)\approx m^2+\frac{1}{2}
(\ka-\p)^j\left(\delta^{jl}-{\rm v}^j_{\rm p}{\rm v}^l_{\rm p}\right)(\ka-\p)^l, \quad 
\p\equiv |\p|\n_{\rm p}, 
\label{D_2} \\
\fl 
\mbox {whence: }\; 
\phi^\sigma(\ka,\p)=e^{i((p_a-k)x_a)}\langle\ka\ket{\{p_a, x_a,\sigma\}}=
{N}_\sigma  e^{-(k\zeta_a)}
\stackunder{\tau\to\infty}{\longmapsto}{N}_\sigma  e^{-g_1(kp_a)},\mbox{ or}  
\label{D_3}  \\
\fl 
\bigl(\mbox{for $g_2=0$, $\tau = m^2g_1$}\bigr) = 
\frac{\aleph(\tau)}{m^2}e^{-g_1(kp_a)}\longmapsto 
(\mbox{near the maximum:}\;|\ka-\p|\ll E_{\rm p}) 
\label{D_3_00}  \\
\fl 
\longmapsto 
\frac{(2\pi)^2}{m^2}\frac{{\cal I}(\tau)}{h(\tau^2)\,e^\tau}\,
e^{-((\ka-\p)\vec{\rm T}(\ka-\p))}
\stackunder{g_1\to\infty}{\longrightarrow}
2m(2\pi)^3\,\frac{E_{\rm p}}{m}
\left[\frac{|\vec{\rm T}|}{\pi^{3}}\right]^{1/2} e^{-((\ka-\p)\vec{\rm T}(\ka-\p))}  
\label {D_3_0}  \\
\fl 
=2m(2\pi)^3\,\frac{E_{\rm p}}{m}
\left\{\left[\frac{g_1}{2\pi}\frac{m^2}{E^2_{\rm p}}\right]^{1/2}
\exp\left[-\frac {g_1}{2}\frac{m^2}{E^2_{\rm p}}
\left({\rm k}_\parallel-{\rm p}_\parallel\right)^2\right]\right\}
\left\{\frac{g_1}{2\pi}\,e^{-(g_1/2)(\ka_\perp-\p_\perp)^2}\right\}
\label {D_3_01}  \\
\fl 
\stackunder{g_1\to\infty}{\longrightarrow}
(2\pi)^3\,2E_{\rm p}\,\delta\left({\rm k}_\parallel-{\rm p}_\parallel\right)
\delta_2\left(\ka_\perp-\p_\perp\right)=(2\pi)^3\, 2E_{\rm p}\,\delta_3(\ka-\p),
\quad \ka=\n_{\rm p}{\rm k}_\parallel+\ka_\perp,
\label {D_3_1}  \\
\fl 
(\ka-\p)^2\!=\left({\rm k}_\parallel-{\rm p}_\parallel\right)^2\!+
(\ka_\perp\!-\p_\perp)^2,\;\mbox{ with: }\;\p_\perp\mapsto 0, \;\;\;
{\rm p}_\parallel=|\p|,\;\;\; {\rm k}_\parallel=\left(\ka\cdot\n_{\rm p}\right),
\label {D_3_2}  \\
\fl 
\mbox{and for: }
\left(\vec{\rm T}\right)^{jl}=
\frac{g_1}2\left(\delta^{jl}-{\rm v}^j_{\rm p}{\rm v}^l_{\rm p}\right), \quad 
|\vec{\rm T}|=\det\{\vec{\rm T}\}=\left(\frac{g_1}2\right)^3(1-\vs^2_{\rm p})=
\left(\frac{g_1}2\right)^3\frac{m^2}{E^2_{\rm p}}, 
\label{D_4} 
\eea
may be easy obtained by direct calculation as rotationally invariant determinant of 
separable positively defined operator (matrix). Surely the {\sl nonzero ``independent''} 
limit $m\to 0$ of (\ref{2_28_4}) or (\ref{D_3_01}) already implies $\sigma=0$ i.e. plane 
wave (\ref{2_29}), (\ref{D_3_1}). Whereas the usual solution to free massless KG equation 
(\ref{3_07}) with $\sigma\neq 0$, $z_a=x_a+i\zeta_a(p_a,\sigma)$ requires changed 
normalization: 
\bea
\fl 
m^2\psi_\sigma(\p_a,x_a-x)\stackunder{m\to 0}{\longmapsto}
(-i)\aleph(0)D^-_{0}\!\left(x-z_a\right)=
(-1) \aleph(0)(2\pi)^{-2}\left(x-z_a\right)^{-2}. 
\label{D_4_0}
\eea
As it is easy to see, the   (\ref{D_3_00}) admits also the limit $\sigma\to 0$, 
$g_1\to\infty$ jointly with $m\to 0$ for fixed $\tau=m^2g_1$, $p^\nu_a=(p^0,\p)$, 
$p^0=E_{\rm p}\mapsto |\p|={\rm p}$, $k^\nu=(k^0,\ka)$, 
$k^0=E_{\rm k}\mapsto |\ka|={\rm k}$, so, that $p^\nu_a\mapsto{\rm p}(1,\n_{\rm p})$, 
$k^\nu\mapsto{\rm k}(1,\n_{\rm k})$, and with the help of the relations:
\bea
&&\!\!\!\!\!\!\!\!\!\!\!\!\!\!\!\!\!\!\!\!\!\!\!
\lim_{g_1\to\infty} g_1 e^{-g_1Y}=\delta(Y),\quad (Y\geqslant 0),\;\;
\mbox{ where, for }\; g_1\gg g_2:
\label{D_7_0} \\
&&\!\!\!\!\!\!\!\!\!\!\!\!\!\!\!\!\!\!\!\!\!\!\!
(k\zeta_a)= g_1(kp_a)\equiv g_1\left[E_{\rm k}E_{\rm p}-(\ka\cdot\p)\right] \approx 
g_1{\rm k}{\rm p}\left[1-(\n_{\rm k}\cdot\n_{\rm p})\right]+
\frac{\tau}2 \left(\frac{\rm k}{\rm p}+\frac{\rm p}{\rm k}\right),
\label{D_7_1} \\
&&\!\!\!\!\!\!\!\!\!\!\!\!\!\!\!\!\!\!\!\!\!\!\!
\delta\left(1-(\n_{\rm k}\cdot\n_{\rm p})\right)\longmapsto 
2\pi\,\delta_\Omega(\n_{\rm k},\,\n_{\rm p}), \;
\mbox{ under }\; d\Omega(\n_{\rm k}), 
\label{D_7_0_0} 
\eea
it leads to:
\bea
&&\!\!\!\!\!\!\!\!\!\!\!\!\!\!\!\!\!\!\!\!\!\!\!
\phi^\sigma(\ka,\p)=\frac{\aleph(\tau)}{\tau}g_1e^{-g_1(kp_a)}
\stackrel{m\to 0}{\stackunder{\sigma\to 0}{\longmapsto}}\,\phi_\tau(\ka,\p), 
\label{D_7_02} \\
&&\!\!\!\!\!\!\!\!\!\!\!\!\!\!\!\!\!\!\!\!\!\!\!
\phi_\tau(\ka,\p)=2\pi\,\frac{\aleph(\tau)}{\tau}\,
\exp\left[-\,\frac {\tau}{2}\left(\frac{\rm k}{\rm p}+\frac{\rm p}{\rm k}\right)\right]
\,\frac{\delta_\Omega(\n_{\rm k},\,\n_{\rm p})}{{\rm k}{\rm p}},
\label{D_7} \\
&&\!\!\!\!\!\!\!\!\!\!\!\!\!\!\!\!\!\!\!\!\!\!\!
\mbox{or: }\;\phi_\tau(\ka,\p)=(2\pi)^3\,
\frac{{\cal I}(\tau)}{\tau\,e^{\tau}\,h(\tau^2)}\,
\exp\left[-\,\frac {\tau}{2{\rm k}{\rm p}} \left({\rm k}-{\rm p}\right)^2\right]
\,\frac{\delta_\Omega(\n_{\rm k},\,\n_{\rm p})}{{\rm k}{\rm p}},
\label{D_7_00} \\
&&\!\!\!\!\!\!\!\!\!\!\!\!\!\!\!\!\!\!\!\!\!\!\!
\mbox{giving: }\;
\phi_\tau(\ka,\p)\stackunder{\tau\gg 1}{\longmapsto}
(2\pi)^3\left(\frac{2\tau}{\pi}\right)^{1/2}
\exp\left[-\,\frac {\tau}{2{\rm k}{\rm p}} \left({\rm k}-{\rm p}\right)^2\right]\,
\frac{\delta_\Omega(\n_{\rm k},\,\n_{\rm p})}{{\rm k}{\rm p}}, 
\label{D_8} 
\eea
for $\tau\gg 1$ with the use of   (\ref{2_20}) -- (\ref{eq:sc_limit_0}), or 
(\ref{A_7}), (\ref{2_15_36}). According to (\ref{2_26}), (\ref{2_25}), for 
$\tau\to\infty$ this returns us to   (\ref{D_3_1}). 
For $\tau\ll 1$   (\ref{D_7}), (\ref{2_21})  gives: 
\bea
\phi_\tau(\ka,\p)\stackunder{\tau\ll 1}{\longmapsto} 
2\pi\,\frac{\aleph(0)}{\tau}\,
\frac{\delta_\Omega(\n_{\rm k},\,\n_{\rm p})}{{\rm k}{\rm p}}.
\label{D_9}
\eea 
Since $d\Omega(\n_{\rm k})\equiv d^2\ka_\perp/({\rm k}{\rm k}_\parallel)$, then for 
$\n_{\rm k}\simeq \n_{\rm p}$, ${\rm k}_\parallel\simeq{\rm k}$ in   
(\ref{D_7}) -- (\ref{D_9}) we can replace: 
\bea
\frac{\delta_\Omega(\n_{\rm k},\,\n_{\rm p})}{{\rm k}{\rm p}}\longmapsto
\frac{\rm k}{\rm p}\,\delta_2\left(\ka_\perp\right). 
\label{D_9_0}
\eea
The obtained longitudinal in fact non Gaussian profile with arbitrary $\tau$ in 
(\ref{D_7}), (\ref{D_7_00}), (\ref{D_8}) reflects above mentioned freedom of profile 
function of wave packet appearing for interpolating wave packet of massless particle 
unlike the massive one in (\ref{D_3_01}), (\ref{D_3_1}). 

In spite of the singularity of the measure, the limit (\ref{D_7}) is uniform with respect 
to integral of Lorentz invariant Fourier-representation 
(\ref{eq:scalar_wp_wf})$\,\mapsto\,$(\ref{2_18}), (\ref{2_19}). So, the dimensionless 
function (\ref{2_15_34}) of dimensionless argument (\ref{2_15_35}) for fixed 
$\tau=m^2g_1$, $\underline{x}=x_a-x$, $p^\nu_a\mapsto{\rm p}(1,\n_{\rm p})$ with 
$p^2_a=0$, $p^0={\rm p}$, $(p_a \underline{x})= 
{\rm p}\left(\underline{x}^0-(\n_{\rm p}\cdot\underline{\x})\right)\equiv 
{\rm p}\left(\underline{x}^0-\underline{\rm x}_\parallel\right)\equiv 
{\rm p}\underline{x}^-$: 
\bea
\fl 
\psi_\sigma(\p_a,\underline{x})\stackrel{m\to 0}{\stackunder{\sigma\to 0}{\longmapsto}}
\psi^\tau(\p_a,\underline{x})=
\frac{{\cal I}(\tau)}{h(\tau^2)}\,h\!\left(\tau^2-2i\tau(p_a \underline{x})\right)=
\frac{\aleph(\tau)}{(2\pi)^2}\,h\!\left(\tau^2-2i\tau(p_a \underline{x})\right),  
\label{D_10} 
\eea
is still a Fourier-image (\ref{eq:scalar_wp_wf}) of limit $\phi_\tau(\ka,\p)$ 
(\ref{D_7}), (\ref{D_7_00}) of its momentum profile (\ref{eq:phidef}). Substitution of 
above to representation (\ref{eq:scalar_wp_wf}) leads to the integral:
\bea
&&\!\!\!\!\!\!\!\!\!\!\!\!\!\!\!\!\!\!\!\!\!\!\!\! 
\psi^\tau(\p,\underline{x})=
\frac{\aleph(\tau)}{(2\pi)^2\, 2{\rm p}\tau}\int\limits^\infty_0\! d{\rm k}\,
\exp\left[i{\rm k}\left(\underline{x}^0-(\n_{\rm p}\cdot\underline{\x})\right)-\,
\frac{\tau}{2}\left(\frac{\rm k}{\rm p}+\frac{\rm p}{\rm k}\right)\right], 
\label{D_11}
\eea 
which due to (\ref{A_7}) with ${\rm k}=2{\rm p}\tau t$ coincides with (\ref{D_10}). 
This reflects a drastic change of Lorentz symmetry \cite{Nvj,Wnb} for massless states 
since the problem in fact is reduced from 3+1 to 1+1 dimensions. Indeed, substituting the 
massless wave packet (\ref{4_7}) into 3D scalar product (\ref{2_30}) one  
immediately shows the necessity of condition $\n_{{\rm p}_b}=\n_{{\rm p}_a}$ for its 
existence. Thus the inner product for these light-cone wave packets (\ref{4_7}) should be 
of the following 1D form for $\overline{x}^\nu=(x^0,{\rm x}_\parallel)$: 
\bea
\fl 
\left(F^{(\tau_b)}_{p_bx_b},F^{(\tau_a)}_{p_ax_a}\right)\!= 
\!\!\!\int\limits^\infty_{-\infty}\!\!\! 
d{\rm x}_\parallel F^{*(\tau_b)}_{p_bx_b}(\overline{x})
(i\!\stackrel{\leftrightarrow}{\partial_x^{0}})F^{(\tau_a)}_{p_ax_a}(\overline{x})=
{\cal M}_{ba}\,e^{i(p_bx_b)-i(p_ax_a)}\left(-\partial_{Z^{ba}}\right)h(Z^{ba}), 
\label{D_12} \\
\fl 
\mbox{where: }\,
Z^{ba}=\tau^2_b+\tau^2_a+\tau_b\tau_a\left(\frac{{\rm p}_b}{{\rm p}_a}+
\frac{{\rm p}_a}{{\rm p}_b}\right)-2i(\tau_b{\rm p}_b+\tau_a{\rm p}_a)(x_b-x_a)^-
\stackunder{b=a}{\longmapsto} 4\tau^2,
\label{D_12_0} \\
\fl 
{\cal M}_{ba}=\frac{\aleph(\tau_b)\aleph(\tau_a)}{(2\pi)^3}
2\left(2+\frac{\tau_b{\rm p}_b}{\tau_a{\rm p}_a}+
\frac{\tau_a{\rm p}_a}{\tau_b{\rm p}_b}\right)\stackunder{b=a}{\longmapsto}
8\,\frac{\aleph^2(\tau)}{(2\pi)^3}, 
\label{D_12_1} \\
\fl 
\mbox{so the norm: }\;\left(F^{(\tau)}_{p_ax_a},F^{(\tau)}_{p_ax_a}\right)= 
-\,\frac{\aleph^2(\tau)}{\tau (2\pi)^3}\frac{\partial}{\partial\tau}h(4\tau^2)=
\frac{\aleph^2(\tau)}{\tau^2}\frac{K_2(2\tau)}{(2\pi)^3}
\stackunder{\tau\to\infty}{\longrightarrow}2\sqrt{\pi\tau},
\label{D_13} 
\eea 
due to (\ref{eq:sc_limit_0}). This dimensionless norm differs from them in 3+1 D case 
(\ref{3_6}), (\ref{3_6_0}). 
The approximation (\ref{D_3_0}), (\ref{D_3_01}) also admits the joint limit 
$\sigma\to 0$, $m\to 0$ with fixed $\tau$: 
\bea
&&\!\!\!\!\!\!\!\!\!\!\!\!\!\!\!\!\!\!\!\!\!\!\!
\mbox{(\ref{D_3_01}) }
\stackrel{m\to 0}{\stackunder{\sigma\to 0}{\longmapsto}}
\frac{(2\pi)^3}{\tau\,e^\tau}\frac{{\cal I}(\tau)}{h(\tau^2)} 
\exp\left[-\,\frac {\tau}{2{\rm p}^2}\left({\rm k}-{\rm p}\right)^2\right]
\delta_2\left(\ka_\perp\right).
\label{D_14} 
\eea
For $\tau\to\infty$ it also returns to   (\ref{D_3_1}). This approximation, unlike 
$\phi_\tau(\ka,\p)$ (\ref{D_7}) with correct Fourier-image (\ref{D_11}) = (\ref{D_10}), 
is very rough and leads to divergent integral. 
 
For the non relativistic limit: $c\to\infty$, with $x^0_a=ct_a$, $m\mapsto mc$, 
$\vs^2_{\rm p}\mapsto \vs^2_{\rm p}/c^2\to 0$, and from (\ref{2_303}), (\ref{2_28_2})  
or (\ref{D_2}) -- (\ref{D_3_0}), (\ref{D_4}),  one has: 
$\left(\vec{\rm T}\right)^{jl}\mapsto (g_1/2)\delta^{jl}$, 
$c(k^0-p^0)=E_{\rm k}-E_{\rm p}=\varepsilon_{\rm k}-\varepsilon_{\rm p}$, 
$E_{\rm p}=mc^2+\varepsilon_{\rm p}$, $\varepsilon_{\rm p}=\p^2/(2m)$. 
Whence,   (\ref{eq:gauss_p}), (\ref{sgm_p}) imply: $g_1=\sigma^{-2}$ for 
$\sigma=\sigma_p$, with: 
\bea
\fl 
\phi^\sigma(\ka,\p)\stackunder{c\to\infty}{\longmapsto}
(2\pi)^3\,2mc \left\{\left(\frac{g_1}{2\pi}\right)^{3/2}e^{-(g_1/2)(\ka-\p)^2}\right\}=
(2\pi)^3\,2mc\, \frac{<\ka |\{\p,\vec{0},\sigma\}>}{(2\sigma\sqrt{\pi})^{3/2}},
\label{D_5} \\
\fl 
\mbox{and thus: }\;
\lim_{c\to\infty}\frac{\langle\ka\ket{\{p_a, x_a,\sigma\}}}{2mc}\,= 
\frac{(2\pi)^3}{(2\sigma\sqrt{\pi})^{3/2}}\,
e^{it_a (\varepsilon_{\rm k}-\varepsilon_{\rm p})}<\ka|\{\p_a,\x_a,\sigma\}>. 
\label{D_6} 
\eea
\section{Some useful intermediate results, formulas and definitions.}\label{ap:B}
For $k^0=E_{\rm k}>0$, $(k\widehat{s})=0$, $\xi=\pm 1$, $p^2_a=m^2$, 
$(p_a\widehat{s}_a)=0$, with arbitrary ${\cal S}$, one has: 
\bea
\fl 
(\gamma k)^2=k^2=m^2,\;\;\;(\vec{\gamma}\!\cdot\!\ka)^2=-\ka^2,\;\;\; 
\widehat{\rm E}(\ka)=\gamma^0\left(\vec{\gamma}\!\cdot\!\ka\right)+\gamma^0 m=
\widehat{\rm E}^\dagger(\ka),
\label {A_00} \\
\fl 
(\widehat{\rm E}(\xi \ka))^2 =E^2_{\rm k}=\ka^2+m^2, \quad  
(\gamma k)+\xi m=\bigl[E_{\rm k}+\xi\widehat{\rm E}(\xi\ka)\bigr]\gamma^0=
\gamma^0\bigl[E_{\rm k}+\xi\widehat{\rm E}(-\xi\ka)\bigr], 
\label {A_0} \\
\fl 
\left[(\gamma k)+\xi m\right]\gamma^0\left[(\gamma k)+\xi m\right]=
2E_{\rm k}\left[(\gamma k)+\xi m\right],
\label{A_2} \\
\fl 
\left[(\gamma k)+\xi m\right]\gamma^0\xi(\gamma k)\left[(\gamma k)+\xi m\right]=
2E_{\rm k}\left[(\gamma k)+\xi m\right]m,
\label{A_3} \\
\fl 
\left[(\gamma k)+\xi m\right]\gamma^0\gamma^5(\gamma\widehat{s})
\left[(\gamma k)+\xi m\right]=
2E_{\rm k}\left[(\gamma k)+\xi m\right]\gamma^5(\gamma\widehat{s}),
\label{A_4} \\
\fl 
\mathrm{Sp}\!\left\{\!\left[(\gamma k)+m \xi\right]\gamma^0\gamma^5
\frac{(\gamma{\cal S})(\gamma k)-(k{\cal S})}{m\xi}\left[(\gamma k)+m \xi\right]
\left[(\gamma p_a)+m \xi\right]
\left[{\rm I}\pm \gamma^5(\gamma\widehat{s}_a)\right]\!\right\}  
\label{A_4_0} \\
\fl 
=\pm 8k^0
\left\{[{\rm g}_{\mu\alpha}{\rm g}_{\nu\beta}-{\rm g}_{\mu\nu}{\rm g}_{\alpha\beta}]
\widehat{s}^\mu_a{\cal S}^\nu(k^\alpha k^\beta+k^\alpha p^\beta_a)\right\}  
\label{A_4_1} \\
\fl 
=\pm 8k^0\left\{-(\widehat{s}_a{\cal S})[m^2+(kp_a)]+(k\widehat{s}_a)(k{\cal S})+
(k\widehat{s}_a)(p_a{\cal S})\right\},\;\mbox { where }\, \pm 1=2s_a. 
\label{A_4_2}
\eea
Here $\gamma^5=i\gamma^0\gamma^1\gamma^2\gamma^3$. 
For Dirac fields in (\ref{3_3})--(\ref{eq:spinor_wp_wf_end}) \cite{blt}, for $k^0=E_{\rm k}>0$: 
\bea
\fl 
\frac 1i S^{-\xi}(x-y)=
\!\int\!\frac{d^3{\rm k} }{(2\pi)^3 2k^0}\sum_{r=\pm 1/2}
{\cal U}^{\xi}_{{\rm k},r}(x)\,\overline{\cal U}^{\xi}_{{\rm k},r}(y)= 
\bigl[i(\gamma\partial_x)+m\bigr]\frac 1i D^{-\xi}_m(x-y),
\label {A_1} \\
\fl 
\{\psi(x),\overline{\psi}(y)\}=\frac 1i S(x-y)=\frac 1i \sum_{\xi=\pm}S^{-\xi}(x-y), 
\quad  S(x-y)\bigr|_{x^0=y^0}=i\gamma^0\delta_3(\x-\y).
\label {A_1_0} \\ 
\fl 
\langle 0|{\rm T}\left(\psi(x)\overline{\psi}(y)\right)|0\rangle=\frac 1i S^c(x-y)=
\sum_{\xi=\pm}\theta\left(\xi(x^0-y^0)\right)\frac {\xi}i \,S^{-\xi}(x-y)=
\label {A_1_1} \\
\fl 
=\frac 1i \int\frac{d^4 q}{(2\pi)^4}e^{-i(q(x-y))}\left(\frac{(\gamma q)+m}
{m^2-q^2-i0}\right)=\bigl[i(\gamma\partial_x)+m\bigr]\frac 1i D^{c}_m(x-y).
\label {A_1_2} 
\eea
For vector field  
${\cal A}^{\nu}_{{\rm k},\lambda}(x)=\epsilon^\nu_{(\lambda)}(k)f_{\rm k}(x)$ 
with polarization $\epsilon^\nu_{(\lambda)}(k)$ in (\ref{3_07_01})--(\ref{3_07_03}): 
\bea
\fl 
\left(\epsilon^*_{(\lambda)}(k)\epsilon_{(\sigma)}(k)\right)=-\delta_{\lambda\sigma}\,, 
\quad  \sum^3_{\lambda=1} 
\epsilon^\mu_{(\lambda)}(k)\epsilon^{*\nu}_{(\lambda)}(k)=
-{\rm g}^{\mu\nu}+\frac{k^\mu k^\nu}{m^2}, \quad k^0=E_{\rm k}>0, 
\label{A_5} \\
\fl 
\frac 1i  {\cal D}^{\mu\nu\, -}(x-y)=
-\!\int\!\frac{d^3{\rm k} }{(2\pi)^3 2k^0}\sum^3_{\lambda=1}
{\cal A}^{\mu}_{{\rm k},\lambda}(x)\,{\cal A}^{*\nu}_{{\rm k},\lambda}(y)=\!
\left({\rm g}^{\mu\nu}+\frac{\partial^\mu_x\partial^\nu_x}{m^2}\right)\!
\frac 1i D^-_m(x-y), 
\label{A_6} \\
\fl 
\mbox{where: }\;\;
D^\mp_m(x)=\pm \lim_{V^+\ni\zeta\to 0}
i\frac{m^2}{(2\pi)^2}\,h\left(-m^2(x\mp i\zeta)^2\right),\;\mbox{ for }\; 
\zeta^2,\,\zeta^0\longrightarrow +0,
\label{2_23_0} \\
\fl 
\mbox {with: }\; -Z/m^2=(x\mp i\zeta)^2 = x^2\mp 2i (x\zeta)-\zeta^2
\mapsto x^2\mp 2i x^0\zeta^0\mapsto x^2\mp i0\,\varepsilon(x^0), 
\label{A_7_0} \\
\fl 
D^c_m(x)=\theta(x^0)D^-_m(x)+\theta(-x^0)D^-_m(-x)=D^c_m(-x)=
i\frac{m^2}{(2\pi)^2}\,h\left(m^2\left(i0-x^2\right)\right),
\label{2_30_01} \\
\fl 
\mbox {with: }\; -Z/m^2=x^2-i0,\;\mbox{ so: }\; D^-_m(x)\equiv{\rm D}_m(x^0;\x),\quad 
D^c_m(x)={\rm D}_m(|x^0|;\x), 
\label{A_7_1}
\eea
for the analytic function $h(Z)$, defined for the main branch $\sqrt{Z}>0$ at $Z>0$, 
as \cite{oksak,b_er}:
\bea
&&
h(Z)=\frac {K_1(\sqrt{Z})}{\sqrt{Z}}=
\int\limits^\infty_0\!dt\exp\left\{-\frac 1{4t}-tZ\right\}   
\label{A_7} \\
&& 
\equiv
\int\limits^\infty_0 \!dt e^{f(t)}\approx 
\left[\frac{2\pi}{-f^{\prime\prime}(\overline{t})}\right]^{1/2}\!e^{f(\overline{t})}=
\sqrt{\frac{\pi}{2}}\,Z^{-3/4}e^{-\sqrt{Z}}, 
\label{2_15_36} 
\eea
for $|Z|\to\infty$, $|{\rm arg}\,Z|<3\pi$ \cite{b_er}, where for the saddle point: 
$f^\prime(\overline{t})=0$, $1/\,\overline{t}=2\sqrt{Z}$.

The following relations are also useful for our aims for 
$\varrho^\lambda=(\varrho^0,\vec{\rho})$, $\xi,\eta =\pm$:
\bea
\fl 
\frac{1}i D^{-\xi}_m(x-y)=\frac{\xi}i D^{-}_m\left(\xi(x-y)\right)=\xi \!\int\!
\frac{d^4 k}{(2\pi)^3}\,\theta(k^0)\,\delta(k^2-m^2)\, e^{-i\xi\left(k(x-y)\right)},
\label{E_00} \\
\fl 
\frac{1}i D^{c}_m(x-y)=\sum_{\xi=\pm}\theta\left(\xi(x^0-y^0)\right) 
\frac{\xi}i\,D^{-\xi}_m(x-y),  \qquad 
\theta(t)=\frac 1{2i\pi}\!
\int\limits^\infty_{-\infty}\!\! d\omega\frac{e^{i\omega t}}{\omega-i0}, 
\label{E_01} \\
\fl 
\int\! d^3\rho\,\frac{1}iD^{-\eta}_m\!\left(x-\varrho\right)\,
(i\!\stackrel{\leftrightarrow}{\partial^0_\varrho})\,
\frac{1}iD^{-\xi}_m\!\left(\varrho-y\right)=
\delta_{\eta\xi}\,\frac{1}i D^{-\xi}_m\!\left(x-y\right).  
\label{3_19_1} 
\eea
Let's consider integrals over $\y_\perp$ of the massless propagators. For 
$(\gamma\partial_x)\equiv\gamma^0\partial^0_x+(\vec{\gamma}\!\cdot\!\vec{\nabla}_{\rm x})$,  
with $p^\nu_a={\rm p}(1,\n_{\rm p})$ they will depend only on 2-dimensional variable  
$\overline{x}^\nu=(x^0,{\rm x}_\parallel)$, with 
${\rm x}_\parallel=(\n_{\rm p}\!\cdot\!\x)$, for 
$\x_\perp=\x-\n_{\rm p}{\rm x}_\parallel$, 
$\partial_{{\rm x}_\parallel}=(\n_{\rm p}\!\cdot\!\vec{\nabla}_{\rm x})$, 
$\gamma_\parallel=(\n_{\rm p}\!\cdot\!\vec{\gamma})$, 
$(\overline{\gamma}\overline{x})=\gamma^0x^0-\gamma_\parallel{\rm x}_\parallel$, 
$(\overline{\gamma}\partial_{\overline{x}})=\gamma^0\partial^0_x+
\gamma_\parallel\partial_{{\rm x}_\parallel}$, 
$(x-y)^2=(x^0-y^0)^2-({\rm x}_\parallel-{\rm y}_\parallel)^2-(\x_\perp-\y_\perp)^2$, 
where $(\gamma\partial_x)\mapsto (\overline{\gamma}\partial_{\overline{x}})$:    
\bea
\fl 
D^c_{(2)}(\overline{x}-\overline{y})=\!\!\int\!\! d^2{\rm y}_\perp D^c_{(4)}(x-y)=
\!\!\int\!\!\frac{i(2\pi)^{-2}d^2{\rm y}_\perp}{i0-(x-y)^2}=
\frac{1}{4\pi i}\ln\left[\frac{i0-(\overline{x}-\overline{y})^2}{{\rm R}^2_\perp}\right]\!, 
\label{E_1} \\
\fl 
S^c_{(2)}(\overline{x}-\overline{y})=\!\!\int\!\! d^2{\rm y}_\perp S^c_{(4)}(x-y)= 
i(\gamma\partial_x)D^c_{(2)}(\overline{x}-\overline{y})=
\frac{(\overline{\gamma},\overline{x}-\overline{y})}
{2\pi[(\overline{x}-\overline{y})^2-i0]}.  
\label{E_2} 
\eea
Here ${\rm R}^2_\perp\mapsto \mu^{-2}$ is a scale of infrared regularization for 
massless 1+1-dimensional case \cite{oksak}. Note, that according to 
\cite{blt,oksak,b_er,g_r,g_sh}, for ${\rm N}=2\Lambda+2$- dimensional 
case it is easy to observe the inversion of (\ref{E_1}) for the causal propagator with 
$\varrho=x-y$: 
\bea
\fl 
D^c_{({\rm N})m}(\varrho)=\!\int \!\frac{d^{\rm N}q}{(2\pi)^{\rm N}}
\frac{e^{-i(q\varrho)}}{(m^2-q^2-i0)}=
i\left(\frac{m}{(i0-\varrho^2)^{1/2}}\right)^\Lambda 
\frac{K_{\pm\Lambda}\left(m(i0-\varrho^2)^{1/2}\right)}{(2\pi)^{\Lambda+1}},
\label{E_3}\\
\fl 
\mbox{so: }\,
D^c_{(2)m}(\varrho)=\frac{i}{2\pi}K_0\left(m(i0-\varrho^2)^{1/2}\right), \quad 
D^c_{({\rm N+2L})m}(\varrho)=
\left(\frac 1{\pi}\,\frac{\partial}{\partial\varrho^2}\right)^{\rm L}
D^c_{({\rm N})m}(\varrho),
\label{E_4}
\eea
for any integer L and Macdonald function $K_\Lambda(z)$ \cite{b_er} with respectively 
changed dimension of $\varrho$- space. 

So called Grimus-Stockinger theorem \cite{GS}, for $\vec{\rm R} = {\rm R}\n$, $\n^2=1$, 
and a sufficiently smooth (non oscillating) function $\Phi(\q) \in C^3$ decreases at 
least like $1/\q^2$ together with its first and second derivatives, gives the leading 
asymptotic behavior with ${\rm R}=|\vec{\rm R}| \rightarrow \infty$ for the integral:
\bea
\fl 
{\cal J}(\pm \vec{\rm R}) = \int \frac{d^3 q}{(2\pi)^3}
\frac{e^{\pm i(\q\cdot\vec{\rm R})}\Phi(\q)}{(\q^2 - {\cal K}^2 - i0)}
\biggr|_{{\rm R}\to\infty}=\frac{e^{i{\cal K}{\rm R}}}{4\pi {\rm R}}
\Phi\left(\pm{\cal K}\n\right)\left[1+ O({\rm R}^{-1/2})\right], 
\label{E_5}
\eea
where ${\cal K}^2>0$ (otherwise for ${\cal K}^2<0$ it fails at least as $O({\rm R}^{-2})$). 
Its most simple explanation for analytic $\Phi(\q)$ is given in \S 130 of \cite{LL} or 
in \cite{Mess,Tl,g_w} with the help of the relation in the sense of distributions for 
$\q={\rm q}\vec{\omega}$: 
\bea
&&\!\!\!\!\!\!\!\!\!\!\!\!\!\!\!\!\!\!\!\!
e^{\pm i(\q\cdot\vec{\rm R})}\biggr|_{{\rm R}\to\infty}=
\frac{2\pi i}{{\rm qR}}\bigl\{e^{-i{\rm qR}}\delta_\Omega(\vec{\omega},\,\mp\n)-
e^{i {\rm qR}}\delta_\Omega(\vec{\omega},\,\pm\n)\bigr\}+O({\rm R}^{-2})=
\label{B_as} \\
&&\!\!\!\!\!\!\!\!\!\!\!\!\!\!\!\!\!\!\!\!
=\pm \frac{2\pi i}{{\rm qR}}\bigl\{e^{\mp i{\rm qR}}\delta_\Omega(\vec{\omega},\,-\n)-
e^{\pm i{\rm qR}}\delta_\Omega(\vec{\omega},\,\n)\bigr\}+O({\rm R}^{-2}). 
\label{B_as_0}
\eea  
In fact Jacob-Sachs theorem \cite{bth,J_S} states, that for sufficiently smooth (non 
oscillating) function $\Psi(q)=\Psi(q_0;\q)$, distinct from zero only within certain 
finite bounds on $q^2=q^2_0-\q^2$ for $M^2_1<q^2<M^2_2$, $q_0>0$, where it is taken to be 
infinitely differentiable, the integral $I(T)$ has the following asymptotic behavior at 
$T\to +\infty$: 
\bea
&&\!\!\!\!\!\!\!\!\!\!\!\!\!\!\!\!\!\!\!\!
I(T)=\!\int\!\frac{dq_0}{2\pi} \,\frac{e^{-iTq_0}\;\Psi(q)}
{2E_{{\rm q}j}(E_{{\rm q}j}-q^0-i0)}\biggr|_{T\to +\infty}\!\!
\longmapsto\, 
i\,\frac{e^{-iTE_{{\rm q}j}}}{2E_{{\rm q}j}}\Psi(E_{{\rm q}j};\q). 
\label{E_6}
\eea

\section{Calculation of two-packet overlap function.}\label{ap:C}
Calculation of overlap function $\widehat{V}_{\{C\}}(q)$ (\ref{3_20_2}) is simplified by 
the following hint. 
Substitution of representation (\ref{2_19}) in explicitly invariant form (\ref{2_18}), 
(\ref{eq:scalar_wp_wf}) with definitions (\ref{3_25}) -- (\ref{3_27}) $\forall\, q$ 
leads to the following structure of integral in (\ref{3_20_2}) with some function 
$\widehat{\Phi}(a,b,\ldots)$: 
\bea
\fl 
\widehat{\cal I}(q)=\!\!\int\!\! d^4l\, \theta(-l^0)\,\delta(l^2-m^2_\mu)\!\!
\int\!\!d^4r\,\theta(r^0)\delta(r^2-m^2_\pi)\delta_4(q-l-r)
\widehat{\Phi}((lQ),(rQ^\prime),\ldots),
\label{B_1_0} \\
\fl 
\mbox{for: }\,  
r=\eta_2 p+\kappa,\quad l=\eta_1 p-\kappa,\quad d^4ld^4r=d^4pd^4\kappa,\,\mbox{ with: }\,
\eta_1+\eta_2\equiv 1, \;\; p=l+r,
\label{B_2} \\
\fl 
\eta_2-\eta_1=\frac{m^2_\pi-m^2_\mu}{q^2},\quad \sigma=\frac{\Delta(q)}{4q^2}, \quad 
\delta(\omega)\delta(\upsilon)=2\delta(\omega-\upsilon)\delta(\omega+\upsilon), \;\;
\forall\, \omega,\,\upsilon,\;\mbox{ it}
\label{B_2_0} \\
\fl 
\mbox{reads: }
\widehat{\cal I}(q)=
\frac{1}2\!\int\!\!d^4\kappa\,\delta((q\kappa))\delta(\kappa^2+\sigma)
\theta(\kappa^0-\eta_1 q^0)\theta(\kappa^0+\eta_2 q^0)
\Phi((\kappa Q),(\kappa Q^\prime),\ldots).
\label{B_3} 
\eea
$q^2>0$ gives $\eta_2>0$, $\sigma>0$, $\kappa^2<0$, $\kappa^0\mapsto 0$, $q^0>0$, 
$\eta_1<0$. In the rest frame of time-like vector $q^\mu_*=(q^0_*,\q_*\!=0)$ there are the 
following invariant substitutions, with any external vectors $Q$, $Q^\prime$, for  
$\widehat{\cal I}(q)\mapsto\underline{\cal I}(q)$:   
\bea
\fl 
q^0_*\mapsto\sqrt{q^2},\quad Q^0_*\mapsto\frac{(qQ)}{\sqrt{q^2}}, \quad 
\vec{\mathrm Q}^2_*\mapsto\frac{(qQ)^2-q^2Q^2}{q^2},\quad 
(\kappa Q)\mapsto -(\vec{\kappa}\cdot \vec{\mathrm Q}_*),\quad 
\vec{\rm n}^2_\kappa=1,
\label{B_4} \\
\fl 
\mbox{with: }\;
\delta((q\kappa))\,\delta(\kappa^2+\sigma)\longmapsto
\frac{\delta(\kappa^0)}{\sqrt{q^2}}\,\delta(\sigma-\vec{\kappa}^2),
\;\mbox{ for: }\;
d^4\kappa=d\kappa^0\frac{|\vec{\kappa}|}2 d\vec{\kappa}^2 d\Omega_3(\vec{\rm n}_\kappa), 
\label{B_5} \\
\fl 
\mbox{giving: }\;
\underline{\cal I}(q)=\frac{\Theta[\Delta,q]}4\sqrt{\frac{\sigma}{q^2}}
\!\int \!d\Omega_3(\vec{\rm n}_\kappa)
\Phi\left(-(\vec{\kappa}\cdot \vec{\mathrm Q}_*),
-(\vec{\kappa}\cdot \vec{\mathrm Q}^\prime_*),\ldots\right), \;\;
\vec{\kappa}=\vec{\rm n}_\kappa\sqrt{\sigma},
\label{B_6} \\
\fl 
\mbox{with: }\;
\int \! d\Omega_N(\n)\,{\rm n}^{\alpha_1}{\rm n}^{\alpha_2}\ldots
{\rm n}^{\alpha_{2l-1}} {\rm n}^{\alpha_{2l}}=
\frac{\Omega_N}{C_N(l)}\sum\limits^{(2l-1)!!}_{\wp=1}
\{\delta\ldots\delta\}^{\{\alpha_1\alpha_2\ldots\alpha_{2l-1}\alpha_{2l}\}}_\wp, 
\label{B_6_1} \\
\fl 
\mbox{where: }\;
\int \! d\Omega_N(\n)=\Omega_N=\frac{2\pi^{N/2}}{\Gamma\left(N/2\right)},
\quad C_N(l)=\prod\limits^{l-1}_{k=0}(N+2k)=\frac{2^l\Gamma(l+N/2)}{\Gamma(N/2)}, 
\label{B_6_2}
\eea
for $N$- dimensional space, with $C_N(0)=1$, $C_N(1)=N$, $C_3(l)=(2l+1)!!$, 
$\Omega_3=4\pi$, and sum stands for permutations $\wp$ symmetrizing all the indices 
$\alpha_1\ldots\alpha_{2l}$. Fron (\ref{B_3}) for $m_\pi>m_\mu$ with the conditions 
(\ref{3_25})--(\ref{3_27}) for (\ref{3_24}):
\begin{equation}
\Theta[\Delta,q]=\theta(\Delta)\theta(q^2)\theta(q^0), \qquad  
\theta(\Delta)=\theta[(m_\pi-m_\mu)^2-q^2]. 
\label{C_THet} 
\end{equation}
The transverseness of $\kappa$ with $q$ becomes very convenient when 
$\widehat{\cal I}^\mu(q)$ contains tensor structure, because for transverse tensor 
$q^2\Pi^{\mu\nu}_q={\rm g}^{\mu\nu}q^2-q^\mu q^\nu$, $\Pi^{\mu\nu}_q Q_\nu=Q^\mu_{\Pi_q}$, 
$(qQ_{\Pi_q})\equiv 0$: 
\bea
\fl 
\Pi^{\mu\nu}_q=\Pi^{\mu}_q{}_\lambda\Pi^{\lambda\nu}_q, \quad 
q^2Q^2-(qQ)^2=q^2 Q_\mu\Pi^{\mu\nu}_q Q_\nu=q^2(QQ_\Pi)= q^2Q^2_\Pi \leftarrowtail 
-q^2\vec{\mathrm Q}^2_*, 
\label{B_7_0} \\
\fl 
\widehat{\cal I}^\mu(q)=
\frac{1}2\!\int\!\!d^4\kappa\,\kappa^\mu
\delta((q\kappa))\delta(\kappa^2+\sigma)
\theta(\kappa^0+\eta_2 q^0)\theta(\kappa^0-\eta_1 q^0)\Phi((\kappa Q))=
Q^\mu_\Pi \widehat{\cal I}_1(q), 
\label{B_7} 
\eea
and so on, where calculation of $\widehat{\cal I}_1(q)$ is reduced to previous case by 
multiplying on $Q_\mu$. With the more complicated tensor structure, choosing for 
$q^2,q^0>0$, $Q^2_\Pi<0$ one of $\vec{\mathrm Q}^\perp_*=0$, it is useful to 
transcribe the measure (\ref{B_5}) for $\vec{\kappa}^2=\kappa^2_3+\vec{\kappa}^2_\perp$, 
$\kappa^\mu=(\kappa^0,\vec{\kappa}_\perp,\kappa^3)\mapsto\kappa^\mu(\kappa^3,\varphi)$ as:  
\bea
\fl 
d^4\kappa=d\kappa^0\,d\kappa^3\,\frac 12\, d\vec{\kappa}^2_\perp\, d\varphi, \quad  
\int\!d\vec{\kappa}^2_\perp\,\delta(\sigma-\kappa^2_3-\vec{\kappa}^2_\perp)=
\theta(\sigma-\kappa^2_3), \quad  |\vec{\kappa}_\perp|=\sqrt{\sigma-\kappa^2_3},
\nonumber \\ 
\fl 
\mbox{so, that: }\,
\underline{\cal I}^{\mu\nu}(q)= \frac{\Theta[\Delta,q]}{4\sqrt{q^2}}\!
\!\int\limits^{\sqrt{\sigma}}_{-\sqrt{\sigma}}\!\!\!d\kappa_3\!
\int\limits^{2\pi}_0\!d\varphi\,\kappa^\mu\,\kappa^\nu\, 
\Phi(\kappa_3{\mathrm Q}^3_{*},\ldots),\,\mbox{ where: }
\int\limits^{2\pi}_0\!d\varphi\,\kappa^\mu\,\kappa^\nu\,\Phi,  
\label{B_8} 
\eea
is immediately expressed only via $\Pi^{\mu\nu}_q$ itself and convolutions of tensors 
$\Pi^{\mu\alpha}_q$, $\Pi^{\nu\beta}_q$ with all existing external vectors $Q,Q^\prime$.  
Similarly for tensor structures of higher rank.

For $q^2<0$, $\Delta(q)>0$, $\sigma=-|\sigma|$ the Breit frame of space-like vector 
${}_*\!q^\mu=(0,\vec{0}_\perp,{}_*\!{\rm q}^3)$, with 
$\kappa^\mu=(\kappa^0,\vec{\kappa}_\perp,\kappa^3)\mapsto\kappa^\mu(\kappa^0,\varphi)$, 
$\kappa^0>0$, $\vec{\kappa}_\perp=|\vec{\kappa}_\perp|\vec{\rm n}_\perp$, 
$d\varphi=d\Omega_2(\vec{\rm n}_\perp)$, leads to the folowing invariant substitutions with 
any external vector $Q$ and integrals (\ref{B_3}), (\ref{B_7}) with   
$\widehat{\cal I}(q)\mapsto \widetilde{\cal I}(q)$:
\bea
\fl 
{}_*\!{\rm q}^3 \mapsto\sqrt{-q^2},\quad {}_*\!{\rm Q}_3\mapsto\frac{(qQ)}{\sqrt{-q^2}}, 
\quad {}_*\!Q^2_0-{}_*\!\vec{\mathrm Q}^2_\perp=Q^2+{}_*\!{\rm Q}^2_3\mapsto 
Q^2-\frac{(qQ)^2}{q^2}=Q^2_\Pi,
\label{B_9} \\
\fl 
\delta((q\kappa))\,\delta(\kappa^2+\sigma)\mapsto
\frac{\delta(\kappa^3)}{\sqrt{-q^2}}\,\delta(\kappa^2_0-|\sigma|-\vec{\kappa}^2_\perp),
\;\mbox{ for: }\;
d^4\kappa=d\kappa^0\,d\kappa^3\,\frac {d\vec{\kappa}^2_\perp}2\,
d\Omega_2(\vec{\rm n}_\perp),
\label{B_10} \\
\fl 
\int\!d\vec{\kappa}^2_\perp\,\delta(\kappa^2_0-|\sigma|-\vec{\kappa}^2_\perp)=
\theta(\kappa^2_0-|\sigma|),\;\mbox{ and: }\; 
|\vec{\kappa}_\perp|=\sqrt{\kappa^2_0-|\sigma|},
\nonumber \\
\fl 
\mbox{giving: }\;
\widetilde{\cal I}(q)= \frac{\theta(-q^2)}{4\sqrt{-q^2}}
\!\int\limits^{\infty}_{\sqrt{|\sigma|}}\!\!d\kappa_0\!
\int\limits^{2\pi}_0\!d\Omega_2(\vec{\rm n}_\perp)\,
\Phi\biggl((\kappa Q)\rightarrowtail  
\kappa^0{}_*\!Q^0-(\vec{\kappa}_\perp\!\cdot\!{}_*\!\vec{\mathrm Q}_\perp),\ldots\biggr).  
\label{B_11} 
\eea
For $\widehat{\Phi}=\exp\left\{i(rZ_\pi)-i(lZ^*_\mu)\right\}\rightarrowtail
\Phi=\exp\left\{i(qZ_\eta)\right\}\exp\left\{i(\kappa Z_-)\right\}$ with the help of 
(\ref{B_6_1}), (\ref{B_6_2}) and modified Bessel function $I_{\lambda}(r)$ \cite{b_er},   
for any complex vector $\vec{\rm R}_N\neq \vec{\rm R}^*_N$, it leads to expressions:  
\bea
\fl 
\int\! d\Omega_N(\n)\exp\{(\n\cdot\vec{\rm R}_N)\}=2\pi^{N/2}
\sum^\infty_{l=0}\frac{\left(\vec{\rm R}^2_N\right)^{l}}{2^{2l}l!\Gamma(l+N/2)} 
\label{B_12} \\
\fl 
=2\pi^{N/2}\!\left(\frac{2}{\left(\vec{\rm R}^2_N\right)^{1/2}}\right)^
{{\textstyle \frac{N}2}-1}\!\!\!
I_{{\textstyle \frac{N}2}-1}\left(\left(\vec{\rm R}^2_N\right)^{1/2}\right)
\Longrightarrow 
\left\{\!\! \begin{array}{c} \displaystyle 
4\pi\frac{\sinh \left(\vec{\rm R}^2_N\right)^{1/2}}{\left(\vec{\rm R}^2_N\right)^{1/2}},
\;\, (N=3), \\
2\pi\, I_0\left(\left(\vec{\rm R}^2_N\right)^{1/2} \right), \;\, (N=2),  
\end{array}\!\! \right. 
\label{B_13} 
\eea
with the complex vectors:
\bea
\fl 
\vec{\rm R}_3=-i\vec{\rm Z}_{*-}\sqrt{\sigma}\,
\stackunder{\vec{\rm Y}_-\to\,0}{\longmapsto}\,\vec{\zeta_{*+}}\sqrt{\sigma},\quad 
\vec{\rm R}_2=-i{}_*\!\vec{\rm Z}^\perp_-\sqrt{\kappa^2_0-|\sigma|}\,
\stackunder{\vec{\rm Y}_-\to\, 0}{\longmapsto}\,
{}_*\vec{\zeta^\perp_+}\sqrt{\kappa^2_0-|\sigma|}: 
\label{B_13_1} \\
\fl 
\left. \!\!
\begin{array}{c} \underline{\cal I}(q) \\ \\
\widetilde{\cal I}(q)
\end{array}\!\!\right\}=2\pi\frac{ e^{i(qZ_\eta)} }4 \left\{\!\!
\begin{array}{c} 
\displaystyle \Theta[\Delta,q]\,2\left[-q^2\vec{\rm Z}^2_{*-}\right]^{-1/2}
\sinh\left(\chi_\Delta\left[-q^2\vec{\rm Z}^2_{*-}\right]^{1/2}\right), \;\;  
\sqrt{\sigma}=\chi_\Delta\sqrt{q^2}, \\ \!\!\!\!\!\!
\!\!\!\displaystyle \frac{\theta(-q^2)}{\sqrt{-q^2}}
\!\int\limits^{\infty}_{\sqrt{|\sigma|}}\!\!d\kappa_0
\exp\left\{i\kappa_0{}_*\!Z^0_-\right\}
I_0\left(\left[-({}_*\!\vec{\rm Z}^\perp_-)^2\right]^{1/2}
\sqrt{\kappa^2_0-|\sigma|}\right).  
\end{array}\!\! \right.
\label{B_13_2}
\eea
By using (\ref{B_9}) together with the discussed below integral (\ref{C_0}) and its 
analytic continuations (\ref{C_0_1}), (\ref{C_0_2}), one finds the overlap function 
$\widehat{V}_{\{C\}}(q)$, if the same main branch ${\rm Re}\,W>0$ of square root in both 
pieces (\ref{B_13_2}) of $\widehat{\cal I}(q)$ is used for the value 
\bea
\fl 
W({\mathrm w})={\mathrm w}^{1/2}=
\left[q^2Z^2_- -(qZ_-)^2\right]^{1/2}\equiv\left[q^2Z^2_{\Pi-}\right]^{1/2}
\leftarrowtail \left[(-q^2)(-Z^2_{\Pi-})\right]^{1/2},  
\label{B_14_00} 
\eea
\bea
\fl 
\mbox{as: }\; 
\widehat{V}_{\{C\}}(q)=\frac{N_{\sigma\mu}N_{\sigma\pi}}{(2\pi)^2}\,
\widehat{\cal I}(q),\;\;\mbox{ with: }\;\;
\widehat{\cal I}(q)=\underline{\cal I}(q)+\widetilde{\cal I}(q)\equiv
2\pi\frac{ e^{i(qZ_\eta)} }4 
\label{B_14_0} \\
\fl 
\times
\left\{\!\Theta[\Delta,q]\,2
\frac{\sinh\left(\chi_\Delta\left[q^2Z^2_{\Pi-}\right]^{1/2}\right)}
{\left[q^2Z^2_{\Pi-}\right]^{1/2}}+\theta(-q^2)
\frac{\exp\left(\chi_\Delta\left[(-q^2)(-Z^2_{\Pi-})\right]^{1/2}\right)}
{\left[(-q^2)(-Z^2_{\Pi-})\right]^{1/2}}\right\}  
\label{B_14} \\
\fl 
= 2\pi\frac{ e^{i(qZ_\eta)} }4\,\frac{\theta(\Delta)}{W}
\biggl\{\exp\left(\chi_\Delta W\right)-\theta(q^2)
\exp\left(-\varepsilon(q^0)\chi_\Delta W\right)\biggr\}, 
\label{B_15} 
\eea
since for $g_a\to+\infty$, i.e. $\zeta^2_+,\zeta^0_+\to+\infty$, or equivalently 
$Y_-\to 0$, the argument of this square root:   
${\mathrm w}=q^2Z^2_{\Pi_q-}\mapsto {\mathrm w}_\zeta=-q^2\zeta^2_{\Pi_q+}>0$ for both 
cases $q^2\gtrless 0$, because only either $q$ or $\zeta_{\Pi_q+}$ (\ref{3_32_0}), 
is a time-like vector. Here of course $\varepsilon(q^0)={\rm sign}(q^0)$ with 
$\theta(\Delta)$ from (\ref{C_THet}). 
This essential difference of overlap function (\ref{B_1_0}) from usual two-particle phase 
volume \cite{b_k} is due to opposite sign in $\theta(-l^0)$. When $g_a\to+\infty$ only the 
first exponential in the difference (\ref{B_15}) contributes for both cases $q^2\gtrless 0$. 
Conversely, any two initial (or final) wave packets lead to (\ref{B_15}) with common factor 
(\ref{C_THet}), i.e. for $q^2>0$, $q^0>0$, but with 
$\theta(\Delta)\mapsto\theta[q^2-(m_\pi+m_\mu)^2]$ as for the usual phase volume.   

To obtain the answer (\ref{B_14}) from (\ref{B_13_2}) the formula \# 6.646.2 
\cite{g_r},  or \# 4.17(5) \cite{tit_1} is used with the main branch of square roots,  
at first for $\alpha>0$, $Re\,\gamma>|Re\,\beta|$, in the form:
\bea
&&\!\!\!\!\!\!\!\!\!\!\!\!\!\!\!\!\!\!\!\!\!\!\!
\int\limits^{\infty}_{\alpha}\!dx\,e^{-{\gamma}x}
I_0\left({\beta}\sqrt{x^2-\alpha^2}\right)=
\frac{\exp\left\{-\alpha(\gamma^2-\beta^2)^{1/2}\right\}}
{(\gamma^2-\beta^2)^{1/2}},   
\;\;\mbox{ because,}
\label{C_0} 
\eea
for time-like $\zeta_+$:  
$Re\,\gamma=Re\,(-i{}_*\!Z^0_-)={}_*\zeta^0_+>\sqrt{({}_*\vec{\zeta}^\perp_+)^2}
\geqslant|Re\,[-({}_*\!\vec{\rm Z}^\perp_-)^2]^{1/2}|=|Re\,\beta|$, when $g_a\to\infty$, 
or equivalently $Y_-\to 0$. 
Than that answer is understood in the sense of analytic continuation with respect to 
both variables $\beta$ and $\gamma$. For $\beta=\pm ib$ it is given by the formula 
\# 4.15(9) \cite{tit_1} with $\alpha>0$, $Re\,\gamma>|Im\,b|$: 
\bea
\fl 
\int\limits^{\infty}_{\alpha}\!dx\,e^{-{\gamma}x}
J_0\left(b\sqrt{x^2-\alpha^2}\right)=
\frac{\exp\left\{-\alpha(\gamma^2+b^2)^{1/2}\right\}}{(\gamma^2+b^2)^{1/2}},
\;\;\mbox{ whence, for }\; \gamma = \pm iy:   
\label{C_0_1} \\
\fl 
\int\limits^{\infty}_{\alpha}\!\!dx\,e^{\mp iyx}\!
J_0\!\left(b\sqrt{x^2-\alpha^2}\right)\!=
\theta(b-y)\frac{\exp\left(\!-\alpha\sqrt{b^2-y^2}\right)}{\sqrt{b^2-y^2}}+
\theta(y-b)\frac{\exp\left(\!\mp i\alpha\sqrt{y^2-b^2}\right)}{\pm i\sqrt{y^2-b^2}},
\nonumber \\
\fl 
\mbox{with: }\; b,y>0, \mbox{ using: }\;
(\gamma^2+b^2)^{1/2}=\theta(b-y)\sqrt{b^2-y^2}\pm i\theta(y-b)\sqrt{y^2-b^2}. 
\label{C_0_2}
\eea
The first formula (\ref{C_0_2}) may be checked independently with the help of relations 
\# 1.13(48), \# 2.13(47) \cite{tit_1}. Then, relation \# 1.13(49) \cite{tit_1} connects 
it with the used below representation (\ref{3_28_0}) in the form \# 1.13(47) \cite{tit_1}. 
The corresponding analytic continuation from the 
narrow-packet domain $g_a\to+\infty$ to the domain of localization 
$g_a\to 0$, ${}_*\zeta^0_+\to +0$, ${}_*\!Z^\lambda_-\to {}_*\!Y^\lambda_-$, 
in the chosen reference frame, can go along the pass, with 
$({}_*\vec{\zeta}^\perp_+\!\cdot\!{}_*\!\vec{\rm Y}^\perp_-)=0$, whence:   
$Im\,[\mp({}_*\!\vec{\rm Z}^\perp_-)^2]^{1/2}=0$, with 
$|Re\,\beta|=|Re\,[-({}_*\!\vec{\rm Z}^\perp_-)^2]^{1/2}|=
\sqrt{({}_*\vec{\zeta}^\perp_+)^2-({}_*\!\vec{\rm Y}^\perp_-)^2}\longmapsto
Re\,[({}_*\!\vec{\rm Z}^\perp_-)^2]^{1/2}=
\sqrt{({}_*\!\vec{\rm Y}^\perp_-)^2-({}_*\vec{\zeta}^\perp_+)^2}=Re\,b$. 

\section{Narrow-packet approximation of exact overlap function.}\label{ap:D}
To see how the plane-wave limit (\ref{3_22})--(\ref{3_23}) of function (\ref{3_24}) is 
realized, we put below $g_{2a}=0$ for $a=\pi,\mu$ and $g_a=g_{1}(m_a,\sigma_a)$, and note, 
that for definitions (\ref{B_7_0}), (\ref{3_25})--(\ref{3_27}), for $q^2\gtrless 0$, 
with $P=p_\pi+p_\mu$, $k=p_\pi-p_\mu=k_{\{C\}}$, $ \chi_0=(kP)/(2q^2)$, 
$g_\pm=g_\pi\pm g_\mu$, $\zeta^2_+,\zeta^0_+>0$, and: 
\bea
&&\!\!\!\!\!\!\!\!\!\!\!\!\!\!\!\!\!\!\!\!\!\!\!
\zeta_\pm=\frac{g_+}2\!\left\{\!\!\begin{array}{c} 
P+\varsigma k \\ k+\varsigma P \end{array}\!\!\right\}\!, \quad 
\varsigma=\frac{g_-}{g_+}, \quad 
\zeta_\eta=\frac{\zeta_-}2+\chi_0\zeta_+, \quad 
Y_\eta=\frac{Y_+}2+\chi_0 Y_-,
\label{3_35} \\
&&\!\!\!\!\!\!\!\!\!\!\!\!\!\!\!\!\!\!\!\!\!\!\!
q^2\Pi^{\lambda\nu}_q={\rm g}^{\lambda\nu}q^2-q^\lambda q^\nu, \quad 
Z^\lambda_{\Pi_q-} \equiv (\Pi_q Z_-)^\lambda=Y^\lambda_{\Pi_q-}+i\zeta^\lambda_{\Pi_q+},
\label{C_1} \\
&&\!\!\!\!\!\!\!\!\!\!\!\!\!\!\!\!\!\!\!\!\!\!\!
q^2 Z^2_{\Pi_q-}=q^2Z^2_- -(qZ_-)^2, \quad
Z^2_{\Pi_q-}=-\zeta^2_{\Pi_q+}+2i(\zeta_{\Pi_q+}Y_{-})+Y^2_{\Pi_q-},
\label{3_32} \\
&&\!\!\!\!\!\!\!\!\!\!\!\!\!\!\!\!\!\!\!\!\!\!\!
(qY_{\Pi_q-})=(q\zeta_{\Pi_q+})=0, \quad 
\zeta^2_{\Pi_q+}\equiv(\zeta_+\Pi_q\zeta_+) \lessgtr 0, \quad 
-q^2\zeta^2_{\Pi_q+}>0,
\label{3_32_0} \\
&&\!\!\!\!\!\!\!\!\!\!\!\!\!\!\!\!\!\!\!\!\!\!\!
{\cal N}_{\Pi_q}=\frac{\zeta_{\Pi_q+}}{[-q^2 \zeta^2_{\Pi_q+}]^{1/2}},\quad  
q^2{\cal N}^2_{\Pi_q}=-1, \quad  (q{\cal N}_{\Pi_q})=0,\;\mbox{ for }\;
g_{a},g_+ \to+\infty:  
\label{3_34} \\
&&\!\!\!\!\!\!\!\!\!\!\!\!\!\!\!\!\!\!\!\!\!\!\!
W=\left[q^2Z^2_{\Pi_q-}\right]^{1/2}\approx
\left[-q^2 \zeta^2_{\Pi_q+}\right]^{1/2}\!+i\,({\cal N}_{\Pi_q}Y_-)q^2\!+
\frac{Y^2_{\Pi_q-}+q^2({\cal N}_{\Pi_q}Y_{-})^2}
{2\left[-q^2 \zeta^2_{\Pi_q+}\right]^{1/2}}\,q^2,    
\label{3_33} 
\eea
and from (\ref{B_14_0})--(\ref{B_15}), the narrow-wave-packet approximation for overlap 
function (\ref{3_24}), for $\Delta=\Delta(q)$, follows as: 
\bea
\fl 
\widehat{V}_{\{C\}}(q)\approx  
(2\pi)^4\,\theta(\Delta)
\left[\frac{m^2_\pi m^2_\mu(g_\pi g_\mu)^3}{[-q^2\zeta^2_{\Pi_q+}](2\pi)^4}\right]^{1/2}
\!\! e^{i\Phi(q)} e^{{\rm L}(q)}e^{\Sigma_q(Y_-)},\mbox{ with: }
\chi_\Delta=\frac{\Delta^{1/2}(q)}{2q^2}, 
\label{3_36} \\
\fl 
2\Phi(q)=(qY_+)+\left(Q(q)Y_-\right)\!, \;\;
Q(q)\!=2\!\left[q \chi_0+{\cal N}_\Pi q^2\chi_\Delta\right]\!, \;\;
Q^2(q)\!=2(m^2_\pi+m^2_\mu)-q^2\!, 
\label{3_37} \\
\fl 
\Phi(k)=\Phi_{\{C\}}(k)=(p_\pi Y_\pi)-(p_\mu Y_\mu), \quad Q(k)=P,\quad 
2Y^\lambda_{\{C\}}=Y^\lambda_+ +\partial^\lambda_q(Q(q)Y_-)\bigr|_{q=k}, 
\label{3_37_A} \\
\fl 
{\rm L}(q)=
g_\pi m^2_\pi +g_\mu m^2_\mu -(q\zeta_\eta)+
\chi_\Delta [-q^2 \zeta^2_{\Pi_q+}]^{1/2}\!\!, 
\quad {\rm L}(k)=0, \quad \partial^\lambda_q{\rm L}(q)\bigr|_{q=k}\!\!=0,
\label{3_38} \\
\fl 
\mbox {where for the extremum point $q\mapsto k$: } \Delta(q)\mapsto
\Delta(k)=-k^2 P^2_{\Pi_k}=(kP)^2-k^2P^2,
\label{3_38_D} \\
\fl 
\Sigma_q(Y_-)=
\frac{Y^2_{\Pi-}+q^2({\cal N}_{\Pi}Y_{-})^2}{2[-q^2\zeta^2_{\Pi+}]^{1/2}}\,q^2
\chi_\Delta \stackunder{q\to k}{\longmapsto}
\Sigma_k(Y_-)=\frac{Y^2_{\Pi-}+k^2({\cal N}_{\Pi}Y_{-})^2}{2g_+},\;\;
\Pi_q\!\stackunder{q\to k}{\longrightarrow} \Pi_k,
\label{3_38_S} \\
\fl 
\mbox {and for: }\, 
\Pi_k P\!= P_{\Pi_k},\;\;\; \zeta_{\Pi_q\pm}\!\mapsto\!\frac{g_\pm}2 P_{\Pi_k},\;\;\;
{\cal N}_{\Pi_q} \mapsto\!\frac{P_{\Pi_k}}{\sqrt{\Delta(k)}}, \;\;\;
\kappa=q-k,\;\;\; u_a\!=\frac{p_a}{m_a}\!,
\label{3_38_0} \\
\fl 
\mbox{one has: }\;
\Phi(q)\approx \Phi(k)+\left(\kappa Y_{\{C\}}\right),\quad 
{\rm L}(q)\approx -(\kappa\vec{\cal T}\kappa), \quad 
{\cal T}^{\beta\lambda}=
-\,\frac 12\partial^\beta_q\partial^\lambda_q{\rm L}(q)\bigr|_{q=k}, 
\label{3_38_00} \\
\fl 
\widehat{V}_{\{C\}}(q)\approx  
(2\pi)^4\,\theta(\Delta(q))\, e^{i\Phi(k)}\,e^{\Sigma_k(Y_-)}\,
\left[\frac{(g_\pi g_\mu)^3\,(2\pi)^{-4}}{g_+^2[(u_\pi u_\mu)^2-1]}\right]^{1/2} 
e^{i\left(\kappa Y_{\{C\}}\right)}e^{-(\kappa\vec{\cal T}\kappa)}, 
\label{3_40} \\
\fl 
\mbox{with: }\;|\vec{\cal T}|\equiv \det\{{\cal T}^{\beta\lambda}\}=
\frac{(g_\pi g_\mu)^3\,2^{-4}}{g_+^2[(u_\pi u_\mu)^2-1]}, \quad 
\lim_{{\cal T}\to +\infty}
\left[\frac{|\vec{\cal T}|}{\pi^{4}}\right]^{1/2} e^{-(\kappa\vec{\cal T}\kappa)}
=\delta_4(\kappa),
\label{3_41} \\
\fl 
\mbox{for: }
\frac{4}{g_+}{\cal T}^{\beta\lambda}=
\frac{k^{\beta}k^{\lambda}}{k^2}{\rm T}_{0}+
\widehat{\Pi}^{\beta\lambda}_\perp {\rm T}_{\rm d}- 
\frac{P_{\Pi}^{\beta}P_{\Pi}^{\lambda}}{P_{\Pi}^2}{\rm T}_{\rm dd}-
\frac{k^{\beta}P_{\Pi}^{\lambda}+k^{\lambda}P_{\Pi}^{\beta}}{P_{\Pi}^2}
{\rm T}_{0\rm d},
\label{3_39} \\
\fl 
\mbox{with: }\;
\widehat{\Pi}^{\beta\lambda}_\perp=
\frac{P_{\Pi_k}^{\beta}P_{\Pi_k}^{\lambda}}{P_{\Pi_k}^2}-\Pi^{\beta\lambda}_k\equiv 
-(\Pi_{P_{\Pi}}\Pi_k)^{\beta\lambda},\;\mbox{ and where for: }\;
\left.\!\!\begin{array}{c}a \\ b\end{array}\!\!\right\}=\frac{m^2_\pi \pm m^2_\mu}{k^2}, 
\label{3_39_0} 
\eea
since $k^2(a-b)>0$, there are such invariant functions:
\bea
\fl 
{\rm T}_{0}=
\frac{(k^2)^2}{\Delta(k)}\left[(1+b^2)(a+\varsigma b)-2b(b+\varsigma a)\right], 
\quad {\rm T}_{\rm d}=\frac{(1-\varsigma^2)}{2}>0,
\label{3_43} \\
\fl 
{\rm T}_{\rm dd}=(a+\varsigma b),\quad {\rm T}_{0\rm d}=b(a+\varsigma b)-(b+\varsigma a),
\;\mbox{ that }\;{\rm T}_{0},\,{\rm T}_{\rm dd}\gtrless 0, \;\mbox{ with }\;k^2\gtrless 0, 
\label{3_44} \\
\fl 
\Delta(k)=
-k^2 P^2_{\Pi_k}=4m^2_\pi m^2_\mu\left[\left(u_\pi u_\mu\right)^2-1\right]\longmapsto 
\left\{\!\!
\begin{array}{c} 
4m^2_\pi m^2_\mu\left(\vs_\mu-\vs_\pi\right)^2,\;\;|\p_a|\ll m_a, \\
4\p^2_\pi\p^2_\mu\left[1-(\n_{\pi}\!\cdot\n_{\mu})\right]^2\!,\;\; m_a\to 0.
\end{array}\right.
\label{C_00} 
\eea
Because for $k^2\gtrless 0$ it follows $P^2_{\Pi_k}\lessgtr 0$, one has at the respective 
rest frames (*) of vector $k$ {\sl or} vector $P_{\Pi}$: 
$k^\lambda_*=(\sqrt{k^2},\vec{0})$, $P^\lambda_{*\Pi}=(0,\vec{0}_\perp,{\rm P}^3_{*\Pi})$ 
with $({\rm P}^3_{*\Pi})^2=\vec{\rm P}^2_{*\Pi}=-P^2_{\Pi}=-(P\Pi_k P)>0$ for 
$k^2>0$, {\sl or} ${}_*\!P^\lambda_{\Pi}=(\sqrt{P^2_{\Pi}},\vec{0})$, 
${}_*\!k^\lambda=(0,\vec{0}_\perp,{}_*\!{\rm k}^3)$ for  
$({}_*\!{\rm k}^3)^2={}_*\!\vec{\rm k}^2=-k^2>0$ otherwise, and respectively: 
\bea
&&\!\!\!\!\!\!\!\!\!\!\!\!\!\!\!\!\!\!\!\!\!\!\!
\frac{4}{g_+}{\cal T}^{00}\overset{*}{=}{\rm T}_{0}\, ,\;
\mbox{ {\sl or} }\;\overset{*}{=} -{\rm T}_{\rm dd}\,;
\qquad 
\frac{4}{g_+}{\cal T}^{33}\overset{*}{=}{\rm T}_{\rm dd}\, ,\;
\mbox{ {\sl or} } \overset{*}{=} -{\rm T}_{0}\,;
\label{3_43_1} \\
&&\!\!\!\!\!\!\!\!\!\!\!\!\!\!\!\!\!\!\!\!\!\!\!
\frac{4}{g_+} {\cal T}^{11}\overset{*}{=}\frac{4}{g_+} {\cal T}^{22}
\overset{*}{=}{\rm T}_{\rm d}\,; \qquad 
\frac{4}{g_+} {\cal T}^{03}=\frac{4}{g_+}{\cal T}^{30}\overset{*}{=}
\frac{k^2}{\sqrt{\Delta(k)}}{\rm T}_{0\rm d}\,; 
\label{3_44_1} 
\eea
where for {\sl both} cases, with (\ref{3_39_0}):   
$\widehat{\Pi}^{\beta\lambda}_\perp\overset{*}{\mapsto}\delta^{\beta\lambda}_\perp$, 
i.e. it is $\neq 0$ for $\beta, \lambda=1,2$ only. Thus      
\bea
\fl 
|\vec{\cal T}|=
\varepsilon_{\beta\lambda\nu\sigma}{\cal T}^{0\beta}
{\cal T}^{1\lambda}{\cal T}^{2\nu}{\cal T}^{3\sigma}
\overset{*}{=}\left|\begin{array}{cccc}
{\cal T}^{00} & 0 & 0 & {\cal T}^{03} \\ 0 & {\cal T}^{11} & 0 & 0 \\
0 & 0 & {\cal T}^{22} & 0 \\ {\cal T}^{30} & 0 & 0 & {\cal T}^{33} \end{array}\right|
\overset{*}{=} {\cal T}^{11}{\cal T}^{22}
\left[{\cal T}^{00}{\cal T}^{33}-{\cal T}^{2}_{03}\right], 
\label{3_42} 
\eea
for {\sl both} cases $k^2\gtrless 0$ is reduced to the same value (\ref{3_41}). So, 
the Lorentz invariant form $(\kappa\vec{\cal T}\kappa)$ is strongly positively defined, 
since ${\cal T}^{00},\,{\cal T}^{11},\,{\cal T}^{22},\,{\cal T}^{33},\,|{\cal T}|>0$ 
for both cases.  

\section{One-packet representation of two-packet state}\label{sec:one_p_rep}

To trace the another limiting properties of the on-shell composite wave function 
(\ref{3_20_0}) -- (\ref{3_20_1}) it is convenient to use for the last multipliers of 
  (\ref{3_24}) the integral representation \# 6.677.6 \cite{g_r} (or \# 1.13(47) 
\cite{tit_1}), which factorizes its different $q$- dependences for $q^2>0$, 
$q^0>0$, $\Delta(q)>0$, $\chi_\Delta>0$ via Bessel function $J_0(r)$, with the main 
branch of square roots: 
\bea
\fl  
\frac{\sin\left({\alpha}\sqrt{{\beta}^2+y^2}\right)}
{\sqrt{{\beta}^2+y^2}}=\frac 12\! \int\limits^{\alpha}_{-\alpha}\!dx\,e^{iyx}
J_0\left({\beta}\sqrt{{\alpha}^2-x^2}\right),\,
\mbox{ for arbitrary }\,\beta,y\,\mbox{ and }\,{\alpha}>0,
\label{3_28_0} \\
\fl  
\frac{\sinh\left(\chi_\Delta \left[q^2Z^2_- -(qZ_-)^2\right]^{1/2}\right)}
{\left[q^2Z^2_- -(qZ_-)^2\right]^{1/2}}=
\frac{1}{2}\!\!\int\limits^{\chi_\Delta}_{-\chi_\Delta}\!\! d\chi\,e^{i\chi(qZ_-)}
J_0\left(\left[-q^2Z^2_-\right]^{1/2} \left(\chi^2_\Delta-\chi^2\right)^{1/2}\right).  
\label{3_28}
\eea
Therefore the function (\ref{3_20_1}) with $q^2=m^2_j$ becomes Lorentz covariant linear 
superposition of interpolating wave packets (\ref{2_18}), (\ref{3_19_0}) with the same 
fixed mass $m_j$ but with various centers $Y_\chi$, various on-shell momentums 
$p^2_{\chi j}= m^2_j$, $p_{\chi j}=m_j\upsilon_\chi$, 
$\upsilon_\chi=\zeta_{\chi}/\sqrt{\zeta^2_{\chi}}$, and various {\sl invariant widths}  
$\sigma_{\chi j}$, defined by e.g. $g_{\chi j}\mapsto 1/\sigma^2_{\chi j}$, with 
$\tau^2_{\chi j}=m^2_j\zeta^2_\chi=m^4_j g^2_{\chi j}$:
\bea
\fl  
{\cal G}_{\{C\}j}(\varrho)= i\frac{N_{\sigma\mu}N_{\sigma\pi}}{8\pi}\!\!
\int\limits^{\chi_\Delta}_{-\chi_\Delta}\!\!\frac{d\chi}{N_{\chi j}}\,
J_0\left(\left[-m^2_jZ^2_-\right]^{1/2} \left(\chi^2_\Delta-\chi^2\right)^{1/2}\right)
\frac {N_{\chi j}}{i} D^{-}_{m_j}(\varrho-Z_\chi),  
\label{3_29} \\
\fl  
\frac {N_{\chi j}}i D^{-}_{m_j}(\varrho-Z_\chi)=  
\psi_{\sigma_{\chi j}}(\p_{\chi j},Y_\chi-\varrho)=
e^{i\left(p_{\chi j}Y_\chi\right)} F_{p_{\chi j}Y_\chi}(\varrho) 
\stackunder{\tau_{\chi j}\to\infty}{\longrightarrow}
e^{i\left(p_{\chi j}(Y_\chi-\varrho)\right)}, 
\label{3_30_1} \\
\fl  
\mbox {normalized by: }\; N_{\chi j}=
\frac{(2\pi)^2}{m^2_j}\,\frac{{\cal I}(\tau_{\chi j})}{h(\tau^2_{\chi j})}
\stackunder{\tau_{\chi j} \to\infty}{\longrightarrow}
\frac{2}{m^2_j}\,(2\pi)^{3/2}\tau^{3/2}_{\chi j}\exp\left\{\tau_{\chi j}\right\},\,
\label{3_30_0} \\
\fl  
Z_\chi=Z_\eta+\chi Z_-=\eta_+Z_\pi+\eta_-Z^*_\mu =Y_\chi+i\zeta_\chi, \quad 
\eta_{\pm}=\eta_{\pm}(\chi)=\frac 12\pm(\chi_0+\chi), 
\label{3_30} \\
\fl  
Y_\chi=\eta_+Y_\pi+\eta_-Y_\mu=\frac{Y_+}2+(\chi_0+\chi)Y_-, \quad 
\zeta_\chi=\eta_+\zeta_\pi-\eta_-\zeta_\mu=\frac{\zeta_-}2+(\chi_0+\chi)\zeta_+,
\label{3_30_00} 
\eea
where for $g_{2a}=0$ vector:  
\bea
&&\!\!\!\!\!\!\!\!\!\!\!\!\!\!\!\!\!\!\!
\zeta_\chi\!= 
\frac{g_+}{2}\!\left\{\!P\!\left[\frac{\varsigma}{2}+(\chi_0+\chi)\right]\!
+k\!\left[\frac{1}{2}+\varsigma(\chi_0+\chi)\right]\!\right\},  
\label{3_30_2} 
\eea
and $\tau_{\chi j} \to\infty$ means $g_+\to \infty$. The opposite limit, with independent 
localizations of pion and muon, implies $\sigma_\pi,\sigma_\mu \to\infty$, so, the values 
$g_{1,2\pi},g_{1,2\mu},\zeta_\pi,\zeta_\mu,\zeta_\chi \to 0$, i.e. $\tau_{\chi j}\to 0$. 
This transforms (\ref{3_29}) into superposition of localized packets 
(\ref{eq:limit_sigma_inf}) with different centers only: 
\bea
\fl  
{\cal G}_{\{C\}j}(\varrho)
\stackunder{\sigma_{a}\to\infty}{\longmapsto}
\frac{i}{8\pi}\,\frac{(\aleph(0))^2}{m^2_\pi m^2_\mu}
\!\!\int\limits^{\chi_\Delta}_{-\chi_\Delta}\!\!d\chi\,
J_0\left(\left[-m^2_jY^2_-\right]^{1/2}\left(\chi^2_\Delta-\chi^2\right)^{1/2}\right)
\frac {1}i D^{-}_{m_j}(\varrho-Y_\chi).  
\label{3_29_0}
\eea
Coincidence of centers of $\pi$ and $\mu$ packets in the point $Y_\pi$ leads to localization 
(\ref{eq:limit_sigma_inf}) in this point $Y_\chi \to Y_\pi$ ($Y_-\to 0$), for the full 
on-shell composite wave packet as a single-packet function:   
\bea
\fl 
{\cal G}^{[\sigma_a=\infty]}_{\{C\}j}(\varrho)
\stackunder{Y_-\to\,0}{\longrightarrow} 
\frac{(\aleph(0))^2}{m^2_\pi m^2_\mu}\,
\frac{\chi_\Delta}{4\pi}\,D^{-}_{m_j}(\varrho-Y_\pi)= 
i\,\frac{\aleph(0)}{8\pi}\,\frac{\Delta^{1/2}(m_j)}{m^2_\pi m^2_\mu}\,
\psi_\infty(\p_j,Y_\pi-\varrho).  
\label{3_29_00} 
\eea 
Its further zero-mass limit $m_j\to 0$ repeats (\ref{D_4_0}) and also implies the change 
of normalization. So, the above obtained limiting properties of on-shell composite wave 
function for intermediate neutrino wave packets (\ref{3_20_0}), or (\ref{3_20_G}), 
(\ref{3_20_0G}), accurately replicate these properties following external interpolating 
wave packets, as given by  (\ref{eq:limit_sigma_0}), (\ref{eq:limit_sigma_inf}), 
(\ref{D_4_0}). However some differences appear for another cases. To consider at first 
the limit $m_j\to 0$, we note that the 4-vector (\ref{3_30_2}) is reduced to  
$\zeta_\chi\mapsto(g_+/4)k(1-\varsigma^2)$ only for $\chi_0+\chi\mapsto -\varsigma/2$. 
For $m_j\geqslant 0$ this requires a value of $\chi$, which is unachievable for the 
integrand in (\ref{3_29}). Indeed, with substitution $\chi=\gamma-\chi_\Delta$, 
$\gamma>0$ one has $\chi_0+\chi=\gamma+\overline{\epsilon}$, where, for $m_j\to 0$:   
\bea      
&&\!\!\!\!\!\!\!\!\!\!\!\!\!\!\!\!\!\!\!\!\!\!\!
\overline{\epsilon}\equiv \chi_0-\chi_\Delta\longmapsto \frac 12 
\left(\frac{m^2_\pi+m^2_\mu}{m^2_\pi-m^2_\mu}\right)+O(m^2_j)\approx 1,79, \;\;\;\;
\chi_0,\,\chi_\Delta\to +\infty. 
\label{3_30_3} 
\eea
With the replacing $\chi_0+\chi=\gamma+\overline{\epsilon}$ for $Z_\chi=Z_{(\gamma)}$ 
and with respective redefinition of (\ref{3_30}) -- (\ref{3_30_2}), (\ref{D_4_0}),  
the finite limit of expression (\ref{3_29}) reads: 
\bea
\fl 
{\cal G}_{\{C\}j}(\varrho)\biggr|_{m_j=0}=
\frac{N_{\sigma\mu}N_{\sigma\pi}}{8\pi}\!\int\limits^{\infty}_{0}\!d\gamma\,
J_0\left(\left[-Z^2_-(m^2_\pi-m^2_\mu)\right]^{1/2}\sqrt{\gamma}\right)
D^{-}_{0}(\varrho-Z_{(\gamma)}).   
\label{3_301} 
\eea
Unlike the single-packet case (\ref{D_4_0}), such superposition of these massless 
solutions to KG equation arises without any change of normalization. The integral exists 
for $(Y_-\zeta_+)=0$, when $-Z^2_-\mapsto \zeta^2_+-Y^2_->0$. For the rest frame of 
time-like vector $\zeta^\lambda_+\mapsto(\zeta^0_+,\vec{0})$, that means arbitrary 
$Y^\lambda_-=(0,\vec{\rm Y}_-)$. 
The use of formula \# 6.532.4 \cite{g_r} (or \# 7.14.(58) \cite{b_er}) reduces this to 
the form admitting analytic continuation to arbitrary $Y_-$. Since for $a>0$, $Re\,\beta>0$: 
\bea
&&\!\!\!\!\!\!\!\!\!\!\!\!\!\!\!\!\!\!\!\!\!\!\!
\int\limits^\infty_0\frac {t dt J_0(at)}{t^2+\beta^2}=K_0(a\beta),\;\;\mbox{ then, for }
\;\;{\cal R}=\varrho-\frac{Z_+}2, \quad Z_{\mp}=Y_{\mp}+i\zeta_{\pm},  
\label{3_3_0_3} \\ 
&&\!\!\!\!\!\!\!\!\!\!\!\!\!\!\!\!\!\!\!\!\!\!\!
{\cal L}=\left[(Z_-{\cal R})^2-Z^2_-{\cal R}^2\right]^{1/2}, \quad 
\Lambda_{\pm}=
\left\{(m^2_\pi-m^2_\mu)\left[(Z_-{\cal R})\pm{\cal L}-Z^2_-\overline{\epsilon}
\right]\right\}^{1/2}, 
\label{3_3_0_2} \\
&&\!\!\!\!\!\!\!\!\!\!\!\!\!\!\!\!\!\!\!\!\!\!\!
\mbox{one obtains: }\;
{\cal G}_{\{C\}j}(\varrho)\biggr|_{m_j=0}=
\frac{N_{\sigma\mu}N_{\sigma\pi}}{4(2\pi)^3\,i}\!\left[
\frac{K_0(\Lambda_+)-K_0(\Lambda_-)}{{\cal L}}\right], 
\label{3_3_0_1} 
\eea
which looks similarly finite-difference analog of relation (\ref{E_4}) with N=2, L=1, 
for the step, proportional in (\ref{3_3_0_2}) to the value of 
${\cal L}=[-Z^2_-({\cal R}\Pi_{Z_-}{\cal R})]^{1/2}$ for small ${\cal L}$. 



\section*{References}
\begin{thebibliography}{99}



\bibitem{nn} 
   	Naumov D V and Naumov V A 2010
   	{\it J. Phys.\ G: Nucl.\ Part.\ Phys.} 
	{\bf 37}  105014  
   	(arXiv:hep-ph/1008.0306v2)

\item[]  
  	Naumov D V  and Naumov V A 2010
 	{\it Russ.\ Phys.\ J.}
 	{\bf 53} 549

\bibitem{nsh} 
   	Naumov V A and Shkirmanov D S  2015
 	{\it Mod.\ Phys.\ Lett.}
 	{\bf A 30}, No. 24 1550110.  
 	(arXiv: 1409.4669v2 [hep-ph]) 


\bibitem{dn} 
   	Naumov D V 2013
  	{\it Phys.\ Part.\ Nucl.\ Lett.} 
  	{\bf 10}  642-650
   	(arXiv:hep-ph/1309.1717)

\bibitem{n_sh} 
   	Naumov V A  and Shkirmanov D S  2013
	{\it Eur.\ Phys.\ J.} {\bf C 73}   2627


\bibitem{NN_shk} 
	Naumov D V , Naumov V A and  Shkirmanov D S 2017
	{\it Phys.\ Part.\ Nucl.} 
	{\bf 48} No 1  12-20 
	(arXiv: 1507.04573 [hep-ph])  


 \item[] 
	Naumov D V, Naumov V A and  Shkirmanov D S 2017 
	{\it Phys.\ Part.\ Nucl.} 
	{\bf 48} No 6  1007-1010 


\bibitem{kt_1} 
	Korenblit S E  and Taychenachev D V  2015
	{\it Mod.\ Phys.\ Let.}
	{\bf A 30} No 14  1550074
	(arXiv: 1401.4031v4 [math-ph]) 

\bibitem{LL}
  	Landau L D and Lifshitz E M 1977
	{\it Quantum Mechanics}
	(Pergamon Press Ltd.) 

\bibitem{Mess}
	Messiah A 1961
	{\it Quantum Mechanics}
	(John Wiley \& Sons, New York) 

\bibitem{Tl} 
	Taylor J R 1972
	{\it Scattering Theory}, 
	(John Wiley \& Sons, New York)  

\bibitem{g_w}
	Goldberger M and Watson K  1964
	{\it Collision theory}
	(Wiley, New York) 

\bibitem{feyn}
	Feynman R P 1961
	{\it Quantum Electrodynamics}
	(W A Benjamin, New York) 


\bibitem{blp}
	Beresteckii V B, Lifshitz  E M and Pitaevskii L P 1982
	{\it Quantum Electrodynamics},
	(Butterworth-Heinemann) 

\bibitem{tir}
	Thirring  W E 1958
	{\it Principles of Quantum Electrodynamics}
	(Academic Press, New York)

\bibitem{p_s} 
	Peskin M E and Schroeder D V 1995
	{\it An Introduction to quantum field theory}
	(Addison-Wesley Publishing Company)

\bibitem{vrg} 
	Vergeles S N 2006
	{\it Lectures on Quantum Electrodynamics}
	(M. Fizmatlit)   

\bibitem{schw}
	Schweber S S 1961
	{\it An Introduction to Relativistic Quantum Field Theory}
	(Row, Peterson and Company, Evanston)

\bibitem{blt}
	Bogoliubov N N, Logunov A A and Todorov I T 1969
	{\it Foundations of axiomatic approach in quantum field theory},
	(M. Nauka) 

\bibitem{oksak}
	Bogoliubov N N,  Logunov  A A, Oksak A I and Todorov I T 1990
	{\it General Principles of Quantum Field Theory},
	(Kluwer, Dordrecht et al) 

\bibitem{jost} 
	Jost R 1965
	{\it The General Theory of Quantized Fields}
	(American Matematical Society, Providence)


\bibitem{s_w}
	Streater  R F and Wightman A S  2000
	{\it PCT, Spin and Statistic and all That}, 	 
	(Princeton University Press)

\bibitem{stroc}
	Strocchi F  2013 
	{\it An Introduction to Non-Perturbative Foundations of Quantum Field Theory}, 
	(Oxford Univ. Press)

\bibitem{Dvt}
	DeWitt B S  1965
	{\it Dynamical Theory of Groups and Fields}, 
	(Gordon and Breach, New York)

\bibitem{b_d}
	Bjorken  J D and  Drell  S D 1978
	{\it Relativistic Quantum Theoty, Vol. 1,2}, 
	(Mc Graw-Hill Book Comp.)  

\bibitem{i_z} 
	Itzykson C and Zuber J-B 1978
	{\it Quantum Field Theory, Vol. 1,2},  
	(Mcgraw-Hill Int. Book Comp.)   

\bibitem{Nvj} 
	Novozhilov Yu V 1975
	{\it Introduction to the theory of elementary particles},  
	(Pergamon Press, Oxford-New York-Toronto)

\bibitem{Bil} 
	Bilenky S M  1974 
	{\it Introduction to the Feynman's diagrams technique}, 
	(Pergamon)  

\bibitem{Wnb} 
	Weinberg S  1995
	{\it The Quantum Theory of Fields, Vol 1} 
	(Cambridge University Press) 


\bibitem{kil}
	Kilin S Ya 1995
	{\it Quantum Optics: Fields and Their Detection}
	(Inst of Physics Pub Inc) 

\bibitem{bern}
	Bernardini A E and  De Leo S 2005
	{\it Phys.\ Rev.} 	{\bf D 71}  076008 	(arXiv:hep-ph/0504239) 

\bibitem{gunt}
	Giunti C 2002
	{\it JHEP} {\bf 11} 017

\bibitem{rich}
	Rich J 1993
	{\it Phys.\ Rev.} {\bf D 48} 4318-4325

\bibitem{bth}
	Beuthe M 2003
	{\it Phys.\ Rep.} {\bf 375} 105-218
	(arXiv:hep-ph/0109119)  
 \item[] 
    Beuthe M 2002
    {\it Phys. Rev.} {\bf D 66} 013003 (arXiv: 0202068v2 [hep-ph])

\bibitem{ahm} 
	Akhmedov E Kh and Kopp J 2010
	{\it JHEP} {\bf 04}  008 
	(arXiv:1001.4815 [hep-ph])

\item[]
  Akhmedov E Kh and Smirnov A Yu 2009
  {\it Yad.\ Fiz.} {\bf 72} 1417 [{\it Phys.\ Atom.\ Nucl.} {\bf 72} 1363]
  (arXiv:0905.1903 [hep-ph])



\bibitem{Dl_1}
	Dolgov  A D, Okun L B, Rotaev M V and Schepkin M G  2004
	{\it Oscillations of neutrinos produced by a beam of electrons}, 
	(arXiv: hep-ph/0407189v2)
 
\item[] 
	Dolgov  A D, Lychkovskiy O V,  Mamonov A A, Okun L B, Rotaev M V  and  Schepkin M G  2005
	{\it Nucl.\ Phys.}  {\bf B 729} 79
	(arXiv: hep-ph/0505251)

\bibitem{Nis} 
	Nishi C C 2006
	{\it Phys.\ Rev.} {\bf D 73} 053013 
	(arXiv: hep-ph/0506109v3)

\bibitem{kbz} 
	Kobzarev I Yu, Martemyanov  B V,  Okun L B and Shchepkin M G 1982 
	{\it Sov.\ J.\ Nucl.\ Phys.} {\bf 35}  708 
	[Yad. Fiz. {\bf 35} 1210 (1982)]

\bibitem{GS}
	Grimus W and Stockinger P 1996
  	{\it Phys.\ Rev.}  {\bf D 54} 3414
  	(arXiv:hep-ph/9603430)

\bibitem{NN_anp}
	Naumov V A 2016  
	\emph{Neutrinos in Physics and Astrophysics}
	(Lectures given in Moscow Institute of Physics and Technology (State University) 
    and in JINR University Center in 2007-2016) Dubna JINR Fall Term 2016
	(http://theor.jinr.ru/~vnaumov/Eng/JINR\_Lectures/Lectures\_files/NPA2017.pdf)

\bibitem{NN_20}
    Naumov D V, Naumov  V A 2020
    {\it Phys.\ Part.\ Nucl.} 
    {\bf 51} No 1  1--106.

\bibitem{Card}
	Cardall C Y and  Chung D J H 1999 
	{\it Phys.\ Rev.} {\bf D 60}  073012
	(arXiv: 9904291[hep-ph])

\bibitem{bGG}
	Bilenky S M,  Giunti C and Grimus W  1999
	{\it Prog.\ Part.\ Nucl.\ Phys.} {\bf 43} 86 
	(arXiv: hep-ph/9812360 v4) 

\bibitem{Dvor}
	Dvornikov M S  2011
	{\it Field theory description of neutrino oscillations, in Neutrinos: Properties, 
    Sources and Detection},
 	(Nova Science Publishers, New York) 
	(arXiv: 1011.4300 v2 [hep-ph]) 



\bibitem{bl1} 
	Blasone M, Henning P A and Vitiello G 1999
	{\it Phys. Lett.} {\bf B 451} 140-145
	(arXiv:hep-th/9803157)

\bibitem{bl2}
	Blasone M, Pacheco P P and Tseung  H W C 2003
	{\it Phys.\ Rev.} {\bf D 67} 073011

\bibitem{bl3} 
	Fujii K, Habe C, and Yabuki T 2001
	{\it Phys.\ Rev.} {\bf D 64} 013011

\bibitem{fujii}  
	Fujii K and Toyota  N 2015
	{\it Expectation values of flavour-neutrino numbers with respect to neutrino-source 
     hadron states}, 
	(arXiv: 1408.1518v2 [hep-ph])

\bibitem{al_h_w} 
     Al-Hashimi M H, Wiese U-J  2009
     {\it Ann. \ Phys.} {\bf 324} 2599  (arXiv: 0907.5178 [quant-ph])

\bibitem{b_k} 
     Byckling E and Kajantie  K 1973
	{\it Particle Kinematics}, 
	(John Wiley \& Sons, London)

\bibitem{b_er} 
	Bateman G and Erdelyi A 1953
	{\it Higher transcendential functions. Vol 2}, 
	(McGraw-Hill, New York) 

\bibitem{g_r}
	Gradshteyn I S and Ryzhik I M 2007
	{\it Tables of Integrals, Series, and Products} 7th edition, ? 
	(Academic Press, San Diego U.S.A.)

\bibitem{tit_1}  
	Bateman G and Erdelyi A 1954
	{\it Tables of integral transforms. Vol. 1}, 
	(New York)  

\bibitem{g_sh} 
	Gel'fand I M and  Shilov G E 1959
	{\it Generalized Functions. Vol. 1}
	(Academic Press, New York-London)

\bibitem{kt_3}
	Korenblit S E and  Taychenachev D V 2013
	{\it Phys.\ Part.\ Nucl.\ Lett.}
	{\bf 10} No. 7  634. 
	(arXiv: 1304.5192v1 [hep-th]) 

\bibitem{J_S}
	Jacob R and Sachs R G 1961
	{\it Phys.\ Rev.} {\bf 121} 350   

\bibitem{anom}  
	Mention G  et al 2011
	{\it Phys.\ Rev.} {\bf D 83}  073006
	(arXiv: 1101.2755 [hep-ex])

\end{thebibliography}
\end{document}